\documentclass{aa}  
\usepackage{graphicx}
\usepackage{txfonts}
\usepackage{natbib}
\usepackage{hyperref}
\usepackage{{booktabs}}

\usepackage{xcolor,colortbl}
\usepackage{enumitem}

\defcitealias{van_beeck_2021}{VB21}
\defcitealias{pedersen_2021}{PD21}

\definecolor{Gray}{gray}{0.85}
\definecolor{LightCyan}{rgb}{0.88,1,1}

\newcolumntype{a}{>{\columncolor{LightCyan}}c}
\newcolumntype{b}{>{\columncolor{Gray}}c}

\newcommand\orc[1]{\href{https://orcid.org/#1}{\includegraphics[width=3mm]{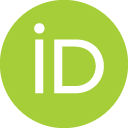}}}

\begin{document}

   \title{Beyond prewhitening: detection of gravity modes and their period spacings in slowly pulsating B stars using the multitaper F-test}

   \author{Aarya A. Patil
          \inst{1}\fnmsep\thanks{LSST-DA Catalyst Fellow}\orc{0000-0002-7626-506X}
          \and
          Conny Aerts\inst{1,2,3}\orc{0000-0003-1822-7126}
          \and
          Nikki Y. N. Wang\inst{1}
          \and
          Jordan Van Beeck\orc{0000-0002-5082-3887}\inst{4}
          \and
          May G. Pedersen\orc{0000-0002-7950-0061}\inst{5,6}
          }

   \institute{Max-Planck-Institut für Astronomie, Königstuhl 17, D-69117 
              Heidelberg, Germany\\
              \email{patil@mpia.de}
              \and
              Institute of Astronomy, Department of Physics \& Astronomy, KU Leuven, Celestijnenlaan 200 D, 3001 Leuven, Belgium\\ \email{conny.aerts@kuleuven.be}
              \and
              Department of Astrophysics, IMAPP, Radboud University Nijmegen, PO Box 9010, 6500 GL Nijmegen, The Netherlands
              \and
              Heidelberg Institute for Theoretical Studies, Schloss-Wolfsbrunnenweg 35, 69118 Heidelberg, Germany
              \and
              Sydney Institute for Astronomy, School of Physics, University of Sydney, Sydney, NSW 2006, Australia
             }

   \date{}
 
  \abstract
   {Gravity modes in main-sequence stars have traditionally been studied using a prewhitening approach, which iteratively identifies modes in the Fourier domain and subsequently tunes their frequencies, amplitudes, and phases through time-domain regression. While effective, this method becomes inefficient when analysing large volumes of long time-series data and often relies on subjective stopping criteria to determine the number of iterations.
   We aim to perform frequency extraction of gravity modes in 
   slowly pulsating B (SPB) stars using a statistically robust, data-driven approach based on advanced power spectrum and harmonic analysis techniques.
   Our approach employs the multitaper non-uniform fast Fourier transform, \texttt{mtNUFFT}, a power spectrum estimator that addresses several statistical limitations of traditional methods such as the Lomb-Scargle periodogram. We apply its extension, the multitaper F-test, to extract coherent gravity modes from 4-year {\it Kepler\/} light curves of SPB stars and to search for period spacing patterns among the extracted modes.
   The multitaper F-test enables fast and accurate extraction of the properties of gravity modes with quasi-infinite lifetimes, preferentially selecting modes that exhibit purely periodic behaviour. Although the method typically extracts fewer frequencies than conventional prewhitening, it recovers most known modes and, in some cases, reveals new ones. We also find evidence for gravity modes with long but finite lifetimes, and detect more than one period spacing pattern in some of the studied SPB stars.
   Overall, the multitaper F-test offers a more objective and statistically sound alternative to prewhitening. It scales efficiently to large datasets containing thousands of pulsators, and has the potential to facilitate mode identification and to distinguish between the different excitation mechanisms operating in SPB stars.}

  
   \titlerunning{Multitaper F-test to detect gravity modes in SPB stars}
   \authorrunning{A.\ Patil et al.}

   \maketitle
%

\section{Introduction}
The power spectra of slowly pulsating B (SPB) stars reveal rich patterns of high-radial-order, low-degree ($l \lessapprox 2$) gravity (g) mode oscillations. These stars have masses in the range of approximately 3–9$M_{\odot}$ and cover the entire main-sequence evolution \citep{aerts_2010_book}. The g-mode pulsations in these stars are driven by the $\kappa$-mechanism (also called the heat-engine mechanism), which operates in the iron-group opacity bump within the stellar radiative envelopes \citep{Pamyatnykh1999}. While we know that rotation has a substantial effect on their mode excitation \citep{Townsend2005a,Townsend2005b,Szewczuk2017},
the impact of other internal physical aspects is not well understood \citep[e.g.,][]{Rehm2024}. Moreover, we know that current instability computations predict overly restricted mode excitation regions in the Hertzsprung–Russell diagram, that is, the regions where stars are theoretically expected to exhibit excited modes. This became clear since the {\it Gaia\/} mission delivered light curves for thousands of g-mode pulsators in the main-sequence phase \citep{DeRidder2023}. These numerous newly discovered pulsators have well-determined astrophysical properties by now \citep{Aerts2023,HeyAerts2024}, leading to the conclusion that g-mode pulsators cover the entire mass range from 1.3\,M$_\odot$ to about 9\,M$_\odot$ along the main sequence \citep{Mombarg2024,Aerts2025}. 

Aside from the incomplete understanding of mode excitation in BAF-type stars, angular momentum transport inside such stars and their successors at different evolutionary stages remains poorly understood.
On the one hand, \citet{Aerts2019} deduced that
angular momentum transport happens more efficiently than predicted by current theories when stars have a convective core. On the other hand, if g~modes themselves play a role in angular momentum transport, it seems to happen too efficiently and on too short time scales in SPB stars \citep{Townsend2018}. Proper measurements of non-linear resonant mode coupling \citep{VanBeeck2024} and g~mode lifetimes in SPB stars are essential to improve the interpretation of mode excitation and to evaluate the role of these modes in active transport processes. 

The coherent, heat-driven g modes in SPB stars are assumed to have quasi-infinite lifetimes, producing sharp, nearly delta-function-like peaks in the frequency domain signal (power spectrum). Such signals differ from the broader Lorentzian profiles characteristic of damped modes observed in sun-like stars \citep{GarciaBallot2019} or in red giants \citep{hekker_2017} with a substantial convective envelope. However, modern space photometry has revealed additional 
stochastic low-frequency variability alongside the coherent modes in several SPB stars and in more massive OB-type stars \citep{Blomme_2011, Neiner2012, Tkachenko_2014,  AertsRogers2015, Aerts_2018, Ramiaramanantsoa_2018, 2019ApJ...872L...9P, Bowman2019, 2020A&A...639A..81B, Bowman2020, Szewczuk2021, 2022MNRAS.509.4246E, 2024A&A...692A..49B, 2024ApJ...969...81Z, 2024ApJS..275....2S, 2025A&A...697A.152K, Pedersen2025}. This low-frequency variability has been interpreted as either damped gravity or gravito-inertial waves excited stochastically at the convective core boundary \citep{Edelmann2019,Ratnasingam2020,Horst2020,Rathish2023,Vanon2023} or through sub-surface convective motions \citep{Cantiello2009,Lecoanet2019,Schultz2022,Anders2023}. A third possible explanation is variability induced by stellar winds \citep{krticka_2018, krticka_2021}.

High-resolution spectroscopic measurements of macroturbulence provide another manifestation of such low-frequency variability \citep{SSD2017}. Yet the observed macroturbulent spectral line broadening cannot be fully attributed to sub-surface convection across the entire SPB mass range, because B stars of later spectral type lack the appropriate sub-surface convection zones \citep{Cantiello2019}. However, a plausible explanation for the low-frequency variability observed in such SPB stars is the presence of internal gravity waves. These waves can tunnel through the evanescent zone adjacent to the mode cavity to be able to reach the surface \citep{Serebriakova2024}. Clearly, there is a need to unravel which of the detected periodic peaks in the frequency spectra of SPB pulsators are due to coherent modes and which ones can be ascribed to internal gravity waves travelling through parts of the interior all the way up to the surface. 

Traditionally, the Lomb-Scargle (LS) periodogram is applied to space-based photometric time-series of pulsating stars to estimate their power spectra, followed by an iterative ``pre-whitening'' procedure to extract individual frequencies of modes whose lifetimes are much longer than the time-series data. The standard prewhitening approach in the literature \citep[see e.g.,\ ][for ground- and space-based photometric light curves, respectively]{breger_1993,Antoci2019} iteratively combines frequency extraction in the Fourier domain with least-squares fitting in the time domain to optimize the description of the light curve. The reason for optimization in the time domain is that the LS periodogram estimates the total power in a time-series that can be attributed to a given frequency without distinguishing between a purely periodic and a quasi-periodic signal. The \texttt{mtNUFFT} framework introduced by \citet{patil_2024}, when combined with the multitaper F-test \citep{patil_2025}, makes this distinction. Thus, the \texttt{mtNUFFT}/F-test methodology appears particularly well-suited for detecting coherent (undamped) modes with quasi-infinite lifetimes and long-lived yet damped gravity waves, potentially outperforming traditional techniques for detecting and distinguishing these different types of signals. The current paper exploits this potential for the case of SPB pulsators.

\section{Targets and starting point of our approach}

This study is a pilot bridging statistical techniques in time-series analysis and the need to extract mode frequencies and mode lifetimes while avoiding data manipulation in an iterative process. For various reasons, the class of SPB pulsators is an optimal population to test our approach via the multitaper F-test. They have 4 years-long {\it Kepler\/} light curves of high quality that serve as the best available observational basis for detailed asteroseismology \citep{Gilliland2010,Papics2014,Papics2015}. The SPB pulsators have numerous coherent g~modes of low degree and consecutive radial order with beating patterns of years-long duration. The numerous excited modes can be dissected and turned into period spacing patterns \citep{2010Natur.464..259D, Papics2014, Papics2015, Papics2017, pedersen_2021, 2018ApJ...854..168Z, Szewczuk2018, Szewczuk2021,Szewkzuk2022}. 
 
The morphology of diagrams that plot the period spacings versus the periods of modes having consecutive radial order for the same degree and azimuthal order, allows us to determine stellar internal rotation in a model-independent way. In particular, one estimates the slopes of these period spacing patterns to determine the internal rotation frequency in the region between the convective core and the radiative envelope. This can be done numerically using the methods developed by \citet{reeth_2015a,VanReeth2016} and by \citet{Christophe2018}, delivering at once the identification of the degree and azimuthal order of the modes. Moreover, constraints on internal mixing can be achieved using structures and dips in the observed patterns post mode identification, as predicted from theory by \cite{miglio2008,bouabid2013,Pedersen2018,Aerts2018,2019A&A...628A..76M}. Finally, internal magnetic fields in these B stars leave measureable imprints on their g-mode frequencies, with the extent of frequency variations depending on both magnetic field strength and stellar rotation rate \citep{Aerts2021-GIW,Lecoanet2022}. 
  
While the forward asteroseismic modelling methods based on g~modes in main-sequence stars are well developed and their potential for probing the internal physics is clear \citep{Aerts2021}, applications to SPB stars remain limited. This is largely because prewhitening applied to {\it Kepler\/} photometry delivers a lot of frequencies but filtering out the coherent modes and constructing period spacing patterns is challenging. An important aspect of this challenge is that some of the modes are not coherent but rather seem to be damped travelling low-frequency waves or modes with lifetimes shorter than or similar to the duration of the observations. Moreover, some of the investigated SPB stars show outbursts in their light curves possibly due to nonlinearly coupled modes \citep[][referred to as \citetalias{van_beeck_2021} hereafter]{van_beeck_2021}, as has also been observed in fast rotating pulsating Be stars \citep{Neiner2009,Neiner2020}. Due to these challenges, the sample of SPB and Be stars with detailed asteroseismic modelling is too limited and we do not have unbiased population studies yet to help improve stellar evolution models \citep{Neiner2012,pedersen_2021,Pedersen2022}. Thousands of pulsating B stars with excellent space photometric light curves remain uninterpreted due to the lack of mode identification \citep[e.g.,][]{Balona2011,HeyAerts2024, 2022MNRAS.515..828S}.

In this work, we try a different approach to extract mode frequencies, assess mode lifetimes, and search for period spacing patterns in SPB pulsators. We do so by applying the \texttt{mtNUFFT}/F-test method introduced in \cite{patil_2025}. We test and evaluate the method on 38 SPB stars whose 4-year {\it Kepler\/} light curves were constructed by \citet{PhD-MayPedersen}. \citet{PhD-MayPedersen} had a total sample of 60 SPB stars for which she performed lightcurve and Fourier analysis to look for period spacing patterns. She found period spacing patterns for 38 of the 60 stars, which we use in this paper. She used prewhitening to extract all the frequencies whose amplitudes exceed a local signal-to-noise ratio (SNR) of 4 in the LS periodogram and hunt for period spacing patterns, succeeding for 38 of the 60 SPB pulsators. She subsequently forward modelled 26 of the pulsators in \citet[][hereafter PD21]{pedersen_2021}. Several of the 38 stars were identified as having multiple period spacing patterns, but \citet{PhD-MayPedersen} decided to focus on one pattern for each star and only model the identified dipole period spacing patterns in \citetalias{pedersen_2021}.

The frequency extraction for the 38 SPB stars was revisited by \citetalias{van_beeck_2021}. They used five different prewhitening strategies to extract frequencies. In contrast to \citetalias{pedersen_2021}, \citetalias{van_beeck_2021} limited the extraction to a maximum of 200 frequencies per star per method, as their purpose was to study the impact of different stopping criteria on the interpretations of the frequency spectrum. Two of their methods were inspired by the approach developed by \citet{reeth_2015a} to hunt for period spacing patterns of g~modes in $\gamma\,$Doradus pulsators, which have longer patterns than SPB stars \citep{reeth_2015b,gang_2020}. 

Here, we in turn revisit the results in \citetalias{van_beeck_2021}, assessing to what level the frequency values resulting from each of their five prewhitening strategies per star are the same or not. We deduce that the two strategies that select frequencies and initialise the optimisation using amplitude significance, which is inferred using either a likelihood ratio test or a Z-test, perform best for 34 of the 38 SPB pulsators treated by \citetalias{van_beeck_2021}. Using the frequencies passing the likelihood ratio test or Z-test leads to systematically better nonlinear harmonic regression models optimising the amplitude, frequency, and phase for each of the periodic signals in the time domain. In particular, these two approaches perform better 
than the traditional method of using an SNR stopping criterion as adopted in most frequency analyses of intermediate- and high-mass pulsators \citep[e.g.,][]{Antoci2019,pedersen_2021}. 

Each prewhitening strategy in \citetalias{van_beeck_2021} yields a separate regression model for every SPB star. For those models with an explained scaled variance (a measure of regression quality, defined in Eq. 3 of \citetalias{van_beeck_2021}) exceeding 90 \%, most extracted frequencies agree within $3\sigma$. Nevertheless, up to a handful of frequencies differ among the five regression models. Moreover, for the stars with regression models explaining less than 90\% of the variance, the number of frequencies differing among the five methods increases as the explained variance decreases.  Given these modest yet clear differences in the regression-model frequencies for each star arising from the five prewhitening approaches, we re-analyse the lightcurves constructed by \citet{PhD-MayPedersen} for the 38 SPB stars with our \texttt{mtNUFFT}/F-test method. We use the exact same light curves as interpreted by \citetalias{pedersen_2021} and \citetalias{van_beeck_2021} as the starting point of our work.

In addition to frequency-by-frequency comparisons, we also compare some period spacing patterns with those in \citetalias{pedersen_2021}. 
These authors went further than \citetalias{van_beeck_2021} in that more than 200 significant frequencies were considered in their regression models for most of the 26 SPB stars they interpreted asteroseismically. We compare the primary period spacing patterns we find with those of \citetalias{pedersen_2021} and also highlight the presence of candidate secondary patterns.

The results discussed below qualitatively show that our method improves upon the computational efficiency of previously used prewhitening approaches. We provide a unique way to pick out g~modes with purely periodic behaviour, that is, modes with quasi-infinite lifetimes. Our results also suggest that some g~modes are damped enough that their power spectra resemble those of damped harmonic oscillators as opposed to being delta functions. The damped nature of the modes will be considered in future work. Finally, we provide a method to detect as many independent frequencies as possible and present a frequency-by-frequency as well as period spacing pattern comparison between our g-mode estimates and those in the literature.

\begin{figure*}[ht]
    \centering
    \includegraphics[width=\linewidth]{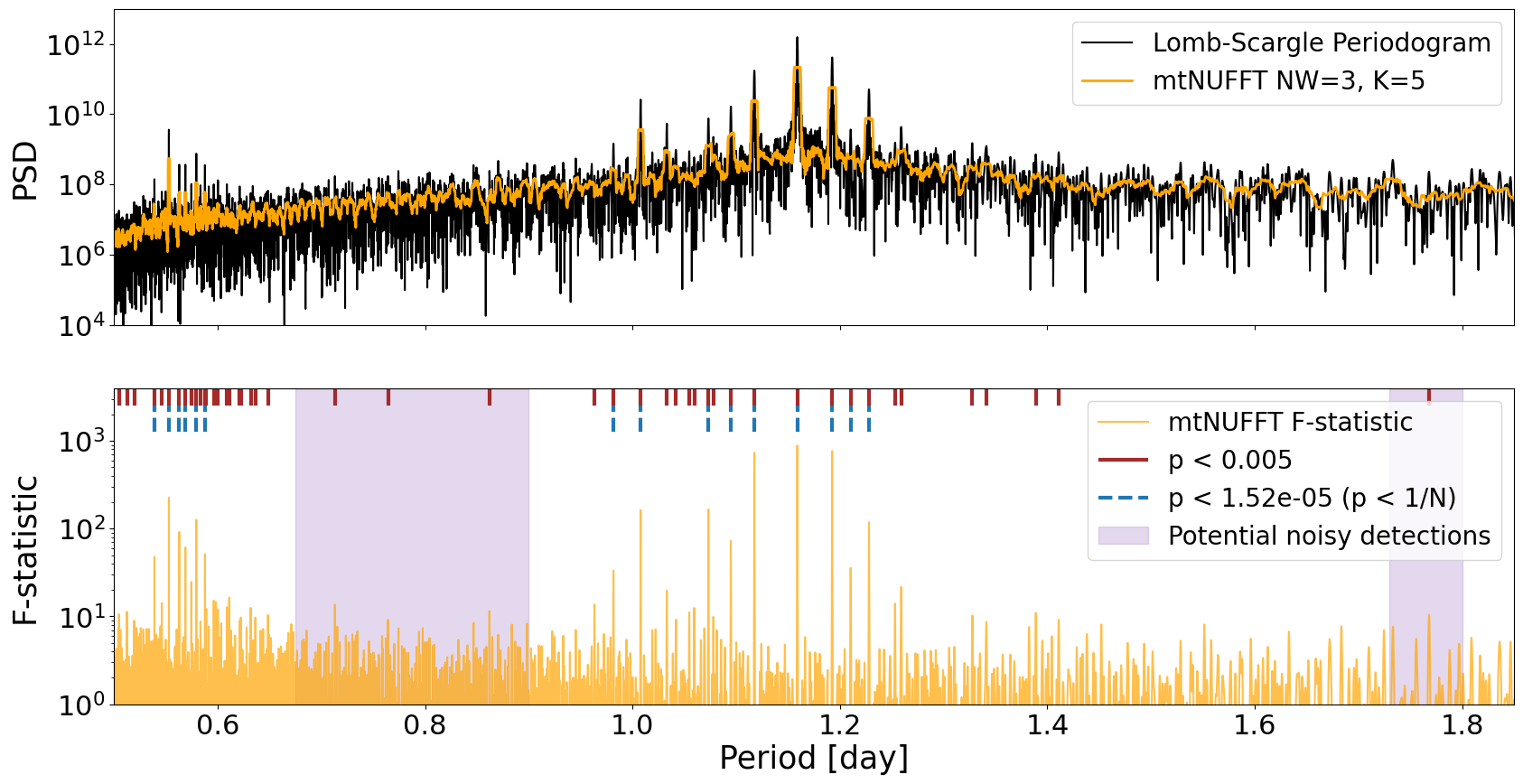}
    \caption{Comparison of F-test detections for the SPB star KIC7760680 using two significance thresholds, $p < 1/N$ and $p < 10^{-3}$.The lower panel illustrates that adopting the more relaxed threshold of $p < 0.001$ rather than $p < 1/N$ increases sensitivity (yielding 14 additional detections in the figure), but some of these, particularly in the low-power region between 0.4 and 0.5 days, appear noise-like (see Section \ref{subsec:multistep}).
    }
    \label{fig:alpha-comp}
\end{figure*}

\begin{figure*}[ht]
    \centering
    \includegraphics[width=\linewidth]{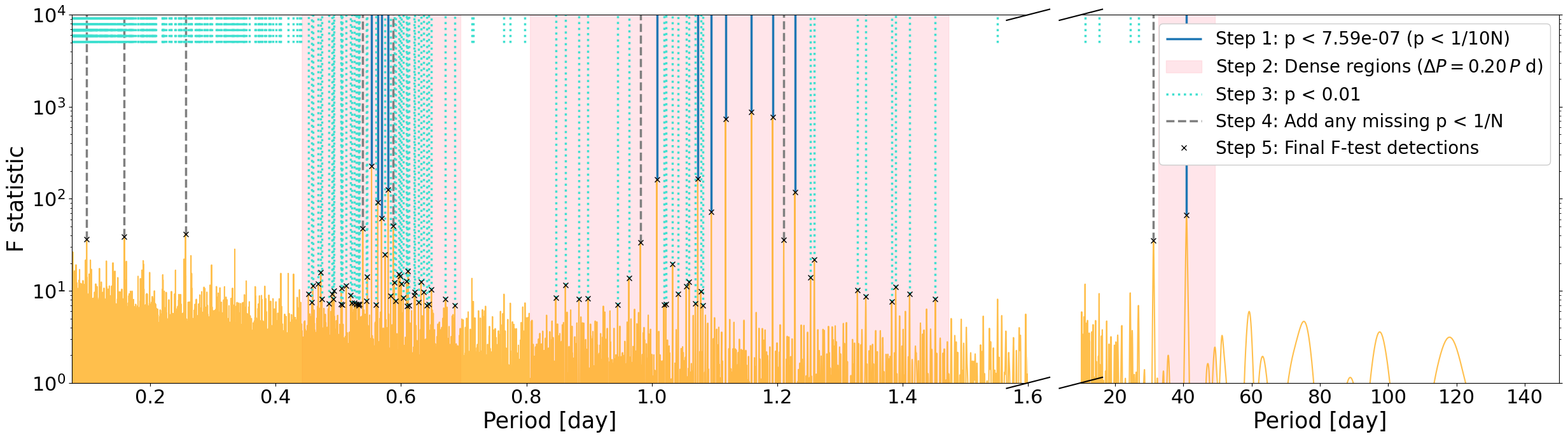}
    \caption{Illustration of the multi-step procedure to detect g~modes in KIC7760680 using the F-test. The five steps are described in Section~\ref{subsec:multistep}.}
    \label{fig:ftest-multi}
\end{figure*}

\section{Statistical analysis method}

\subsection{The \texttt{mtNUFFT} periodogram}
In \cite{patil_2024}, we introduce the \texttt{mtNUFFT} periodogram, which computes the power spectrum underlying a signal $x_n$ sampled at quasi-regular times $t_n, \; n = \{0, \dotsc, N-1\}$ as the average of multiple \textit{tapered} power spectrum estimates. We define it as

\begin{equation}\label{eq:mt_spec}
    \hat{S}^{(\mathrm{mt})}(f) = \frac{1}{K} \sum_{k=0}^{K-1} \hat{S}_k(f).
\end{equation}
Here, $\hat{S}^{(\mathrm{mt})}(f)$ denotes the average of $K$ tapered estimates $\hat{S}_k(f)$ of order $k= 0, \dotsc, K-1$.
These estimates are given by
\begin{equation}
    \hat{S}_k(f) = \left|y_k(f)\right|^2,\label{eq:mt_eigenspec}
\end{equation}
where
\begin{equation}
    y_k(f) = \sum_{n=0}^{N-1} v_{k, n}^{\star} x_n e^{-i 2 \pi f t_n}.\label{eq:mt_eigencoeff}
\end{equation}
Essentially, one estimates the $k$th-eigencoefficient $y_k(f)$ as described in Equation \ref{eq:mt_eigencoeff}, the squared magnitude of which is the eigenspectrum $\hat{S}_k(f)$ in Equation \ref{eq:mt_eigenspec}. To estimate $y_k(f)$, the signal $x_n$ is multiplied by the $k\mathrm{th}$ order taper $v_{k, n}^{\star}$. The $\star$ notation in the taper represents that the regularly-sampled taper $v_{k, n}$ is evaluated at quasi-regular times $t_n$ \citep[refer to][for details on how these tapers are evaluated using interpolation]{springford_2020, patil_2024}. After tapering the signal, one applies the adjoint nonuniform fast Fourier transform (\texttt{NUFFT}) to compute the eigencoefficients $y_k(f)$ on a grid of frequencies $f \in [0, f_\mathrm{Nq}]$, where $f_\mathrm{Nq}$ denotes the Nyquist frequency or the largest frequency measurable in the signal. While the Nyquist frequency is well-defined for regularly-sampled time-series, it is generally replaced by an approximate estimate for quasi-regular or irregular sampling \citep[refer to][for more details]{patil_2024}.

\subsection{DPSS or Slepian tapers}
Slepian sequences \citep{slepian_1978}, denoted here as $v_{k}(N, W)$, are also known as Discrete Prolate Spheroidal Sequences (DPSSs). They are specifically designed to minimize spectral leakage outside a user-defined bandwidth $W$ \citep{thomson_1982}. The sequences minimize leakage by concentrating (power) spectral energy at a given frequency $f$ within the frequency range $\{f - W, f + W\}$. The lowest order DPSS taper $k=0$ results in the most energy concentration within (and the least leakage outside) this range, with higher order tapers having progressively larger leakage. Note that the DPSS tapers are orthonormal, and so the resulting power spectrum estimates are statistically independent and can be averaged. Averaging reduces the variance of the final power spectrum estimate compared to traditional single-taper methods \citep{thomson_1982}. 

In practice, selecting the total number of tapers to be $K \lessapprox 2NW$ ensures minimal out-of-band(width) leakage, and thereby bias, in the averaged power spectrum estimate. Increasing the time-bandwidth product $NW$ (with $N$ the number of time samples) allows the use of more tapers, thereby reducing the variance of the averaged estimate, but at the cost of decreased frequency resolution. The value $NW=1$ corresponds to the Rayleigh resolution $f_\mathcal{R} = 1/T$, whereas $NW > 1$ corresponds to a resolution of $NW \times f_\mathcal{R}$ (a rule-of-thumb is to start with $NW=4$ and tune according to the given problem). We refer the reader to \citet{patil_2024} for more details on the \texttt{mtNUFFT} estimate and ways to choose the parameters $NW$ and $K$. 

\subsection{The multitaper F-test}

In addition to providing a power spectrum estimate with reduced bias and variance compared to the standard LS periodogram approach, \texttt{mtNUFFT} has the ability to compute an F-statistic, which allows one to test whether the power at a given frequency $f$ is predominantly due to a purely periodic signal or quasi-periodic one \citep{thomson_1982}. Here, a purely periodic signal refers to one having coherent frequency $f$, amplitude $A$, and phase $\phi$, i.e., a sinusoid $A \cos(2 \pi f t + \phi)$. The F-statistic follows an F-distribution under the null hypothesis that no coherent signal is present at $f$, and is defined as \citep{patil_2025}

\begin{equation}\label{eq:F-test}
    F(f) = \frac{(K-1) \left|\hat{\mu}(f)\right|^2 \sum\limits_{k=0}^{K-1} \left|U_k(N, W; 0)\right|^2}
    {\left|\sum\limits_{k=0}^{K-1} y_k(f) - \hat{\mu}(f) U_k(N, W; 0)\right|^2}.
\end{equation} 
Here, $U_k(N, W; 0)$\footnote{$U_k(N, W)$ is called the discrete prolate spheroidal wave function.} represents the Discrete Fourier Transform (DFT) of the $k$-th order DPSS taper $v_k(N, W)$ at frequency $f=0$. The term $\hat{\mu}(f)$ is a regression estimate of the complex-valued amplitude of the periodic component at $f$, defined as

\begin{equation}\label{eq:F-test_regr}
    \hat{\mu}(f) = \frac{\sum\limits_{k=0}^{K-1} U_k(N, W; 0) \, y_k(f)} {\sum\limits_{k=0}^{K-1} {U_k(N, W; 0)}^2}.
\end{equation} 
We then compute the real-valued amplitude of the periodic component at $f$ as
\begin{equation}\label{eq:amplitude}
    A = {\mid \hat{\mu}(f) \mid}^2
\end{equation} and the phase as follows
\begin{equation}\label{eq:phase}
    \phi = \arctan \frac{\mathbb{R}e \{\hat{\mu}(f)\}}{\mathbb{I}m\{\hat{\mu}(f)\}}.
\end{equation}

\begin{figure*}[ht]
    \centering
    \includegraphics[width=\linewidth]{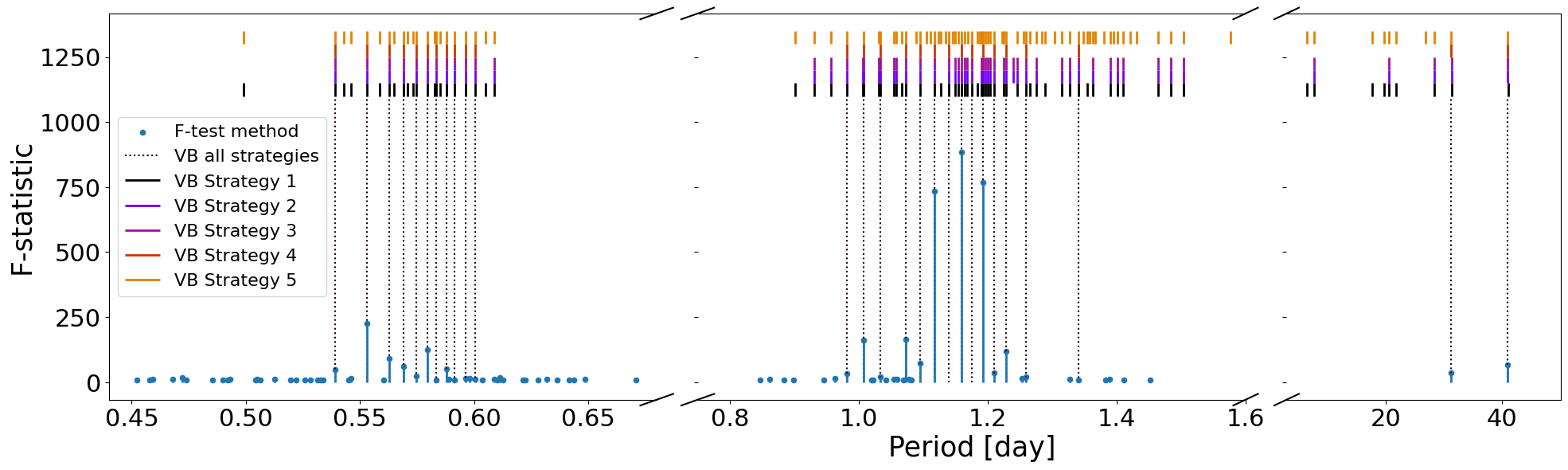}
    \caption{Frequencies extracted using the multi-step \texttt{mtNUFFT}/F-test and their comparison with estimates in \citetalias{van_beeck_2021} for the SPB KIC7760680. The comparison is made with frequencies extracted by each of the five prewhitening strategies, as well as those consistent across all strategies (dotted grey lines). Values of $NW=4$ and $K=7$ with a p-value $p < 0.01$ were used.}
    \label{fig:van_beeck}
\end{figure*}

\begin{table*}[ht!]
\caption{Frequency matches between the periodic g-mode signals detected by the F-test ($NW=4$, $K=7$, $p<0.01$) and those reported using Strategy 3 in \citetalias{van_beeck_2021} for 
KIC7760680. Matches are defined when the frequencies agree within $3\sigma$ (or $6\sigma$ if highlighted in purple). The signals are listed in order of increasing frequency. For each frequency $\hat{f_0}$, we report the corresponding amplitude $A$, phase $\phi$, jackknife uncertainties on the frequency, amplitude, and phase $\sigma_{\hat{f_0}}, \, \sigma_{\widehat{A}}, \, \sigma_{\widehat{\phi}}$ and the F-statistic [columns under `Patil et al. (this paper)']. The corresponding values from \citetalias{van_beeck_2021} are shown in the adjacent columns.The Nonlinear ID column from \citetalias{van_beeck_2021} indicates the order in which each independent frequency was extracted during the prewhitening procedure. If a signal was identified as a combination frequency, this column lists the corresponding combination of independent frequencies, which may indicate a nonlinear resonant mode (see \citetalias{van_beeck_2021}).}
\resizebox{\linewidth}{!}{
\begin{tabular}{crrrrrr|crrrrrl}
\toprule 
\multicolumn{7}{c|}{Patil et al. (this paper)} & \multicolumn{7}{c}{VB21 Strategy 3}\\
Frequency $\hat{f_0}$ & $\sigma_{\hat{f_0}}$ & Amplitude $\widehat{A}$ & $\sigma_{\widehat{A}}$ & Phase $\widehat{\phi}$ & $\sigma_{\widehat{\phi}}$ &  $F$-statistic  & Frequency $\hat{f_0}$ & $\sigma_{\hat{f_0}}$ & Amplitude $\widehat{A}$ & $\sigma_{\widehat{A}}$ & Phase $\widehat{\phi}$ & $\sigma_{\widehat{\phi}}$ & Nonlinear ID \\
($1/\mathrm{day}$) & ($1/\mathrm{day}$) & (ppm) & (ppm) & (rad) & (rad) & & ($1/\mathrm{day}$) & ($1/\mathrm{day}$) & (ppm) & (ppm) & (rad) & (rad) \\

\midrule 0.024414 & 0.000041 & 181.073034 & 5.222283 & 0.170401 & 0.118590 & 66.399528 & 0.024408 & 0.000026 & 201.337980 & 14.076463 & 1.623552 & 0.069915 &  freq\_ 1 - freq\_ 2  \\
0.031997 & 0.000077 & 166.228713 & 6.642989 & 1.879346 & 0.119531 & 35.294917 & 0.031968 & 0.000031 & 168.803104 & 14.076463 & 2.985211 & 0.083390 &  freq\_ 3 - freq\_ 1  \\ 
\rowcolor{blue!10}
0.708723 & 0.000068 & 206.558042 & 2.788264 & 1.544018 & 0.026121 & 9.188998 & 0.709102 & 0.000044 & 120.270005 & 14.076463 & -0.536887 & 0.117041 &  freq\_ 24 - freq\_ 5  \\
\rowcolor{blue!10}
0.719978 & 0.000067 & 212.459463 & 45.382386 & -2.885599 & 0.094390 & 10.976999 & 0.719421 & 0.000073 & 72.770917 & 14.076463 & 2.871236 & 0.193435 &  freq\_ 46  \\
\rowcolor{blue!10}
0.745616 & 0.000017 & 307.311704 & 8.517915 & 2.395007 & 0.121733 & 8.657512 & 0.745761 & 0.000015 & 354.643486 & 14.076463 & -1.400792 & 0.039692 &  freq\_ 11  \\
0.753164 & 0.000117 & 287.157437 & 2.855840 & 0.287936 & 0.114001 & 10.234381 & 0.753171 & 0.000044 & 121.105066 & 14.076463 & 1.690412 & 0.116233 &  freq\_ 26  \\
0.794206 & 0.000052 & 533.514136 & 29.953011 & -0.981846 & 0.127727 & 21.701908 & 0.794263 & 0.000011 & 463.265803 & 14.076463 & -0.782122 & 0.030385 &  freq\_ 9  \\
0.814302 & 0.000003 & 1773.909360 & 58.448621 & 2.262334 & 0.004290 & 118.497266 & 0.814310 & 0.000003 & 1742.943213 & 14.076463 & -0.826346 & 0.008076 &  freq\_ 4  \\
0.826203 & 0.000045 & 467.395425 & 34.159221 & 1.798332 & 0.112823 & 35.698434 & 0.826299 & 0.000021 & 246.865471 & 14.076463 & -0.831270 & 0.057021 &  freq\_ 13  \\
\rowcolor{blue!10}
0.838546 & 0.000003 & 4994.183939 & 59.291025 & -0.689480 & 0.000635 & 767.373999 & 0.838526 & 0.000001 & 5051.359257 & 14.076463 & -0.759907 & 0.002787 &  freq\_ 2  \\
\rowcolor{blue!10}
0.862858 & 0.000003 & 9723.284972 & 37.756610 & -1.131417 & 0.001833 & 883.618296 & 0.862876 & 0.000001 & 9861.385107 & 14.076463 & 2.617689 & 0.001427 &  freq\_ 1  \\
0.894855 & 0.000029 & 3224.571614 & 46.625277 & -2.811816 & 0.083097 & 733.948659 & 0.894848 & 0.000002 & 3411.811513 & 14.076463 & 1.198383 & 0.004126 &  freq\_ 3  \\
0.913386 & 0.000029 & 1029.591640 & 31.683735 & -1.504075 & 0.083445 & 72.418387 & 0.913385 & 0.000006 & 945.192152 & 14.076463 & 3.106222 & 0.014893 &  freq\_ 6  \\
0.931782 & 0.000038 & 715.868768 & 15.866187 & 1.169553 & 0.118725 & 165.600997 & 0.931822 & 0.000009 & 618.398057 & 14.076463 & -2.891580 & 0.022763 &  freq\_ 7  \\
0.948103 & 0.001233 & 200.405627 & 11.312159 & -2.771724 & 0.129250 & 11.146117 & 0.948578 & 0.000071 & 74.589479 & 14.076463 & -1.225614 & 0.188719 &  freq\_ 10 - freq\_ 39  \\
0.967961 & 0.000052 & 553.234919 & 11.769183 & -0.617589 & 0.125204 & 19.687436 & 0.967846 & 0.000010 & 543.280547 & 14.076463 & -0.329278 & 0.025910 &  freq\_ 8  \\
0.992171 & 0.000029 & 1223.301000 & 23.308126 & 1.393588 & 0.092141 & 162.323962 & 0.992183 & 0.000004 & 1391.618410 & 14.076463 & -2.969371 & 0.010115 &  freq\_ 5  \\
1.018693 & 0.000081 & 310.666752 & 9.471978 & -0.424294 & 0.120711 & 33.368551 & 1.018649 & 0.000024 & 219.803346 & 14.076463 & -0.963861 & 0.064041 &  freq\_ 16  \\
1.642273 & 0.000111 & 75.988482 & 2.628093 & 1.081864 & 0.106613 & 12.795054 & 1.642385 & 0.000110 & 48.059778 & 14.076463 & 0.204618 & 0.292895 &  freq\_ 13 + freq\_ 19  \\ 
1.665803 & 0.000808 & 87.815637 & 13.853889 & 1.895834 & 0.118510 & 11.999618 & 1.665832 & 0.000074 & 71.062001 & 14.076463 & -0.668319 & 0.198087 &  freq\_ 1 + freq\_ 27  \\ 1.677092 & 0.000105 & 80.604648 & 4.974847 & -0.502215 & 0.096165 & 15.194213 & 1.677221 & 0.000090 & 58.473812 & 14.076463 & 1.498597 & 0.240731 &  freq\_ 1 + freq\_ 4  \\
1.690659 & 0.000034 & 73.816269 & 10.001070 & -2.120619 & 0.020230 & 7.708872 & 1.690679 & 0.000101 & 52.427522 & 14.076463 & 0.524468 & 0.268494 &  freq\_ 11 + freq\_ 52  \\ 1.701336 & 0.000045 & 151.914070 & 2.984866 & -2.413831 & 0.113666 & 50.981831 & 1.701408 & 0.000041 & 130.015238 & 14.076463 & 0.670278 & 0.108268 &  freq\_ 1 + freq\_ 2  \\ 1.713849 & 0.000079 & 73.420379 & 1.597691 & -1.610655 & 0.026821 & 8.764348 & 1.713733 & 0.000075 & 70.503079 & 14.076463 & 0.387284 & 0.199657 &  freq\_ 1 + freq\_ 12  \\
1.725750 & 0.000029 & 211.006785 & 1.792547 & 0.480859 & 0.092016 & 125.895839 & 1.725747 & 0.000022 & 244.434095 & 14.076463 & -2.108996 & 0.057588 &  2 * freq\_ 1  \\ 
1.740031 & 0.000073 & 77.773098 & 5.179423 & -0.614312 & 0.090580 & 24.791359 & 1.740075 & 0.000057 & 92.246351 & 14.076463 & -1.039813 & 0.152596 &  freq\_ 1 + freq\_ 15  \\ 
1.757712 & 0.000038 & 150.079003 & 4.566146 & 1.817308 & 0.106027 & 61.217982 & 1.757737 & 0.000039 & 135.221683 & 14.076463 & 2.655345 & 0.104099 &  freq\_ 1 + freq\_ 3  \\
1.776312 & 0.000029 & 151.863889 & 3.144222 & -2.831968 & 0.091919 & 91.577278 & 1.776275 & 0.000038 & 139.223832 & 14.076463 & -1.692613 & 0.101107 &  freq\_ 1 + freq\_ 6  \\ 1.808309 & 0.000029 & 473.315500 & 13.365018 & -1.343183 & 0.082974 & 225.384487 & 1.808310 & 0.000011 & 463.596624 & 14.076463 & -0.271033 & 0.030364 &  freq\_ 3 + freq\_ 6  \\ 1.855029 & 0.000029 & 93.273912 & 6.475357 & -0.241015 & 0.078317 & 47.669494 & 1.855055 & 0.000058 & 90.752392 & 14.076463 & -1.479945 & 0.155108 &  freq\_ 1 + freq\_ 5 \\
\bottomrule
\end{tabular}}
\tablefoot{Frequencies are considered matched if $|\hat{f}_{0,{\rm Patil}} - \hat{f}_{0,{\rm VB21}}| \leq 3 \, \sigma_\mathrm{Patil + VB21}$ where $\sigma_\mathrm{Patil + VB21} = \sqrt{\sigma_{\hat{f}_{0,{\rm Patil}}}^2 + \sigma_{\hat{f}_{0,{\rm VB21}}}^2}$. The ones highlighted in purple are the ones that match within $6 \, \sigma_\mathrm{Patil + VB21}$ but are within 0.00072 d$^{-1}$ (the Rayleigh resolution $f_\mathcal{R}$).}
    \label{tab:KIC7760680_match}
\end{table*}

\begin{figure}[ht]
    \centering
    \includegraphics[width=\linewidth]{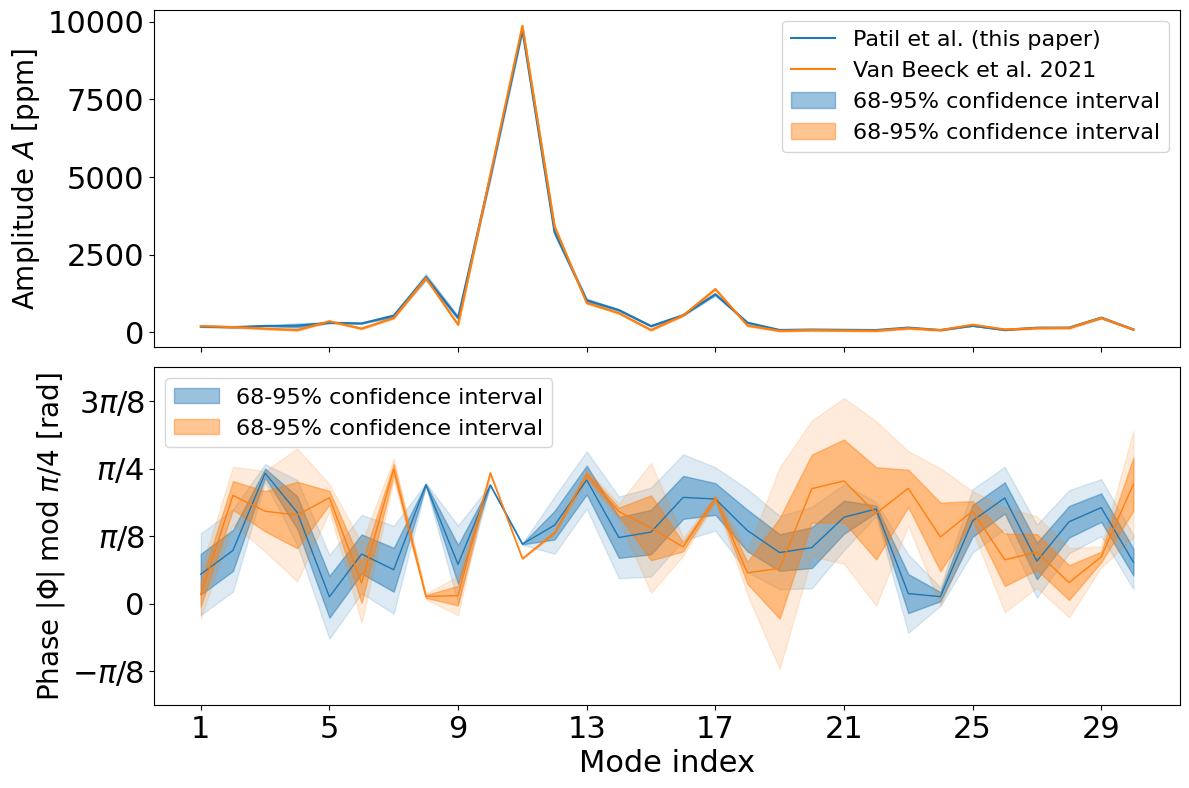}
    \caption{Amplitude and phase comparison of g~modes estimated using the F-test procedure described in this paper and those computed in \citetalias{van_beeck_2021} using Strategy 3 for KIC7760680. The comparison includes modes in our F-test catalogue that have a 3$\sigma$ (or 6$\sigma$) match with those reported by \citetalias{van_beeck_2021}. To highlight relative shifts unambiguously, phase differences are shown as absolute values, mapped modulo $\pi/4$. The 68\% and 98\% confidence intervals are indicated by dark and light shaded regions, respectively, and are computed using the jackknife uncertainty estimates of amplitude and phase. The corresponding amplitude, phase, and uncertainty values are listed in Table \ref{tab:KIC7760680_match}. Note that although the amplitude plot includes the confidence intervals, they are too small to be visible.}
    \label{fig:amp_phase}
\end{figure}

While Equation \ref{eq:F-test} defining the F-statistic at frequency $f$ is rather complicated, the idea behind it is simple. One computes $F(f_0)$, the F-statistic at a given frequency $f_0$, and evaluates how likely it is to be sampled from an F-distribution with degrees of freedom $2$ and $2K-2$. If there is no purely periodic signal present at $f_0$ (the null hypothesis of the F-test), $F(f_0)$ will follow the F-distribution. To see if $F(f_0)$ is sampled from this distribution, we compute the p-value corresponding to this F-statistic, which represents the probability of obtaining another F-statistic at least as extreme as the one collected. If the p-value is below some threshold $\alpha$ (also called confidence level)\footnote{For example, $p < 0.01$ means that only 1\% of the F-statistics under the null hypothesis will be as extreme as the one being tested.}, the F-test at frequency $f_0$ is rejected under the null, that is, we can confidently say that the signal of frequency $f_0$ in the time-series is purely periodic.

\cite{patil_2025} discuss how one can choose the confidence level $\alpha$ when performing multiple F-tests, one for each frequency $f = \{0, \dotsc, f_{Nq} \}$. In particular, their approach relies on \textit{selective inference} procedures such as the Benjamini-Hochberg correction for multi-hypothesis testing, which they apply to disentangle exoplanet transit signals from mixed modes in solar-like oscillators. In contrast, the F-test can be even more powerful when dealing with g~modes as the majority of these are purely periodic. In the next section, we discuss how we apply the F-test to SPB stars and automatically detect g~modes with purely periodic behaviour. 

\subsection{Multi-step F-test to detect g~modes}\label{subsec:multistep}
We know that the multitaper F-test quantifies the degree of strict periodicity in a signal, making it particularly useful for identifying coherent oscillatory modes. In the context of stellar power spectra of SPB stars, such signals ideally correspond to long-lifetime g~modes. However, a key challenge lies in selecting the appropriate significance threshold, $\alpha$, for automatic detection of these modes.

Choosing a conservative threshold (e.g., $\alpha=
0.05$) reduces the rate of false positives but risks missing true g~modes, especially those with lower amplitudes. Conversely, a more relaxed threshold increases sensitivity but also admits false positives; periodic signals that are statistically significant yet likely to arise from structured noise rather than physical modes. This issue is particularly pronounced in the low-power regions of the spectrum, where purely periodic noise can mimic the spectral signature (a delta function in the frequency domain) of a g~mode and still pass the F-test. Examples of such noisy detections are shown in Fig.,\ref{fig:alpha-comp}, which presents the power spectral density (PSD) and the corresponding F-test detections for KIC7760680.

This raises the question: can prior knowledge of g-modes be incorporated to improve detection? A promising approach is to exploit their quasi-uniform period spacing, which causes them to cluster in specific regions of the power spectrum. By combining this structural information with the F-test, we can construct a more robust and targeted detection strategy than relying on the F-test alone, as in \cite{patil_2025}. In that work, the presence of purely periodic signals at different frequencies was assessed by repeated F-tests with selective inference (e.g., the Bonferroni and BHq procedures), which limited the set of detectable signals. Here, we relax this restriction by allowing a larger p-value threshold while using the quasi-uniform period spacing of g-modes as prior information to guide the selection of F-test detections.

We visualise this multi-step procedure in Fig.\,\ref{fig:ftest-multi} and describe it as follows:

\begin{enumerate}[label=Step \arabic*: $\;\;$, wide = 0pt, labelsep=-4pt, leftmargin=*]
    \item $p < 1/M$\\
    First, we compute the eigencoefficients $y_k(f)$ in Equation \ref{eq:mt_eigencoeff} using the adjoint \texttt{NUFTT} on an interpolated frequency grid of 
    \begin{equation}
        f_i = i/M, \; \; i = \{0, 1, \dotsc \}.
    \end{equation} Here, interpolation using a zero-padded time-series with resulting length $M = 20N$ ensures that we do not lose a signal between grid points\footnote{The F-test is very sensitive to frequency shifts and thus needs a fine enough frequency grid to detect all purely periodic signals.}. We then compute the F-statistic and test the presence of purely periodic signals at each $f_i$. This results in the testing of multiple hypotheses $H_i$, which means that we reject the null given a conservative confidence level $\alpha = 1/M$, where $M$ represents the length of the padded time-series.\\
    \item Dense regions given $\Delta P$\\
    In order to automatically detect dense clusters of g~modes in the power spectrum, we mark a specified interval around each F-test detection $f_0$ obtained in step 1. In particular, we choose to mark contiguous dense regions in period space using an interval $P_0 \pm \Delta P \times P_0$, where $P_0 = 1/f_0$. The rationale for this choice is that g~modes in SPB pulsators with the same $(l, m)$ values, in non-rotating and young stars (i.e. without dips from mode trapping caused by chemical gradients), exhibit similar period spacings. One can set the value of $\Delta P$ based on theoretical or empirical arguments. Here, we choose $\Delta P = 0.2$ days as the approximate upper bound (some $12000$ seconds) of typical asymptotic g-mode spacing values $\Delta \Pi_l$ of observed B-type pulsators \citep[]{degroote_2010, moravveji_2015, Aerts2019}. This scaling is motivated by the typical properties of SPB stars, which exhibit slow pulsations with periods between 0.5 and 5 days \citep{aerts_2010_book}. The bound $\Delta P \times P_0$ is also adaptive: it ensures that the marked region is narrower at short periods and broader at long periods, reflecting the expected variability timescales. 
    \item $p < 0.01$ or another chosen threshold\\
    We now find peaks in the entire F-statistic series across frequencies $f_i$ whose p-values are below the threshold $\alpha = 0.01$. This threshold can be modified to change the selectivity of the frequency detection algorithm.\\
    \item $p < 1/N$:
    In order to ensure that we do not miss any significant frequencies that do not necessarily lie in a dense region, we select frequencies whose p-value are below the threshold $p < 1/N$. \cite{thomson_1982} suggests using $p < 1/N$ as a rule of thumb to ensure that the F-test detections are significant.\\
    \item Final F-test detections\\
    In the final step, we generate the list of frequencies detected by this procedure. This list includes frequencies whose F-tests were rejected under the null in Step 1, as well as frequencies detected in Step 3 that belong to any of the dense regions estimated in Step 2, and any significant missing frequencies that were selected in Step 4.
\end{enumerate}

\begin{table*}[ht]
    \centering
    \caption{Number of periodic g-mode signals reported by \citetalias{van_beeck_2021} (column `VB21, Strategy 3') compared with those detected by the F-test (column `F-test, this paper') for the sample of SPB stars analysed here (column `KIC'). The total number of \citetalias{van_beeck_2021} detections ($T_\mathrm{VB}$) is further divided into independent ($I_\mathrm{VB}$) and combination frequencies ($C_\mathrm{VB}$, i.e., potential nonlinear modes), following their classification. Performance of the F-test is evaluated by varying the multitaper parameters $NW$, $K$, and significance threshold $p$ (columns $NW=4, K=7$; $NW=3, K=5$; $p < 0.005$; and $p < 0.01$). For each configuration, we report the total number of detections (`Total' $T$) and the number of $3\sigma$ matches with \citetalias{van_beeck_2021} (`Match’). The matches are separated into independent ($I_m$) and combination ($C_m$) frequencies, with $T_m = I_m + C_m$.}
    \resizebox{\linewidth}{!}{%
    \begin{tabular}{l|aaa|c|ccc|b|aaa|c|ccc|c|ccc}
    \toprule 
    KIC & \multicolumn{19}{c}{Number of g-mode detections} \\
    \midrule
    \rowcolor{LightCyan}
    & \multicolumn{3}{c|}{VB21, Strategy 3} & \multicolumn{16}{c}{F-test, this paper}\\
    \rowcolor{LightCyan}
    & \multicolumn{3}{c|}{} & \multicolumn{8}{c|}{NW = 4, K = 7} & \multicolumn{8}{c}{NW = 3, K = 5}\\
    \rowcolor{LightCyan}
    & \multicolumn{3}{c|}{} & \multicolumn{4}{c}{p < 0.005} & \multicolumn{4}{c|}{p < 0.01} & \multicolumn{4}{c}{p < 0.005} & \multicolumn{4}{c}{p < 0.01}\\
    \midrule
    & Total & Independent & Combination & Total & \multicolumn{3}{c|}{Match} & Total & & Match & & Total & \multicolumn{3}{c|}{Match} & Total & \multicolumn{3}{c}{Match} \\
    & $T_\mathrm{VB}$ & $I_\mathrm{VB}$ & $C_\mathrm{VB}$ & $T$ & $T_m$ & $I_m$ & $C_m$ & $T$ & $T_m$ & $I_m$ & $C_m$ & $T$ & $T_m$ & $I_m$ & $C_m$ & $T$ & $T_m$ & $I_m$ & $C_m$ \\
    \midrule
    3240411 & 178 & 37 & 141 & 419 & 72 & 29 & 43 & 640 & 81 & 30 & 51 & 344 & 64 & 31 & 33 & 558 & 78 & 31 & 47 \\
    3459297 & 180 & 42 & 138 & 159 & 43 & 26 & 17 & 278 & 45 & 26 & 19 & 84 & 32 & 19 & 13 & 134 & 39 & 20 & 19 \\
    3756031 & 182 & 26 & 156 & 206 & 89 & 22 & 67 & 266 & 96 & 23 & 73 & 92 & 50 & 20 & 30 & 123 & 62 & 21 & 41 \\
    3839930 & 190 & 25 & 165 & 113 & 46 & 16 & 30 & 170 & 56 & 16 & 40 & 118 & 39 & 13 & 26 & 175 & 42 & 13 & 29 \\
    3865742 & 190 & 12 & 178 & 200 & 49 & 8 & 41 & 329 & 58 & 8 & 50 & 98 & 37 & 7 & 30 & 142 & 43 & 7 & 36 \\
    4930889 & 175 & 24 & 151 & 228 & 44 & 17 & 27 & 376 & 46 & 17 & 29 & 111 & 35 & 14 & 21 & 154 & 41 & 14 & 27 \\
    4936089 & 181 & 29 & 152 & 131 & 29 & 18 & 11 & 219 & 30 & 18 & 12 & 220 & 19 & 16 & 3 & 350 & 25 & 16 & 9 \\
    \rowcolor{blue!10}
    4939281$^{**}$ & 144 & 28 & 116 & 109 & 8 & 6 & 2 & 175 & 8 & 6 & 2 & 5 & 2 & 2 & 0 & 5 & 2 & 2 & 0 \\
    \rowcolor{blue!20}
    5084439$^{*}$ & 183 & 17 & 166 & 32 & 11 & 3 & 8 & 49 & 13 & 3 & 10 & 45 & 12 & 2 & 10 & 63 & 16 & 3 & 13 \\
    5309849 & 190 & 34 & 156 & 259 & 80 & 28 & 52 & 389 & 90 & 28 & 62 & 193 & 66 & 26 & 40 & 291 & 75 & 26 & 49 \\
    5941844 & 184 & 23 & 161 & 537 & 64 & 18 & 46 & 858 & 75 & 18 & 57 & 141 & 60 & 16 & 44 & 192 & 72 & 17 & 55 \\
    6352430 & 191 & 38 & 153 & 193 & 55 & 27 & 28 & 299 & 70 & 29 & 41 & 207 & 51 & 27 & 24 & 313 & 58 & 27 & 31 \\
    6462033 & 191 & 37 & 154 & 474 & 124 & 32 & 92 & 686 & 131 & 32 & 99 & 463 & 116 & 30 & 86 & 704 & 124 & 30 & 94 \\
    6780397 & 181 & 23 & 158 & 237 & 47 & 15 & 32 & 374 & 51 & 16 & 35 & 106 & 39 & 13 & 26 & 150 & 44 & 14 & 30 \\
    7630417 & 179 & 50 & 129 & 426 & 78 & 34 & 44 & 620 & 85 & 36 & 49 & 260 & 32 & 11 & 21 & 440 & 33 & 11 & 22 \\
    7760680 & 65 & 22 & 43 & 56 & 29 & 14 & 15 & 87 & 30 & 14 & 16 & 52 & 24 & 11 & 13 & 77 & 28 & 13 & 15 \\
    8057661 & 191 & 48 & 143 & 384 & 62 & 26 & 36 & 629 & 69 & 28 & 41 & 469 & 80 & 36 & 44 & 714 & 82 & 36 & 46 \\
    8087269 & 162 & 20 & 142 & 86 & 38 & 16 & 22 & 150 & 49 & 16 & 33 & 94 & 37 & 17 & 20 & 139 & 43 & 17 & 26 \\
    8255796 & 186 & 22 & 164 & 87 & 67 & 20 & 47 & 102 & 72 & 21 & 51 & 50 & 32 & 14 & 18 & 57 & 36 & 14 & 22 \\
    8324482 & 192 & 34 & 158 & 359 & 121 & 31 & 90 & 497 & 128 & 31 & 97 & 290 & 118 & 30 & 88 & 392 & 126 & 31 & 95 \\
    8381949 & 183 & 32 & 151 & 628 & 42 & 21 & 21 & 980 & 43 & 21 & 22 & 640 & 34 & 18 & 16 & 1032 & 36 & 19 & 17 \\
    8459899 & 178 & 24 & 154 & 133 & 67 & 21 & 46 & 183 & 73 & 21 & 52 & 73 & 50 & 17 & 33 & 87 & 55 & 18 & 37 \\
    8714886 & 189 & 33 & 156 & 229 & 76 & 28 & 48 & 323 & 83 & 28 & 55 & 388 & 81 & 30 & 51 & 594 & 82 & 30 & 52 \\
    8766405 & 160 & 18 & 142 & 211 & 48 & 12 & 36 & 300 & 55 & 12 & 43 & 193 & 45 & 11 & 34 & 325 & 48 & 11 & 37 \\
    9020774 & 190 & 26 & 164 & 206 & 36 & 17 & 19 & 358 & 42 & 19 & 23 & 137 & 22 & 13 & 9 & 223 & 27 & 14 & 13 \\
    9227988 & 174 & 16 & 158 & 48 & 37 & 14 & 23 & 63 & 42 & 14 & 28 & 229 & 39 & 13 & 26 & 355 & 45 & 13 & 32 \\
    \rowcolor{blue!10}
    9715425$^{**}$ & 55 & 3 & 52 & 3 & 1 & 1 & 0 & 3 & 1 & 1 & 0 & 1 & 0 & 0 & 0 & 1 & 0 & 0 & 0 \\
    9964614 & 187 & 38 & 149 & 302 & 59 & 19 & 40 & 460 & 62 & 20 & 42 & 153 & 18 & 8 & 10 & 228 & 18 & 8 & 10 \\
    \rowcolor{blue!20}
    10220209$^{*}$ & 186 & 34 & 152 & 1 & 1 & 0 & 1 & 1 & 1 & 0 & 1 & 21 & 7 & 2 & 5 & 32 & 11 & 4 & 7 \\
    10285114 & 165 & 48 & 117 & 130 & 32 & 18 & 14 & 230 & 35 & 19 & 16 & 89 & 17 & 10 & 7 & 158 & 17 & 10 & 7 \\
    10526294 & 78 & 30 & 48 & 83 & 25 & 11 & 14 & 122 & 28 & 13 & 15 & 51 & 34 & 21 & 13 & 69 & 38 & 22 & 16 \\
    10536147 & 190 & 36 & 154 & 105 & 49 & 21 & 28 & 130 & 51 & 22 & 29 & 90 & 37 & 15 & 22 & 132 & 44 & 18 & 26 \\
    10658302 & 172 & 28 & 144 & 137 & 67 & 23 & 44 & 198 & 79 & 24 & 55 & 139 & 54 & 21 & 33 & 209 & 63 & 21 & 42 \\
    11152422 & 167 & 27 & 140 & 94 & 43 & 12 & 31 & 140 & 49 & 12 & 37 & 74 & 28 & 10 & 18 & 100 & 31 & 10 & 21 \\
    11360704 & 176 & 21 & 155 & 114 & 17 & 4 & 13 & 186 & 26 & 5 & 21 & 23 & 4 & 2 & 2 & 39 & 5 & 2 & 3 \\
    11454304 & 179 & 43 & 136 & 285 & 75 & 29 & 46 & 404 & 83 & 32 & 51 & 304 & 62 & 22 & 40 & 457 & 73 & 27 & 46 \\
    11971405 & 175 & 13 & 162 & 35 & 19 & 5 & 14 & 50 & 22 & 5 & 17 & 32 & 12 & 4 & 8 & 47 & 14 & 4 & 10 \\
    12258330 & 189 & 36 & 153 & 482 & 102 & 31 & 71 & 746 & 110 & 31 & 79 & 530 & 87 & 30 & 57 & 857 & 98 & 31 & 67 \\
    \bottomrule
    \end{tabular}
    }
    \tablefoot{The `Total' $T$ column showing F-test detections for $NW=4, K-7$ and $p < 0.01$ is highlighted in grey as it consistently works well. The horizontal highlights refer to the categories of *large gaps in time-series and **outbursting stars, with the former classification defined in this paper and the latter following \citetalias{van_beeck_2021} (see their Table 3).}
    \label{tab:match_allSPB}
\end{table*}
\section{Results}

In Section~\ref{subsec:vanbeeck}, we compare, frequency by frequency, the F-test method proposed in this work with those of \citetalias{van_beeck_2021}. We also examine how variations in multitaper parameters and p-values affect our F-test detections. Finally, we compare the period spacings derived from our mode detections with those reported by \citetalias{pedersen_2021}, highlighting additional spacings and dips that may prove valuable for future new inferences and insights into the internal structure of SPB stars.

\subsection{Comparison with \citet{van_beeck_2021}}\label{subsec:vanbeeck}

We compare the mode frequencies extracted using our \texttt{mtNUFFT}/F-test multi-step procedure (see Section~\ref{subsec:multistep}) with those in \citetalias{van_beeck_2021}. They implement five distinct prewhitening strategies to investigate how mode frequency estimates are affected by (1) the choice of initial parameters for the optimization (i.e., parameter hinting), (2) the stopping criteria, and (3) whether parameter optimization at the end of each iterative step is performed using linear or nonlinear least-squares regression. Instead of iteratively extracting and optimizing the mode frequencies, we estimate them in one iteration based on two criteria
\begin{enumerate}
    \item $NW$ and $K$ for computing the \texttt{mtNUFFT} F-statistic, which is then tested against the $F$ distribution;
    \item The p-value threshold $\alpha$: $p < \alpha$.
\end{enumerate}
In Figure \ref{fig:van_beeck}, we compare our approach given multitaper parameters $NW=4$, $K=7$ and $p < 0.001$ with the \citetalias{van_beeck_2021} strategies for the SPB star KIC7760680. We see that the frequencies consistently found across all five strategies in \citetalias{van_beeck_2021}, that is, those that lie within 3$\sigma$ uncertainties, also coincide with those from the F-test detections. This suggests that the F-test extracts at least those mode frequencies that are in the most conservative \citetalias{van_beeck_2021} strategy 4, a hypothesis we have verified for all 38 SPBs in our sample. In addition, some of the signals we detect coincide with the ones solely detected by strategies 1, 2, 3, as well as some that are undetected by each of the strategies. These frequencies could potentially be new independent modes, manifestations of rotationally-split modes, or combinations of modes that retain their purely periodic behaviour. We also emphasize that any frequency in \citetalias{van_beeck_2021} not picked up by the F-test could hint at a damped nature for those g-modes; the modes potentially have quasi-periodic behaviour that the F-test discards.

We now quantitatively estimate how many of our multi-step F-test mode detections match the estimates in \citetalias{van_beeck_2021}. We compare our results for the frequency values specifically with Strategy 3 in \citetalias{van_beeck_2021}. This strategy chooses the highest amplitude peak in the (residual) LS periodogram at each iterative prewhitening step using a likelihood ratio test and then performs a nonlinear least-squares optimization to achieve the final frequency value. We compare with this strategy since it performs best for 34 out of the 38 pulsators used in this study. Note that \citetalias{van_beeck_2021} uses a Bayesian information criterion (BIC) to evaluate whether the frequency extracted in an iteration should be included in the nested regression model. This evaluation is in the form of a p-value for the likelihood ratio test $p_\mathrm{LRT} > 0.05$.

Table \ref{tab:KIC7760680_match} shows the frequency-by-frequency comparison between our approach and \citetalias{van_beeck_2021} Strategy 3 for KIC7760680. We use $NW=4$, $K=7$ and $p < 0.01$ as mentioned above. The frequency grid step in our \texttt{mtNUFFT}/F-test estimate after padding to a length of $M=20N$ is $0.05 f_\mathcal{R}$ (where $f_\mathcal{R} \equiv 1/T$ or $1/N$ is the Rayleigh resolution). This is smaller than the grid step used in \citetalias{van_beeck_2021}, which is about 0.1 $f_\mathcal{R}$. However, the frequencies we choose upon applying the multi-step F-test procedure have a minimum separation of 0.0018\,d$^{-1}$ (or $2.5 f_\mathcal{R}$) similar to \citetalias{van_beeck_2021} and 3$\sigma$ uncertainties of both approaches are used to find matches. Our approach detects a total of 114 frequencies, which is approximately twice the number extracted by \citetalias{van_beeck_2021}.  Among the 65 frequencies reported by \citetalias{van_beeck_2021}, 25 are found to match those identified in our analysis. Since some of the frequencies with the highest F-statistics were not included within the 3$\sigma$ matching criterion, we relaxed the threshold to 6$\sigma$ uncertainties for both approaches (while still enforcing the $f_\mathcal{R}$ resolution limit). This adjustment yielded five additional matches, which are highlighted in purple in the table. These frequencies exhibit very small jackknife uncertainties owing to their high F-statistics and precise frequency estimates, which explains why their deviations exceed the 3$\sigma$ limit.

In addition to the frequency estimates, Table \ref{tab:KIC7760680_match} lists the 1-$\sigma$ uncertainty on the estimates, which is computed using the jackknife estimator as described in \cite{patil_2025}. Specifically, the uncertainties are obtained 
by sequentially omitting one DPSS taper $v_k$
(i.e., removing a single eigencoefficient $y_k(f)$ at a time) and evaluating the resulting variability in the frequency detections, which provides an estimate of the variance and hence the standard deviation $\sigma$ of the F-test frequencies. The procedure is detailed below:

\begin{enumerate}
    \item Compute the delete-one F-statistic $F_{\setminus j}(f)$ by omitting the eigencoefficient $y_j(f)$ and the corresponding wave function $U_j(N, W; 0)$ from Equations \eqref{eq:F-test} and \eqref{eq:F-test_regr};
    \item Obtain delete-one F-test detections $\hat{f}_{0, {\setminus j}}$ by applying a threshold $p < \alpha$ to each of the delete-one F-statistic estimates $F_{\setminus j}(f)$;
    \item Estimate the jackknife variance of the detection $\hat{f}_{0}$ as
    \begin{equation}\label{eq:jackknife}
        \widehat{\mathrm{Var}}_J\{\hat{f_0}\} = \frac{K-1}{K} \sum_{j=0}^{K-1} \left[\hat{f}_{0, {\setminus j}} -\hat{f}_{0, {\setminus \bullet}}\right]^2 , 
    \end{equation}
\end{enumerate} where $\hat{f}_{0, {\setminus \bullet}}$ is the average of $\hat{f}_{0, {\setminus j}}$. Our jackknifed uncertainty estimates are then  given by $ \sigma_{\hat{f_0}} = \sqrt{\widehat{\mathrm{Var}}_J\{\hat{f_0}\}}$. 
Equation 12 in \cite{patil_2025} provides an analytical expression for the variance, $\mathrm{Var}\,\{\hat{f_0}\}$, of a periodic signal frequency estimate. This variance is typically a few percent larger than the Cramér–Rao bound \citep{rife_1976, thomson_2007}, which represents the theoretical minimum variance achievable. Consequently, the analytical expression serves as a practical upper bound on frequency precision. We use it to validate our jackknife variance estimates, which turn out to be roughly consistent with the \citetalias{van_beeck_2021} estimates. 
Seven of our 26 frequency uncertainties $\sigma_f$ (27\%) in Table \ref{tab:KIC7760680_match} are smaller than those reported in \citetalias{van_beeck_2021}, and the rest are within an order of magnitude. This shows that we do not lose much frequency precision when working in the frequency domain using the F-test, and any resolution lost in the power spectrum due to tapering is recovered by the F-test.

Table \ref{tab:KIC7760680_match} also lists the F-statistic, amplitude $A$ and phase $\phi$ estimates for each frequency, computed using Equations \ref{eq:amplitude} and \ref{eq:phase}), as well as the jackknife uncertainties on amplitude $\sigma_{\widehat{A}}$ and phase $\sigma_{\widehat{\phi}}$ [refer to Equation \eqref{eq:jackknife}].  To validate our results, we compare our amplitude estimates with those from \citetalias{van_beeck_2021} in Fig.~\ref{fig:amp_phase}, finding excellent agreement. The phase estimates ($\phi$) show larger deviations from those of \citetalias{van_beeck_2021}, but generally remain within the 95\% confidence interval. 

We note that the phase of a multiperiodic harmonic signal is inherently challenging to estimate. Our method employs a one-step FFT-based estimation of $\phi$, in contrast to the iterative time-domain prewhitening approach used by \citetalias{van_beeck_2021}. Finally, Table \ref{tab:KIC7760680_match} also includes the independent and combination frequency labels given by \citetalias{van_beeck_2021} for their frequencies. Our approach matches $14$ of the independent and $16$ of the combination frequencies reported by \citetalias{van_beeck_2021}.

We perform such extraction for each of the 38 stars, the results of which are summarized in Table \ref{tab:match_allSPB}. The total number of g-mode detections using our methodology ($T$) are compared with the total number of detections $T_\mathrm{VB}$ in \citetalias{van_beeck_2021}, which is equal to $T_\mathrm{VB} = I_\mathrm{VB} + C_\mathrm{VB}$ (\citetalias{van_beeck_2021} distinguishes between independent and combination frequencies in their frequency detection algorithm). We then find 3$\sigma$ matches between our mode frequencies and those in \citetalias{van_beeck_2021} to compute the matching frequencies $T_m$. These matches frequencies are divided into independent and combination frequencies depending on the non-linear label of the matching frequency in \citetalias{van_beeck_2021} $T_m = I_M + C_m$.

We perform this extraction for each of the 38 stars, and the results are summarized in Table \ref{tab:match_allSPB}. For each star, we compare the total number of g-mode detections obtained with our methodology ($T$) to the total number of detections reported by \citetalias{van_beeck_2021}, denoted as $T_\mathrm{VB}$. Note that \citetalias{van_beeck_2021} distinguishes between independent ($I_\mathrm{VB}$) and combination ($C_\mathrm{VB}$) frequencies in their detection algorithm, such that $T_\mathrm{VB} = I_\mathrm{VB} + C_\mathrm{VB}$. These quantities are listed in the table. We then identify 3$\sigma$ matches between our detected mode frequencies and those reported by \citetalias{van_beeck_2021}, computing the number of matching frequencies $T_m$ for each star. The matched frequencies are further divided into independent and combination frequencies based on the non-linear ID (refer to Table \ref{tab:KIC7760680_match}) of the corresponding frequency in \citetalias{van_beeck_2021}, such that $T_m = I_m + C_m$.

In blue, we highlight the results using $NW=4, K=7$ and $p < 0.01$, which are the same as those shown in Figure \ref{fig:van_beeck} and Table \ref{tab:KIC7760680_match} for KIC7760680. To show how our results vary based on the choice of $NW, K$ and the p-value, we include overall results for different multitaper bandwidths $NW=3$ and $NW=4$ (and thereby the number of tapers $K=2NW-1$) and for different p-values, $p < 0.01$ and $p <0.005$.

Table \ref{tab:match_allSPB} shows that, for a fixed significance threshold of e.g., $p < 0.005$, the number of detections usually (but not always) decreases as the time–bandwidth product $NW$ is reduced. Conversely, when $NW$ is held constant, the number of detections decreases with smaller p-values, reflecting the more conservative nature of stricter thresholds.

\begin{figure}
    \centering
    \includegraphics[width=\linewidth]{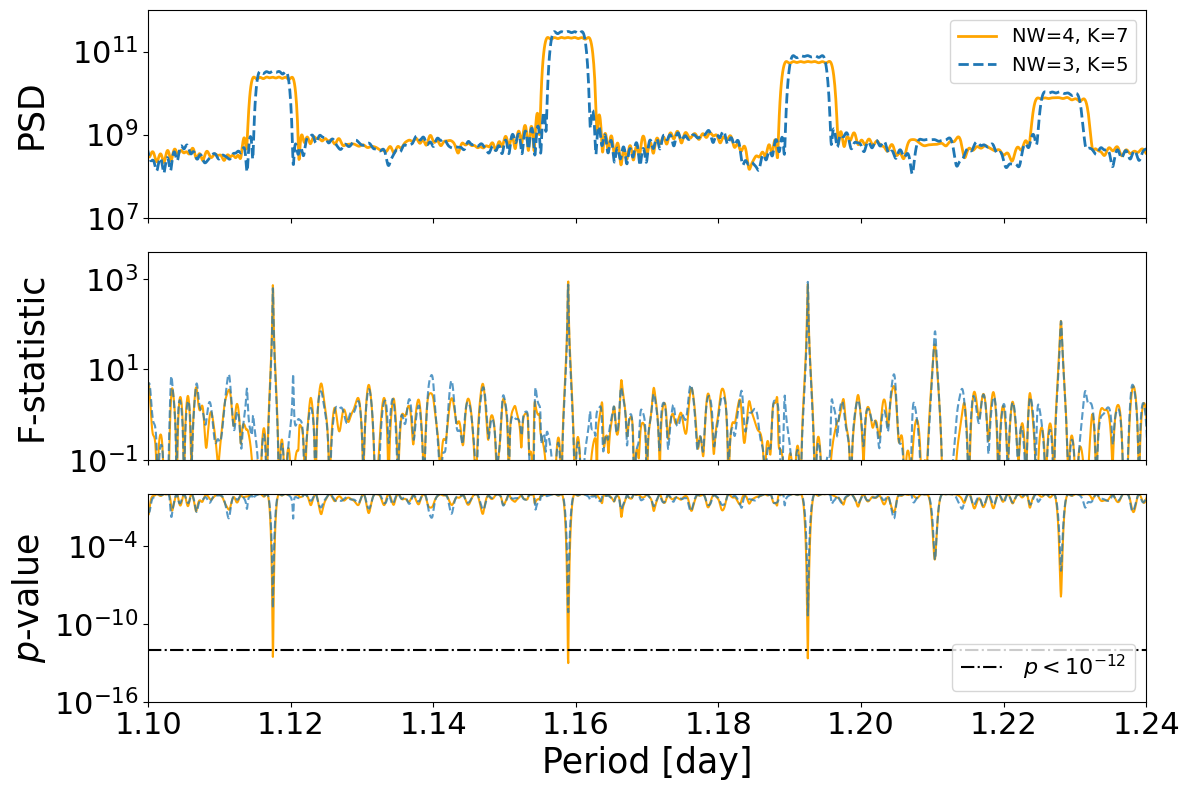}
    \caption{Effect of changing the time–bandwidth product $NW$ and number of tapers $K$ from $NW=4, K=7$ (orange) to $NW=3, K=5$ (blue) for KIC7760680. The top panel shows the \texttt{mtNUFFT} periodograms, while the middle and bottom panels display the corresponding F-statistics and p-values. The bottom panel also includes a $p < 10^{-12}$ threshold, highlighting that $NW=4$ yields three detections, whereas $NW=3$ produces none.}
    \label{fig:nw-change}
\end{figure}

\begin{figure*}[!ht]
    \centering
    \includegraphics[width=\linewidth]{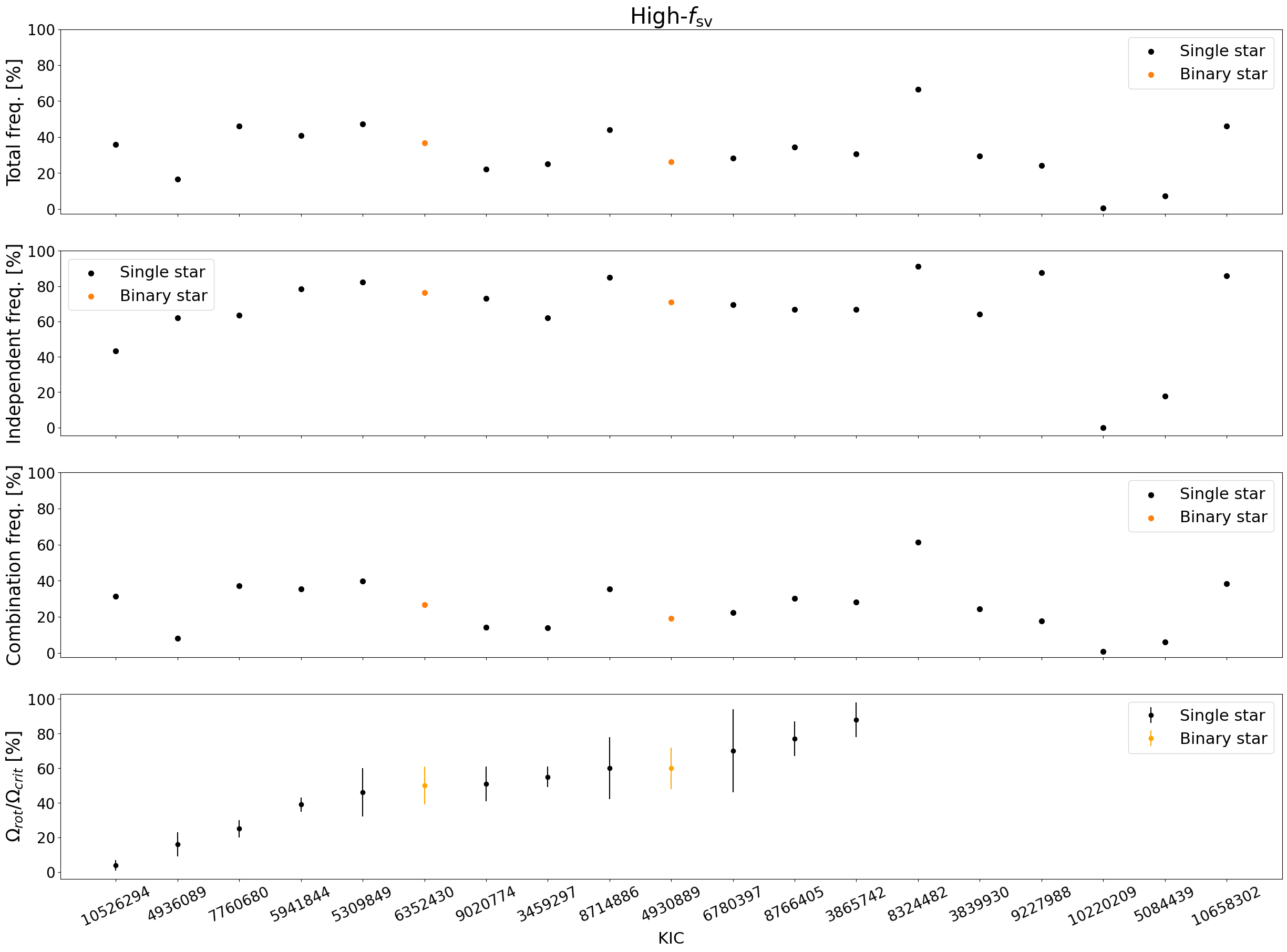}
    \caption{Comparison of our approach with \citetalias{van_beeck_2021} (Strategy 3) for SPBs identified as High-$f_{sv}$ stars. The top and middle panels show the percentage of matched frequencies (total, independent, and combination). The bottom panel illustrates the near-core rotation frequency $\Omega_{\mathrm{rot}}$ relative to the star's critical rotation frequency $\Omega_{\mathrm{crit}}$ deduced by \citetalias{pedersen_2021}. In this panel, we see that the match percentage does not seem to have dependence on the rotation frequency. Single and binary stars \citep{papics_2013, Papics2017} are shown in black and orange, respectively. More of the stars might be binaries, but two of them are confirmed. 
    Also, for some stars we do not have an estimate of the (near-core) stellar rotation frequency.}
    \label{fig:high-fsv}
\end{figure*}

Figure \ref{fig:nw-change} illustrates the effect of changing mutitaper (\texttt{mtNUFFT}) parameters from $NW=4, K=7$ (orange) to $NW=3, K=5$ (blue) for KIC7760680. While both settings recover the same g~modes, reducing $NW$ leads to larger p-values, reflecting reduced robustness of the F-test. At the same time, the narrower spectral concentration with 
$NW=3$ produces slightly sharper PSD peaks, whereas the higher $NW$ smooths the spectrum but yields more statistically significant detections.

Increasing $NW$, and consequently $K$, enhances the robustness of the multitaper F-test against noise, but it reduces the ability to resolve closely spaced periodic signals. In theoretical terms, this increases the degrees of freedom, $2K-2$, of the denominator (noise term) in the variance ratio defining F-statistic (see Equation \ref{eq:F-test}). For SPB stars with sparse power spectra containing only a few isolated g~modes, choosing a moderate-to-large $NW$ improves detection reliability. Conversely, for dense spectra with many closely spaced modes (e.g., KIC 3240411), a smaller $NW$ is preferable to preserve frequency resolution, even if this comes at the cost of higher variance. 

Additionally, for stars such as KIC5084439 and KIC10220209, using $NW = 3$ and $K = 5$ produces better results, likely because these stars contain large gaps in their time series, leading to poorly resolved modes in the data. KIC5084439 has a gap of approximately 190 days, covering more than a quarter of the total time-series length, while KIC10220209 has four gaps of roughly 100 days each, with the largest spanning about 280 days, corresponding to one-fifth of the total time-series duration. Such gaps are considered large because they significantly reduce the effective duty cycle of the observations, thereby decreasing the frequency resolution and introducing strong windowing effects. In other words, the smaller number of available samples, $N$, results in a larger effective bandwidth, $W$, for a given time–bandwidth product, $NW$. To increase the number of detections for such stars, one could consider using a smaller $NW$ together with a less conservative $p$-value threshold. 

We also note that KIC4939281 and KIC9715425 are outbursting stars. These outbursting stars are defined in \citetalias{van_beeck_2021} as those that display outburst-like features in their light curves, characterized by significant deviations from the baseline brightness, which remain visible in the residual light curve after prewhitening. Because the F-test is sensitive to small frequency shifts (or any deviations from strict periodicity), it likely identifies only a limited number of modes in these outbursting cases. Aside from these special instances, $NW=4, K=7$ with $p < 0.01$ yields consistently reliable results across most stars.

\begin{figure*}[!ht]
    \centering
    \includegraphics[width=\linewidth]{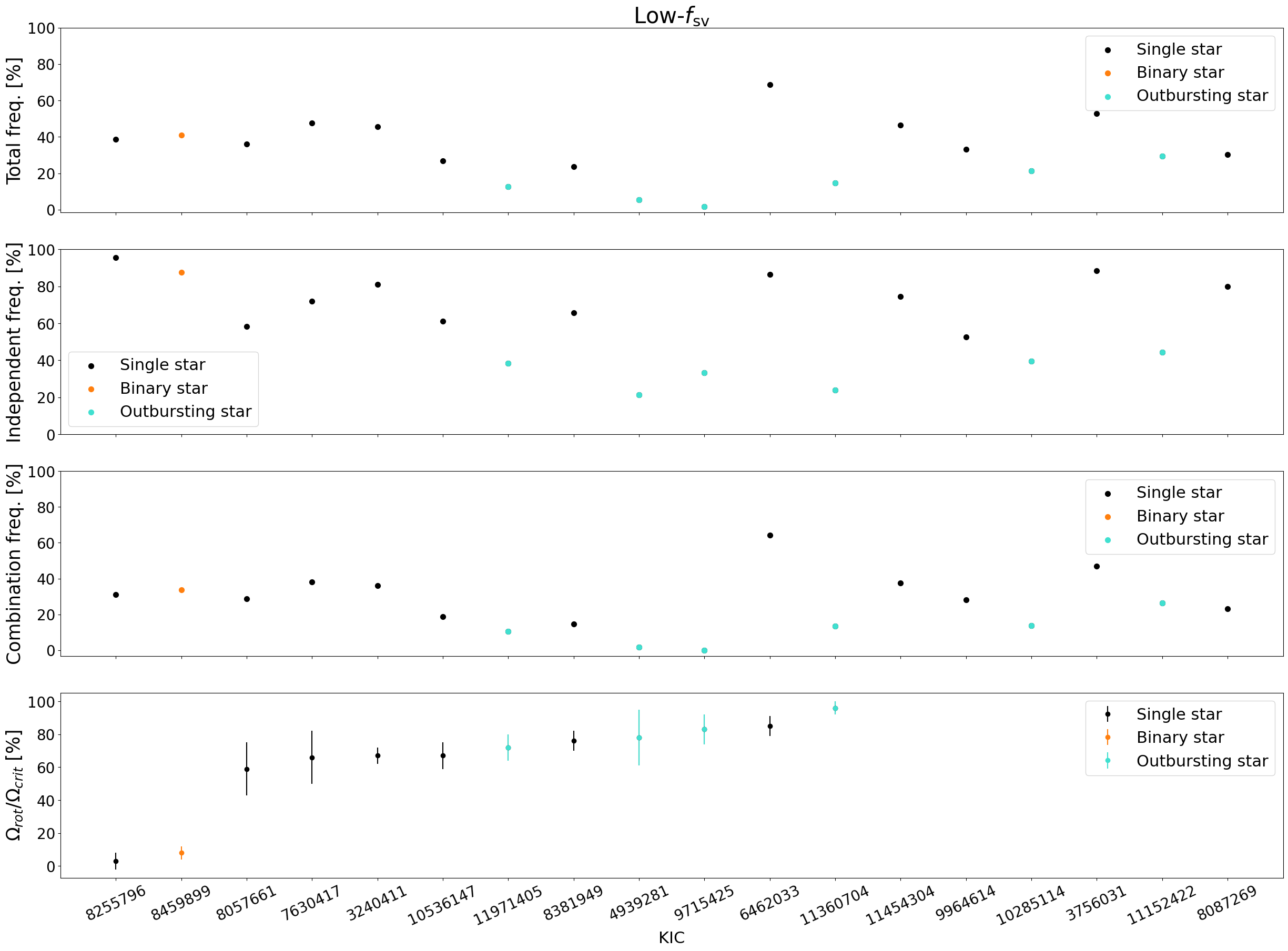}
    \caption{Same as Figure \ref{fig:high-fsv} but for the Low-$f_{sv}$ stars. KIC8459899 is marked in orange as it is suspected to be a binary star \citep{lehmann_2011}.}
    \label{fig:low-fsv}
\end{figure*}

\begin{figure*}
    \centering
    \includegraphics[width=\linewidth]{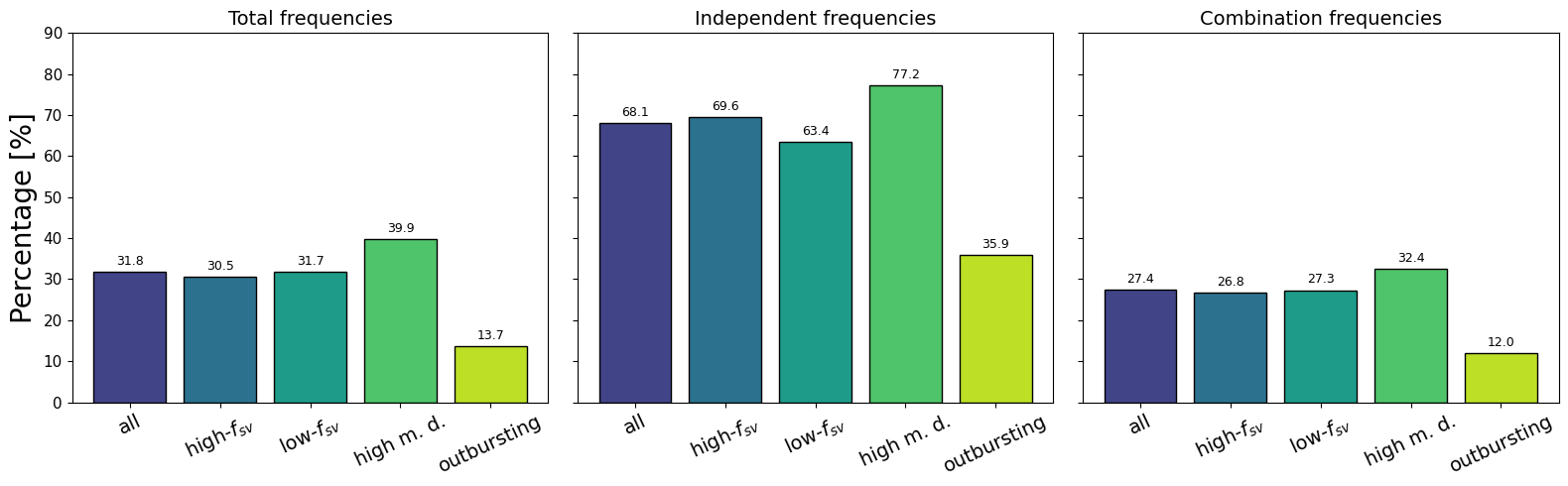}
    \caption{Match percentage between our F-test detections given $NW = 4, K=7$ and $p < 0.1$ and \citetalias{van_beeck_2021}, Strategy 3. The left, middle, and right panels show the percentage of total, independent, and combination matched frequencies, respectively. We consider the sub-classes among the SPB stars defined by \citetalias{van_beeck_2021}, to which we refer for their detailed definitions.}
    \label{fig:perc-comp}
\end{figure*}

Figure \ref{fig:high-fsv} shows the percentage of g-mode matches between our method and Strategy 3 for the 19 SPBs classified as High-$f_{sv}$ in \citetalias{van_beeck_2021}. These stars are characterized by high SNRs in their frequency spectra and, correspondingly, a high $f_{sv}$: a metric quantifying how well the regression models (after prewhitening) describe the light curve of a star. We observe that on average, a higher percentage of independent mode frequencies in \citetalias{van_beeck_2021} have matches with our F-test detections as opposed to combination frequencies. In addition, the percentages are not functions of the rotation frequencies of the stars (see bottom panel of Figure \ref{fig:high-fsv}), i.e., there appears to be no dependence on stellar rotation. Instead, the percentages appear to depend on mode density in the power spectrum, since combination frequencies are more likely to occur in mode-dense regions.  The independent versus combination mode trend is also seen in SPB stars labelled as Low-$f_{sv}$ as seen in Figure \ref{fig:low-fsv}, which include high mode density and outbursting stars.

Note that some stars, e.g., KIC10220209 and KIC5084439 in Figure \ref{fig:high-fsv}, and KIC4939281 and KIC9715425 in Figure \ref{fig:low-fsv}, show very few detections of g-modes (see also Table \ref{tab:match_allSPB}). These stars fall into the categories of having large gaps in their time series or being outbursting stars. For such cases, additional modes can be recovered by adopting a less conservative $p$-value threshold for the F-test, as demonstrated for KIC9715425 in Figure \ref{fig:KIC9715425} in Appendix\,A.

The overall match percentage are similar for High-$f_{sv}$ and Low-$f_{sv}$ stars. These percentages are shown in Figure\,\ref{fig:perc-comp}. We see that the Low-$f_{sv}$ and outbursting stars (as defined in \citetalias{van_beeck_2021}), which have beating cycles that are not well covered by the 4-year Kepler light curves, have fewer mode frequency matches. The SPB stars revealing high mode density in the definition of \citetalias{van_beeck_2021} have the highest matching percentages between the two frequency analysis methods.

\begin{figure*}
   \centering
   \includegraphics[width=\linewidth]{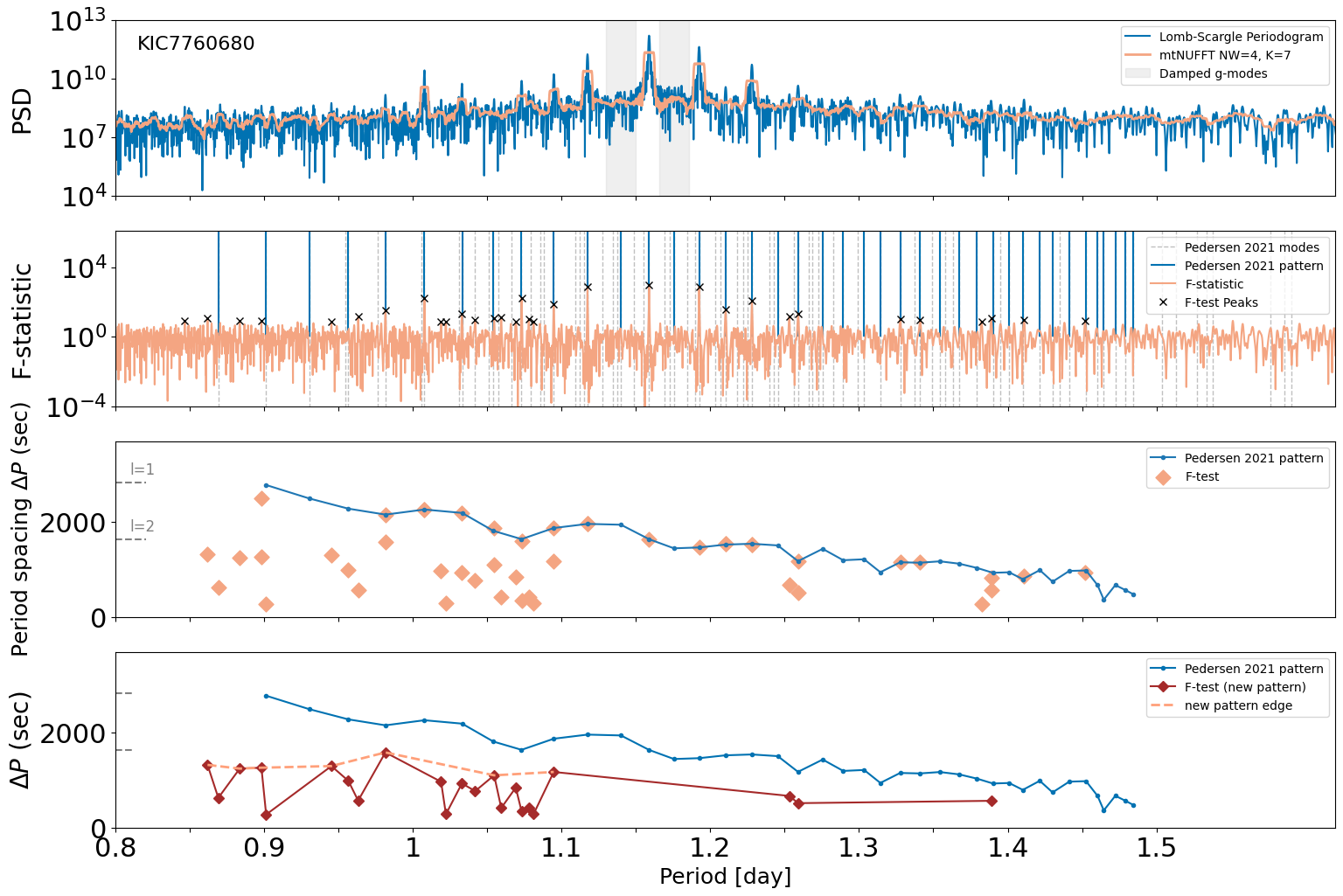}
      \caption{Comparison of our F-test detections with the period-spacing pattern reported by \citetalias{pedersen_2021} for KIC7760680. The top panel shows the mtNNUFFT spectrum (orange) and LS periodogram (blue), suggesting the presence of damped g~modes (grey). The second panel displays the F-statistic (orange) with our detected modes (black crosses) and \citetalias{pedersen_2021} modes (grey dashed lines); modes belonging to the primary period-spacing pattern are highlighted in blue. In the third panel, orange diamonds mark our detections matching the PD21 primary pattern (blue curve) according to the match criteria in Table~\ref{tab:KIC7760680_match}, while additional orange points denote secondary $\Delta P$ sequences from unmatched and nearby \citetalias{pedersen_2021} modes beyond the $3\sigma$ threshold. The bottom panel shows a secondary pattern (brown) near the expected position of $l=2$ modes (grey dashed line on the y-axis), outlined by a dashed orange line.}
         \label{fig:KIC7760680}
\end{figure*}

\begin{figure*}
   \centering
   \includegraphics[width=\linewidth]{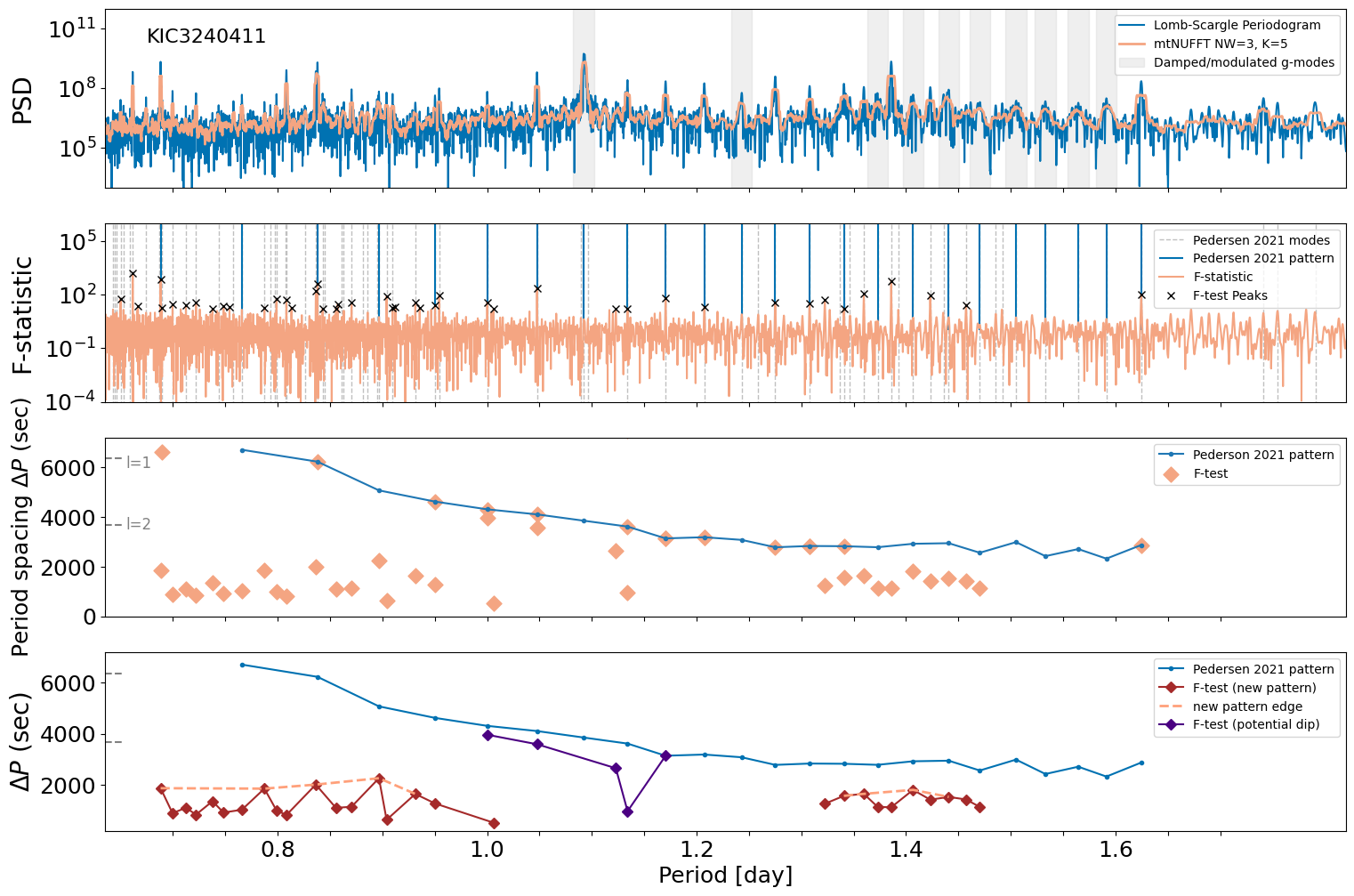}
      \caption{Same as Fig.~\ref{fig:KIC7760680}, but for KIC3240411, with the addition of a potential dip feature in the last panel (shown in violet). Here, the new period-spacing pattern (brown) is divided into separate segments.}
         \label{fig:KIC3240411}
\end{figure*}

\subsection{Comparison with \cite{pedersen_2021}}

We next examine whether our method can recover the frequencies typically used for asteroseismic modelling from period spacing patterns. For this, we compare our results with those of \citetalias{pedersen_2021}, who analyzed 26 of the 38 SPB stars in our sample using Strategy 1 from \citetalias{van_beeck_2021}. Unlike the \citetalias{van_beeck_2021} approach, \citetalias{pedersen_2021} did not impose a limit of 200 significant frequencies but extended their prewhitening analysis to all signals with amplitudes exceeding an SNR threshold of 4.0. In addition, they interpret the frequencies with the purpose of asteroseismic modelling. This requires one to obtain period spacing patterns for mode identification and derivation of the internal rotation frequency. Since \citetalias{van_beeck_2021}'s Strategy 1 (which is used in \citetalias{pedersen_2021}) is known to be less performant than Strategy 3, we do not perform a frequency-by-frequency comparison with \citetalias{pedersen_2021} (as in Section \ref{subsec:vanbeeck}). Instead, we directly compare the period spacing patterns present in our F-test mode detections with those suggested by \citetalias{pedersen_2021}.

To compare our detected period-spacing patterns with those reported by \citetalias{pedersen_2021}, we identify matching modes using the criteria in Table~\ref{tab:KIC7760680_match}: frequency agreement within the Rayleigh limit and within $3\sigma$ uncertainties, accounting for precision differences from the F-test and prewhitening. Figure~\ref{fig:KIC7760680} shows the resulting comparison for KIC7760680. The mtNNUFFT spectrum and corresponding F-statistic reveal modes consistent with the primary period-spacing pattern of \citetalias{pedersen_2021}. Matched detections are plotted as orange diamonds along the PD21 sequence, while additional orange points mark secondary $\Delta P$ sequences derived from consecutive unmatched detections as well as nearby \citetalias{pedersen_2021} modes. \citetalias{pedersen_2021} modes are only included in these secondary sequences if they lie adjacent (beyond the $3\sigma$ match threshold) to one of our unmatched g~modes. A potential secondary pattern (brown) is also automatically identified from modes lying below the primary sequence.

KIC7760680 serves as a benchmark SPB star, exhibiting the longest observed sequence of consecutive dipole $l=1$ g~modes in radial order $n$ (with additional frequencies interspersed). The period-spacing pattern of these modes clearly shows the combined effects of rotation and near-core chemical mixing (as reported by \citet{Papics2015}). Owing to its extensive previous asteroseismic modelling \citep{moravveji_2016,Michielsen2021}, this star provides an ideal test case for assessing our frequency-detection method’s ability to recover period-spacing patterns. As shown in Figure~\ref{fig:KIC7760680}, we recover most of the g~modes reported by \citetalias{pedersen_2021}. We also observe the presence of an additional period-spacing pattern, potentially corresponding to modes of degree $l=2$. Additionally, we note that two missing frequencies adjacent to the main peak in the period-spacing pattern were not detected by the F-test, likely due to low power (unresolved modes) or damping broadening their frequency profiles.

KIC3240411, in contrast, represents a different case. It is the hottest known SPB with a dense and complex power spectrum of dipole modes \citep{Szewczuk2018}. Despite four years of Kepler data, several extracted peaks remain difficult to interpret, with many identified as possible combination frequencies or unresolved signals. This makes KIC3240411 a demanding but valuable testbed to assess whether new approaches can outperform the standard iterative prewhitening procedure in disentangling independent g-mode frequencies. Similar to KIC7760680, we find secondary period spacing patterns as shown in Figure~\ref{fig:KIC3240411}. After the secondary pattern is automatically constructed using modes lying below the \citetalias{pedersen_2021} pattern, we manually pick out features such as a potential dip in the pattern, which could tell us about processes such as mode trapping. \citep{Szewczuk2018} also point multiple possible patterns in this star that are similar to the ones we find.

Note that the g~mode in KIC3240411 with a period of $\approx$1.1\,d has a high amplitude but a weak F-statistic, which is why it is not detected by the F-test. This behaviour suggests the presence of resonant, damped, or modulated g~modes, which we will discuss in detail in a forthcoming paper.

We provide the analogous figures to Figs.~\ref{fig:KIC7760680} and \ref{fig:KIC3240411} for the 24 additional SPB stars treated in \citet{pedersen_2021} in Appendix\,A. The automatic identification of period-spacing patterns, as demonstrated for KIC7760680, differs for other stars depending on the specific mode distributions and morphological features present in their spectra. We apply an automatic approach to first find secondary patterns and manually check them to find general trends and peculiar features. 

Overall, while we find less frequencies than in \citetalias{pedersen_2021}, we present an efficient and more statistically sound approach to deduce mode frequencies and period spacing patterns. Asteroseismic modelling may likely still need prewhitening for sufficient g-mode detection and pattern recognition to secure spacing patterns. Yet, using our method, one could start off with a list of frequencies to help expedite this process and detect g~modes with long lifetimes more securely than combined frequency and time domain analysis in an iterative manner.

\section{Conclusions and outlook}

The new frequency analysis methodology presented in this paper, which is based on the multitaper \texttt{NUFFT} periodogram and its extension with the F-test, is computationally efficient and offers the potential to help understand the content of the frequency spectra of SPB pulsators and the morphology of their period spacing patterns. Such observational information thereby aids in disentangling the excitation mechanisms and behaviour of g~modes in SPB and other main-sequence pulsators, such as $\gamma\,$Doradus stars.

While we tuned our application to the challenging case of g~mode detection, there is no mathematical reason why our method could not also be applied to other types of pulsators that require a prewhitening procedure before asteroseismology can be performed. Prewhitening involves repeated computation of the LS or similar types of computationally efficient periodograms. However, extraction of the highest-amplitude or highest-SNR signal, followed by sinusoidal fitting in the time domain from non-linear regression is cumbersome. It requires fitting the entire multi-sinusoidal model using a non-linear least-squares approach at each step. The computational cost scales roughly as $\mathcal{O}(I N N_f^3)$, where $N$ is the number of time samples, $N_f$ the number of extracted frequencies, and $I$ the number of iterations per fit. This cubic scaling in $N_f$ makes the method prohibitively expensive when hundreds of g-modes are present. By constrast, linear prewhitening (using LS with linear sinusoidal refinements) scales more gently as $\mathcal{O}(N \log N + N_f M + N_f I N)$, where $M$ is the number of frequency bins. While more efficient than non-linear prewhitening, it grows linearly with $N_f$ and remains costly for dense SPB spectra. Above all, use of any of such techniques implies data manipulation by subtracting imperfect regression models numerous times before the final list of significant frequencies is deduced. This `systematic error' is propagated along the asteroseismic modelling process because the frequencies on this list are subsequently used as input to derive interpretable observables for asteroseismic interpretations, such as mode period spacing patterns.

The multitaper F-test is substantially faster because it evaluates significance for all frequencies simultaneously with complexity $\mathcal{O}(K N \log N + M K)$, scaling only with the number of tapers $K$ and remaining independent of $N_f$. This makes the F-test at least an order of magnitude faster than prewhitening approaches. While we find our approach to deliver less frequencies, it is complementary to prewhitening and does not suffer from iterative data manipulation. In this sense, our methodology brings new opportunities for future applications to g-mode pulsators, whose most important asteroseismic observables -- the modes' period spacing patterns -- may suffer from imperfections of the time-consuming prewhitening procedure \citep{BM2021}. Our method is entirely suitable to perform fast and objective frequency extraction, serving as an initial procedure to apply mode identification and asteroseismology to large populations of main-sequence pulsators as those in \citet{Balona2011,Mombarg2024}. This would deliver a complementary approach  to ``boutique modelling'' of particular stars or small samples of pulsators as treated by \citet{pedersen_2021}. Our fast frequency extraction and period search method can potentially pave the way to test and calibrate stellar structure and evolution models from large samples of intermediate- and high-mass pulsators.

The multitaper F-test also provides more information than the power spectrum alone. In addition to precisely identifying statistically significant oscillations in the Fourier domain, it also carries diagnostic value on the underlying physical excitation mechanism. For example, while a purely periodic g~mode will typically yield both high power and a strong F-test detection, a mode with high amplitude but a weak F-statistic may instead suggest the presence of local or partial wave damping in the stellar interior or near the surface of the star. This opens up opportunities for the improvement of non-adiabatic asteroseismology, and we plan to dedicate a separate future study to the level of mode damping occurring in the SPB sample treated here.

\begin{acknowledgements}
AAP is supported by an LSST-DA Catalyst Fellowship; this publication was thus made possible through the support of Grant 62192 from the John Templeton Foundation to LSST-DA.
CA receives funding from the European Research Council (ERC) under the Horizon Europe programme (Synergy Grant agreement N$^\circ$101071505: 4D-STAR). 
JVB acknowledges funding from the European Research Council (Consolidator Grant N$^\circ$101000296: DipolarSound). MGP recognizes support by the Professor Harry Messel Research Fellowship in Physics Endowment, at the University of Sydney.
While partially funded by the European Union, views and opinions expressed are however those of the authors only and do not necessarily reflect those of the European Union or the European Research Council.
Neither the European Union nor the granting authority can be held responsible for them.
\end{acknowledgements}


\bibliography{aa.bib}

@ARTICLE{Aerts2025,
       author = {{Aerts}, Conny and {Van Reeth}, Timothy and {Mombarg}, Joey S.~G. and {Hey}, Daniel},
        title = "{Evolution of the near-core rotation frequency of 2497 intermediate-mass stars from their dominant gravito-inertial mode}",
      journal = {\aap},
         year = 2025,
        month = mar,
       volume = {695},
          eid = {A214},
        pages = {A214},
          doi = {10.1051/0004-6361/202452691},
archivePrefix = {arXiv},
       eprint = {2502.17692},
 primaryClass = {astro-ph.SR},
       adsurl = {https://ui.adsabs.harvard.edu/abs/2025A&A...695A.214A}
}

@ARTICLE{Szewczuk2017,
       author = {{Szewczuk}, Wojciech and {Daszy{\'n}ska-Daszkiewicz}, Jadwiga},
        title = "{Domains of pulsational instability of low-frequency modes in rotating upper main sequence stars}",
      journal = {\mnras},
         year = 2017,
        month = jul,
       volume = {469},
       number = {1},
        pages = {13-46},
          doi = {10.1093/mnras/stx738},
archivePrefix = {arXiv},
       eprint = {1703.08075},
 primaryClass = {astro-ph.SR},
       adsurl = {https://ui.adsabs.harvard.edu/abs/2017MNRAS.469...13S}
}

@ARTICLE{Antoci2019,
       author = {{Antoci}, V. and {Cunha}, M.~S. and {Bowman}, D.~M. and {Murphy}, S.~J. and {Kurtz}, D.~W. and {Bedding}, T.~R. and {Borre}, C.~C. and {Christophe}, S. and {Daszy{\'n}ska-Daszkiewicz}, J. and {Fox-Machado}, L. and {Garc{\'\i}a Hern{\'a}ndez}, A. and {Ghasemi}, H. and {Handberg}, R. and {Hansen}, H. and {Hasanzadeh}, A. and {Houdek}, G. and {Johnston}, C. and {Justesen}, A.~B. and {Kahraman Alicavus}, F. and {Kotysz}, K. and {Latham}, D. and {Matthews}, J.~M. and {M{\o}nster}, J. and {Niemczura}, E. and {Paunzen}, E. and {S{\'a}nchez Arias}, J.~P. and {Pigulski}, A. and {Pepper}, J. and {Richey-Yowell}, T. and {Safari}, H. and {Seager}, S. and {Smalley}, B. and {Shutt}, T. and {S{\'o}dor}, A. and {Su{\'a}rez}, J. -C. and {Tkachenko}, A. and {Wu}, T. and {Zwintz}, K. and {Barcel{\'o} Forteza}, S. and {Brunsden}, E. and {Bogn{\'a}r}, Z. and {Buzasi}, D.~L. and {Chowdhury}, S. and {De Cat}, P. and {Evans}, J.~A. and {Guo}, Z. and {Guzik}, J.~A. and {Jevtic}, N. and {Lampens}, P. and {Lares Martiz}, M. and {Lovekin}, C. and {Li}, G. and {Mirouh}, G.~M. and {Mkrtichian}, D. and {Monteiro}, M.~J.~P.~F.~G. and {Nemec}, J.~M. and {Ouazzani}, R. -M. and {Pascual-Granado}, J. and {Reese}, D.~R. and {Rieutord}, M. and {Rodon}, J.~R. and {Skarka}, M. and {Sowicka}, P. and {Stateva}, I. and {Szab{\'o}}, R. and {Weiss}, W.~W.},
        title = "{The first view of {\ensuremath{\delta}} Scuti and {\ensuremath{\gamma}} Doradus stars with the TESS mission}",
      journal = {\mnras},
         year = 2019,
        month = dec,
       volume = {490},
       number = {3},
        pages = {4040-4059},
          doi = {10.1093/mnras/stz2787},
archivePrefix = {arXiv},
       eprint = {1909.12018},
 primaryClass = {astro-ph.SR},
       adsurl = {https://ui.adsabs.harvard.edu/abs/2019MNRAS.490.4040A}
}

@ARTICLE{VanReeth2016,
       author = {{Van Reeth}, T. and {Tkachenko}, A. and {Aerts}, C.},
        title = "{Interior rotation of a sample of {\ensuremath{\gamma}} Doradus stars from ensemble modelling of their gravity-mode period spacings}",
      journal = {\aap},
         year = 2016,
        month = oct,
       volume = {593},
          eid = {A120},
        pages = {A120},
          doi = {10.1051/0004-6361/201628616},
archivePrefix = {arXiv},
       eprint = {1607.00820},
 primaryClass = {astro-ph.SR},
       adsurl = {https://ui.adsabs.harvard.edu/abs/2016A&A...593A.120V}
}

@ARTICLE{Bowman2019,
       author = {{Bowman}, Dominic M. and {Burssens}, Siemen and {Pedersen}, May G. and {Johnston}, Cole and {Aerts}, Conny and {Buysschaert}, Bram and {Michielsen}, Mathias and {Tkachenko}, Andrew and {Rogers}, Tamara M. and {Edelmann}, Philipp V.~F. and {Ratnasingam}, Rathish P. and {Sim{\'o}n-D{\'\i}az}, Sergio and {Castro}, Norberto and {Moravveji}, Ehsan and {Pope}, Benjamin J.~S. and {White}, Timothy R. and {De Cat}, Peter},
        title = "{Low-frequency gravity waves in blue supergiants revealed by high-precision space photometry}",
      journal = {Nature Astronomy},
         year = 2019,
        month = may,
       volume = {3},
        pages = {760-765},
          doi = {10.1038/s41550-019-0768-1},
archivePrefix = {arXiv},
       eprint = {1905.02120},
 primaryClass = {astro-ph.SR},
       adsurl = {https://ui.adsabs.harvard.edu/abs/2019NatAs...3..760B}
}

@ARTICLE{DeRidder2023,
       author = {{Gaia Collaboration} and {De Ridder}, J. and {Ripepi}, V. and {Aerts}, C. and {Palaversa}, L. },
        title = "{Gaia Data Release 3. Pulsations in main sequence OBAF-type stars}",
      journal = {\aap},
         year = 2023,
        month = jun,
       volume = {674},
          eid = {A36},
        pages = {A36},
          doi = {10.1051/0004-6361/202243767},
archivePrefix = {arXiv},
       eprint = {2206.06075},
 primaryClass = {astro-ph.SR},
       adsurl = {https://ui.adsabs.harvard.edu/abs/2023A&A...674A..36G}
}

@ARTICLE{VanBeeck2024,
       author = {{Van Beeck}, J. and {Van Hoolst}, T. and {Aerts}, C. and {Fuller}, J.},
        title = "{Non-linear three-mode coupling of gravity modes in rotating slowly pulsating B stars. Stationary solutions and modeling potential}",
      journal = {\aap},
         year = 2024,
        month = jul,
       volume = {687},
          eid = {A265},
        pages = {A265},
          doi = {10.1051/0004-6361/202348369},
archivePrefix = {arXiv},
       eprint = {2311.02972},
 primaryClass = {astro-ph.SR},
       adsurl = {https://ui.adsabs.harvard.edu/abs/2024A&A...687A.265V}
}

@ARTICLE{Szewczuk2021,
       author = {{Szewczuk}, Wojciech and {Walczak}, Przemys{\l}aw and {Daszy{\'n}ska-Daszkiewicz}, Jadwiga},
        title = "{Variability of newly identified B-type stars observed by Kepler}",
      journal = {\mnras},
         year = 2021,
        month = jun,
       volume = {503},
       number = {4},
        pages = {5894-5928},
          doi = {10.1093/mnras/stab683},
archivePrefix = {arXiv},
       eprint = {2103.06146},
 primaryClass = {astro-ph.SR},
       adsurl = {https://ui.adsabs.harvard.edu/abs/2021MNRAS.503.5894S}
}

@ARTICLE{Pedersen2022,
       author = {{Pedersen}, May G.},
        title = "{On the Diversity of Mixing and Helium Core Masses of B-type Dwarfs from Gravity-mode Asteroseismology}",
      journal = {\apj},
         year = 2022,
        month = may,
       volume = {930},
       number = {1},
          eid = {94},
        pages = {94},
          doi = {10.3847/1538-4357/ac5b05},
archivePrefix = {arXiv},
       eprint = {2203.02046},
 primaryClass = {astro-ph.SR},
       adsurl = {https://ui.adsabs.harvard.edu/abs/2022ApJ...930...94P}
}

@ARTICLE{Lecoanet2022,
       author = {{Lecoanet}, Daniel and {Bowman}, Dominic M. and {Van Reeth}, Timothy},
        title = "{Asteroseismic inference of the near-core magnetic field strength in the main-sequence B star HD 43317}",
      journal = {\mnras},
         year = 2022,
        month = may,
       volume = {512},
       number = {1},
        pages = {L16-L20},
          doi = {10.1093/mnrasl/slac013},
archivePrefix = {arXiv},
       eprint = {2202.03440},
 primaryClass = {astro-ph.SR},
       adsurl = {https://ui.adsabs.harvard.edu/abs/2022MNRAS.512L..16L}
}

@ARTICLE{Pedersen2018,
       author = {{Pedersen}, M.~G. and {Aerts}, C. and {P{\'a}pics}, P.~I. and {Rogers}, T.~M.},
        title = "{The shape of convective core overshooting from gravity-mode period spacings}",
      journal = {\aap},
         year = 2018,
        month = jul,
       volume = {614},
          eid = {A128},
        pages = {A128},
          doi = {10.1051/0004-6361/201732317},
archivePrefix = {arXiv},
       eprint = {1802.02051},
 primaryClass = {astro-ph.SR},
       adsurl = {https://ui.adsabs.harvard.edu/abs/2018A&A...614A.128P}
}

@ARTICLE{Papics2017,
       author = {{P{\'a}pics}, P.~I. and {Tkachenko}, A. and {Van Reeth}, T. and {Aerts}, C. and {Moravveji}, E. and {Van de Sande}, M. and {De Smedt}, K. and {Bloemen}, S. and {Southworth}, J. and {Debosscher}, J. and {Niemczura}, E. and {Gameiro}, J.~F.},
        title = "{Signatures of internal rotation discovered in the Kepler data of five slowly pulsating B stars}",
      journal = {\aap},
         year = 2017,
        month = feb,
       volume = {598},
          eid = {A74},
        pages = {A74},
          doi = {10.1051/0004-6361/201629814},
archivePrefix = {arXiv},
       eprint = {1611.06955},
 primaryClass = {astro-ph.SR},
       adsurl = {https://ui.adsabs.harvard.edu/abs/2017A&A...598A..74P}
}

@ARTICLE{Pedersen2025,
       author = {{Pedersen}, May G. and {Bildsten}, Lars},
        title = "{Stochastic low-frequency variability of 50 massive stars in the Cygnus OB associations and the Small Magellanic Cloud}",
      journal = {\mnras},
         year = 2025,
        month = may,
       volume = {539},
       number = {3},
        pages = {2742-2764},
          doi = {10.1093/mnras/staf661},
archivePrefix = {arXiv},
       eprint = {2504.15861},
 primaryClass = {astro-ph.SR},
       adsurl = {https://ui.adsabs.harvard.edu/abs/2025MNRAS.539.2742P}
}

@ARTICLE{pedersen_2021,
       author = {{Pedersen}, May G. and {Aerts}, Conny and {P{\'a}pics}, P{\'e}ter I. and {Michielsen}, Mathias and {Gebruers}, Sarah and {Rogers}, Tamara M. and {Molenberghs}, Geert and {Burssens}, Siemen and {Garcia}, Stefano and {Bowman}, Dominic M.},
        title = "{Internal mixing of rotating stars inferred from dipole gravity modes}",
      journal = {Nature Astronomy},
         year = 2021,
        month = jan,
       volume = {5},
        pages = {715-722},
          doi = {10.1038/s41550-021-01351-x},
archivePrefix = {arXiv},
       eprint = {2105.04533},
 primaryClass = {astro-ph.SR},
       adsurl = {https://ui.adsabs.harvard.edu/abs/2021NatAs...5..715P}
}

@ARTICLE{Aerts2018,
       author = {{Aerts}, C. and {Molenberghs}, G. and {Michielsen}, M. and {Pedersen}, M.~G. and {Bj{\"o}rklund}, R. and {Johnston}, C. and {Mombarg}, J.~S.~G. and {Bowman}, D.~M. and {Buysschaert}, B. and {P{\'a}pics}, P.~I. and {Sekaran}, S. and {Sundqvist}, J.~O. and {Tkachenko}, A. and {Truyaert}, K. and {Van Reeth}, T. and {Vermeyen}, E.},
        title = "{Forward Asteroseismic Modeling of Stars with a Convective Core from Gravity-mode Oscillations: Parameter Estimation and Stellar Model Selection}",
      journal = {\apjs},
         year = 2018,
        month = jul,
       volume = {237},
       number = {1},
          eid = {15},
        pages = {15},
          doi = {10.3847/1538-4365/aaccfb},
archivePrefix = {arXiv},
       eprint = {1806.06869},
 primaryClass = {astro-ph.SR},
       adsurl = {https://ui.adsabs.harvard.edu/abs/2018ApJS..237...15A}
}

@ARTICLE{AertsRogers2015,
       author = {{Aerts}, C. and {Rogers}, T.~M.},
        title = "{Observational Signatures of Convectively Driven Waves in Massive Stars}",
      journal = {\apjl},
         year = 2015,
        month = jun,
       volume = {806},
       number = {2},
          eid = {L33},
        pages = {L33},
          doi = {10.1088/2041-8205/806/2/L33},
archivePrefix = {arXiv},
       eprint = {1505.06648},
 primaryClass = {astro-ph.SR},
       adsurl = {https://ui.adsabs.harvard.edu/abs/2015ApJ...806L..33A}
}

@ARTICLE{Neiner2012,
       author = {{Neiner}, C. and {Floquet}, M. and {Samadi}, R. and {Espinosa Lara}, F. and {Fr{\'e}mat}, Y. and {Mathis}, S. and {Leroy}, B. and {de Batz}, B. and {Rainer}, M. and {Poretti}, E. and {Mathias}, P. and {Guarro Fl{\'o}}, J. and {Buil}, C. and {Ribeiro}, J. and {Alecian}, E. and {Andrade}, L. and {Briquet}, M. and {Diago}, P.~D. and {Emilio}, M. and {Fabregat}, J. and {Guti{\'e}rrez-Soto}, J. and {Hubert}, A. -M. and {Janot-Pacheco}, E. and {Martayan}, C. and {Semaan}, T. and {Suso}, J. and {Zorec}, J.},
        title = "{Stochastic gravito-inertial modes discovered by CoRoT in the hot Be star HD 51452}",
      journal = {\aap},
         year = 2012,
        month = oct,
       volume = {546},
          eid = {A47},
        pages = {A47},
          doi = {10.1051/0004-6361/201219820},
       adsurl = {https://ui.adsabs.harvard.edu/abs/2012A&A...546A..47N}
}

@ARTICLE{Christophe2018,
       author = {{Christophe}, S. and {Ballot}, J. and {Ouazzani}, R. -M. and {Antoci}, V. and {Salmon}, S.~J.~A.~J.},
        title = "{Deciphering the oscillation spectrum of {\ensuremath{\gamma}} Doradus and SPB stars}",
      journal = {\aap},
         year = 2018,
        month = oct,
       volume = {618},
          eid = {A47},
        pages = {A47},
          doi = {10.1051/0004-6361/201832782},
archivePrefix = {arXiv},
       eprint = {1807.03707},
 primaryClass = {astro-ph.SR},
       adsurl = {https://ui.adsabs.harvard.edu/abs/2018A&A...618A..47C}
}

@ARTICLE{Balona2011,
       author = {{Balona}, L.~A. and {Pigulski}, A. and {De Cat}, P. and {Handler}, G. and {Guti{\'e}rrez-Soto}, J. and {Engelbrecht}, C.~A. and {Frescura}, F. and {Briquet}, M. and {Cuypers}, J. and {Daszy{\'n}ska-Daszkiewicz}, J. and {Degroote}, P. and {Dukes}, R.~J. and {Garcia}, R.~A. and {Green}, E.~M. and {Heber}, U. and {Kawaler}, S.~D. and {Lehmann}, H. and {Leroy}, B. and {Molenda-{\.Z}aaowicz}, J. and {Neiner}, C. and {Noels}, A. and {Nuspl}, J. and {{\O}stensen}, R. and {Pricopi}, D. and {Roxburgh}, I. and {Salmon}, S. and {Smith}, M.~A. and {Su{\'a}rez}, J.~C. and {Suran}, M. and {Szab{\'o}}, R. and {Uytterhoeven}, K. and {Christensen-Dalsgaard}, J. and {Kjeldsen}, H. and {Caldwell}, D.~A. and {Girouard}, F.~R. and {Sanderfer}, D.~T.},
        title = "{Kepler observations of the variability in B-type stars}",
      journal = {\mnras},
         year = 2011,
        month = jun,
       volume = {413},
       number = {4},
        pages = {2403-2420},
          doi = {10.1111/j.1365-2966.2011.18311.x},
archivePrefix = {arXiv},
       eprint = {1103.0644},
 primaryClass = {astro-ph.SR},
       adsurl = {https://ui.adsabs.harvard.edu/abs/2011MNRAS.413.2403B}
}

@ARTICLE{Papics2014,
       author = {{P{\'a}pics}, P.~I. and {Moravveji}, E. and {Aerts}, C. and {Tkachenko}, A. and {Triana}, S.~A. and {Bloemen}, S. and {Southworth}, J.},
        title = "{KIC 10526294: a slowly rotating B star with rotationally split, quasi-equally spaced gravity modes}",
      journal = {\aap},
         year = 2014,
        month = oct,
       volume = {570},
          eid = {A8},
        pages = {A8},
          doi = {10.1051/0004-6361/201424094},
archivePrefix = {arXiv},
       eprint = {1407.2986},
 primaryClass = {astro-ph.SR},
       adsurl = {https://ui.adsabs.harvard.edu/abs/2014A&A...570A...8P}
}

@ARTICLE{Papics2015,
       author = {{P{\'a}pics}, P.~I. and {Tkachenko}, A. and {Aerts}, C. and {Van Reeth}, T. and {De Smedt}, K. and {Hillen}, M. and {{\O}stensen}, R. and {Moravveji}, E.},
        title = "{Asteroseismic Fingerprints of Rotation and Mixing in the Slowly Pulsating B8 V Star KIC 7760680}",
      journal = {\apjl},
         year = 2015,
        month = apr,
       volume = {803},
       number = {2},
          eid = {L25},
        pages = {L25},
          doi = {10.1088/2041-8205/803/2/L25},
archivePrefix = {arXiv},
       eprint = {1504.00496},
 primaryClass = {astro-ph.SR},
       adsurl = {https://ui.adsabs.harvard.edu/abs/2015ApJ...803L..25P}
}

@ARTICLE{Gilliland2010,
       author = {{Gilliland}, Ronald L. and {Brown}, Timothy M. and {Christensen-Dalsgaard}, J{\o}rgen and {Kjeldsen}, Hans and {Aerts}, Conny and {Appourchaux}, Thierry and {Basu}, Sarbani and {Bedding}, Timothy R. and {Chaplin}, William J. and {Cunha}, Margarida S. and {De Cat}, Peter and {De Ridder}, Joris and {Guzik}, Joyce A. and {Handler}, Gerald and {Kawaler}, Steven and {Kiss}, L{\'a}szl{\'o} and {Kolenberg}, Katrien and {Kurtz}, Donald W. and {Metcalfe}, Travis S. and {Monteiro}, Mario J.~P.~F.~G. and {Szab{\'o}}, Robert and {Arentoft}, Torben and {Balona}, Luis and {Debosscher}, Jonas and {Elsworth}, Yvonne P. and {Quirion}, Pierre-Olivier and {Stello}, Dennis and {Su{\'a}rez}, Juan Carlos and {Borucki}, William J. and {Jenkins}, Jon M. and {Koch}, David and {Kondo}, Yoji and {Latham}, David W. and {Rowe}, Jason F. and {Steffen}, Jason H.},
        title = "{Kepler Asteroseismology Program: Introduction and First Results}",
      journal = {\pasp},
         year = 2010,
        month = feb,
       volume = {122},
       number = {888},
        pages = {131},
          doi = {10.1086/650399},
archivePrefix = {arXiv},
       eprint = {1001.0139},
 primaryClass = {astro-ph.SR},
       adsurl = {https://ui.adsabs.harvard.edu/abs/2010PASP..122..131G}
}

@ARTICLE{Neiner2020,
       author = {{Neiner}, C. and {Lee}, U. and {Mathis}, S. and {Saio}, H. and {Lovekin}, C.~C. and {Augustson}, K.~C.},
        title = "{Transport of angular momentum by stochastically excited waves as an explanation for the outburst of the rapidly rotating Be star HD49330}",
      journal = {\aap},
         year = 2020,
        month = dec,
       volume = {644},
          eid = {A9},
        pages = {A9},
          doi = {10.1051/0004-6361/201935858},
archivePrefix = {arXiv},
       eprint = {2007.08977},
 primaryClass = {astro-ph.SR},
       adsurl = {https://ui.adsabs.harvard.edu/abs/2020A&A...644A...9N}
}

@ARTICLE{Neiner2009,
       author = {{Neiner}, C. and {Guti{\'e}rrez-Soto}, J. and {Baudin}, F. and {de Batz}, B. and {Fr{\'e}mat}, Y. and {Huat}, A.~L. and {Floquet}, M. and {Hubert}, A. -M. and {Leroy}, B. and {Diago}, P.~D. and {Poretti}, E. and {Carrier}, F. and {Rainer}, M. and {Catala}, C. and {Thizy}, O. and {Buil}, C. and {Ribeiro}, J. and {Andrade}, L. and {Emilio}, M. and {Espinosa Lara}, F. and {Fabregat}, J. and {Janot-Pacheco}, E. and {Martayan}, C. and {Semaan}, T. and {Suso}, J. and {Baglin}, A. and {Michel}, E. and {Samadi}, R.},
        title = "{The pulsations of the B5IVe star HD 181231 observed with CoRoT and ground-based spectroscopy}",
      journal = {\aap},
         year = 2009,
        month = oct,
       volume = {506},
       number = {1},
        pages = {143-151},
          doi = {10.1051/0004-6361/200911971},
       adsurl = {https://ui.adsabs.harvard.edu/abs/2009A&A...506..143N}
}

@ARTICLE{SSD2017,
       author = {{Sim{\'o}n-D{\'\i}az}, S. and {Godart}, M. and {Castro}, N. and {Herrero}, A. and {Aerts}, C. and {Puls}, J. and {Telting}, J. and {Grassitelli}, L.},
        title = "{The IACOB project . III. New observational clues to understand macroturbulent broadening in massive O- and B-type stars}",
      journal = {\aap},
         year = 2017,
        month = jan,
       volume = {597},
          eid = {A22},
        pages = {A22},
          doi = {10.1051/0004-6361/201628541},
archivePrefix = {arXiv},
       eprint = {1608.05508},
 primaryClass = {astro-ph.SR},
       adsurl = {https://ui.adsabs.harvard.edu/abs/2017A&A...597A..22S}
}

@ARTICLE{Serebriakova2024,
       author = {{Serebriakova}, Nadya and {Tkachenko}, Andrew and {Aerts}, Conny},
        title = "{The ESO UVES/FEROS Large Programs of TESS OB pulsators: II. The physical origin of macroturbulence}",
      journal = {\aap},
         year = 2024,
        month = dec,
       volume = {692},
          eid = {A245},
        pages = {A245},
          doi = {10.1051/0004-6361/202451581},
archivePrefix = {arXiv},
       eprint = {2408.15888},
 primaryClass = {astro-ph.SR},
       adsurl = {https://ui.adsabs.harvard.edu/abs/2024A&A...692A.245S}
}

@ARTICLE{Vanon2023,
       author = {{Vanon}, R. and {Edelmann}, P.~V.~F. and {Ratnasingam}, R.~P. and {Varghese}, A. and {Rogers}, T.~M.},
        title = "{Three-dimensional Simulations of Massive Stars. II. Age Dependence}",
      journal = {\apj},
         year = 2023,
        month = sep,
       volume = {954},
       number = {2},
          eid = {171},
        pages = {171},
          doi = {10.3847/1538-4357/ace9db},
archivePrefix = {arXiv},
       eprint = {2307.15109},
 primaryClass = {astro-ph.SR},
       adsurl = {https://ui.adsabs.harvard.edu/abs/2023ApJ...954..171V}
}

@ARTICLE{Ratnasingam2020,
       author = {{Ratnasingam}, R.~P. and {Edelmann}, P.~V.~F. and {Rogers}, T.~M.},
        title = "{Two-dimensional simulations of internal gravity waves in the radiation zones of intermediate-mass stars}",
      journal = {\mnras},
         year = 2020,
        month = oct,
       volume = {497},
       number = {4},
        pages = {4231-4245},
          doi = {10.1093/mnras/staa2296},
archivePrefix = {arXiv},
       eprint = {2008.03306},
 primaryClass = {astro-ph.SR},
       adsurl = {https://ui.adsabs.harvard.edu/abs/2020MNRAS.497.4231R}
}

@ARTICLE{Bowman2020,
       author = {{Bowman}, D.~M. and {Burssens}, S. and {Sim{\'o}n-D{\'\i}az}, S. and {Edelmann}, P.~V.~F. and {Rogers}, T.~M. and {Horst}, L. and {R{\"o}pke}, F.~K. and {Aerts}, C.},
        title = "{Photometric detection of internal gravity waves in upper main-sequence stars. II. Combined TESS photometry and high-resolution spectroscopy}",
      journal = {\aap},
         year = 2020,
        month = aug,
       volume = {640},
          eid = {A36},
        pages = {A36},
          doi = {10.1051/0004-6361/202038224},
archivePrefix = {arXiv},
       eprint = {2006.03012},
 primaryClass = {astro-ph.SR},
       adsurl = {https://ui.adsabs.harvard.edu/abs/2020A&A...640A..36B}
}

@ARTICLE{Edelmann2019,
       author = {{Edelmann}, P.~V.~F. and {Ratnasingam}, R.~P. and {Pedersen}, M.~G. and {Bowman}, D.~M. and {Prat}, V. and {Rogers}, T.~M.},
        title = "{Three-dimensional Simulations of Massive Stars. I. Wave Generation and Propagation}",
      journal = {\apj},
         year = 2019,
        month = may,
       volume = {876},
       number = {1},
          eid = {4},
        pages = {4},
          doi = {10.3847/1538-4357/ab12df},
archivePrefix = {arXiv},
       eprint = {1903.09392},
 primaryClass = {astro-ph.SR},
       adsurl = {https://ui.adsabs.harvard.edu/abs/2019ApJ...876....4E}
}

@ARTICLE{Anders2023,
       author = {{Anders}, Evan H. and {Lecoanet}, Daniel and {Cantiello}, Matteo and {Burns}, Keaton J. and {Hyatt}, Benjamin A. and {Kaufman}, Emma and {Townsend}, Richard H.~D. and {Brown}, Benjamin P. and {Vasil}, Geoffrey M. and {Oishi}, Jeffrey S. and {Jermyn}, Adam S.},
        title = "{The photometric variability of massive stars due to gravity waves excited by core convection.}",
      journal = {Nature Astronomy},
         year = 2023,
        month = oct,
       volume = {7},
        pages = {1228-1234},
          doi = {10.1038/s41550-023-02040-7},
archivePrefix = {arXiv},
       eprint = {2306.08023},
 primaryClass = {astro-ph.SR},
       adsurl = {https://ui.adsabs.harvard.edu/abs/2023NatAs...7.1228A}
}

@ARTICLE{Lecoanet2019,
       author = {{Lecoanet}, Daniel and {Cantiello}, Matteo and {Quataert}, Eliot and {Couston}, Louis-Alexandre and {Burns}, Keaton J. and {Pope}, Benjamin J.~S. and {Jermyn}, Adam S. and {Favier}, Benjamin and {Le Bars}, Michael},
        title = "{Low-frequency Variability in Massive Stars: Core Generation or Surface Phenomenon?}",
      journal = {\apjl},
         year = 2019,
        month = nov,
       volume = {886},
       number = {1},
          eid = {L15},
        pages = {L15},
          doi = {10.3847/2041-8213/ab5446},
archivePrefix = {arXiv},
       eprint = {1910.01643},
 primaryClass = {astro-ph.SR},
       adsurl = {https://ui.adsabs.harvard.edu/abs/2019ApJ...886L..15L}
}

@ARTICLE{BM2021,
       author = {{Bowman}, Dominic M. and {Michielsen}, Mathias},
        title = "{Towards a systematic treatment of observational uncertainties in forward asteroseismic modelling of gravity-mode pulsators}",
      journal = {\aap},
         year = 2021,
        month = dec,
       volume = {656},
          eid = {A158},
        pages = {A158},
          doi = {10.1051/0004-6361/202141726},
archivePrefix = {arXiv},
       eprint = {2109.10776},
 primaryClass = {astro-ph.SR},
       adsurl = {https://ui.adsabs.harvard.edu/abs/2021A&A...656A.158B}
}

@ARTICLE{Michielsen2021,
       author = {{Michielsen}, M. and {Aerts}, C. and {Bowman}, D.~M.},
        title = "{Probing the temperature gradient in the core boundary layer of stars with gravito-inertial modes. The case of KIC 7760680}",
      journal = {\aap},
         year = 2021,
        month = jun,
       volume = {650},
          eid = {A175},
        pages = {A175},
          doi = {10.1051/0004-6361/202039926},
archivePrefix = {arXiv},
       eprint = {2104.04531},
 primaryClass = {astro-ph.SR},
       adsurl = {https://ui.adsabs.harvard.edu/abs/2021A&A...650A.175M}
}

@ARTICLE{Cantiello2019,
       author = {{Cantiello}, Matteo and {Braithwaite}, Jonathan},
        title = "{Envelope Convection, Surface Magnetism, and Spots in A and Late B-type Stars}",
      journal = {\apj},
         year = 2019,
        month = sep,
       volume = {883},
       number = {1},
          eid = {106},
        pages = {106},
          doi = {10.3847/1538-4357/ab3924},
archivePrefix = {arXiv},
       eprint = {1904.02161},
 primaryClass = {astro-ph.SR},
       adsurl = {https://ui.adsabs.harvard.edu/abs/2019ApJ...883..106C}
}

@ARTICLE{Cantiello2009,
       author = {{Cantiello}, M. and {Langer}, N. and {Brott}, I. and {de Koter}, A. and {Shore}, S.~N. and {Vink}, J.~S. and {Voegler}, A. and {Lennon}, D.~J. and {Yoon}, S. -C.},
        title = "{Sub-surface convection zones in hot massive stars and their observable consequences}",
      journal = {\aap},
         year = 2009,
        month = may,
       volume = {499},
       number = {1},
        pages = {279-290},
          doi = {10.1051/0004-6361/200911643},
archivePrefix = {arXiv},
       eprint = {0903.2049},
 primaryClass = {astro-ph.SR},
       adsurl = {https://ui.adsabs.harvard.edu/abs/2009A&A...499..279C}
}

@ARTICLE{Townsend2018,
       author = {{Townsend}, R.~H.~D. and {Goldstein}, J. and {Zweibel}, E.~G.},
        title = "{Angular momentum transport by heat-driven g-modes in slowly pulsating B stars}",
      journal = {\mnras},
         year = 2018,
        month = mar,
       volume = {475},
       number = {1},
        pages = {879-893},
          doi = {10.1093/mnras/stx3142},
archivePrefix = {arXiv},
       eprint = {1712.02420},
 primaryClass = {astro-ph.SR},
       adsurl = {https://ui.adsabs.harvard.edu/abs/2018MNRAS.475..879T}
}

@ARTICLE{Townsend2005a,
       author = {{Townsend}, R.~H.~D.},
        title = "{Influence of the Coriolis force on the instability of slowly pulsating B stars}",
      journal = {\mnras},
         year = 2005,
        month = jun,
       volume = {360},
       number = {2},
        pages = {465-476},
          doi = {10.1111/j.1365-2966.2005.09002.x},
archivePrefix = {arXiv},
       eprint = {astro-ph/0503192},
 primaryClass = {astro-ph},
       adsurl = {https://ui.adsabs.harvard.edu/abs/2005MNRAS.360..465T}
}

@ARTICLE{Townsend2005b,
       author = {{Townsend}, R.~H.~D.},
        title = "{Kappa-mechanism excitation of retrograde mixed modes in rotating B-type stars}",
      journal = {\mnras},
         year = 2005,
        month = dec,
       volume = {364},
       number = {2},
        pages = {573-582},
          doi = {10.1111/j.1365-2966.2005.09585.x},
archivePrefix = {arXiv},
       eprint = {astro-ph/0506580},
 primaryClass = {astro-ph},
       adsurl = {https://ui.adsabs.harvard.edu/abs/2005MNRAS.364..573T}
}

@ARTICLE{Rehm2024,
       author = {{Rehm}, Rebecca and {Mombarg}, Joey S.~G. and {Aerts}, Conny and {Michielsen}, Mathias and {Burssens}, Siemen and {Townsend}, Richard H.~D.},
        title = "{The impact of radiative levitation on mode excitation of main-sequence B-type pulsators}",
      journal = {\aap},
         year = 2024,
        month = jul,
       volume = {687},
          eid = {A175},
        pages = {A175},
          doi = {10.1051/0004-6361/202449624},
archivePrefix = {arXiv},
       eprint = {2405.08864},
 primaryClass = {astro-ph.SR},
       adsurl = {https://ui.adsabs.harvard.edu/abs/2024A&A...687A.175R}
}

@ARTICLE{HeyAerts2024,
       author = {{Hey}, Daniel and {Aerts}, Conny},
        title = "{Confronting sparse Gaia DR3 photometry with TESS for a sample of around 60 000 OBAF-type pulsators}",
      journal = {\aap},
         year = 2024,
        month = aug,
       volume = {688},
          eid = {A93},
        pages = {A93},
          doi = {10.1051/0004-6361/202450489},
archivePrefix = {arXiv},
       eprint = {2405.01539},
 primaryClass = {astro-ph.SR},
       adsurl = {https://ui.adsabs.harvard.edu/abs/2024A&A...688A..93H}
}

@ARTICLE{Mombarg2024,
       author = {{Mombarg}, Joey S.~G. and {Aerts}, Conny and {Van Reeth}, Timothy and {Hey}, Daniel},
        title = "{Estimates of (convective core) masses, radii, and relative ages for {\ensuremath{\sim}}14 000 Gaia-discovered gravity-mode pulsators monitored by TESS}",
      journal = {\aap},
         year = 2024,
        month = nov,
       volume = {691},
          eid = {A131},
        pages = {A131},
          doi = {10.1051/0004-6361/202451651},
archivePrefix = {arXiv},
       eprint = {2410.05367},
 primaryClass = {astro-ph.SR},
       adsurl = {https://ui.adsabs.harvard.edu/abs/2024A&A...691A.131M}
}

@ARTICLE{Pamyatnykh1999,
       author = {{Pamyatnykh}, A.~A.},
        title = "{Pulsational Instability Domains in the Upper Main Sequence}",
      journal = {\actaa},
         year = 1999,
        month = jun,
       volume = {49},
        pages = {119-148},
       adsurl = {https://ui.adsabs.harvard.edu/abs/1999AcA....49..119P}
}

@ARTICLE{Aerts2021,
       author = {{Aerts}, C.},
        title = "{Probing the interior physics of stars through asteroseismology}",
      journal = {Reviews of Modern Physics},
         year = 2021,
        month = jan,
       volume = {93},
       number = {1},
          eid = {015001},
        pages = {015001},
          doi = {10.1103/RevModPhys.93.015001},
archivePrefix = {arXiv},
       eprint = {1912.12300},
 primaryClass = {astro-ph.SR},
       adsurl = {https://ui.adsabs.harvard.edu/abs/2021RvMP...93a5001A}
}

@ARTICLE{Aerts2019,
   author = {{Aerts}, C. and {Mathis}, S. and {Rogers}, T.~M.},
    title = "{Angular Momentum Transport in Stellar Interiors}",
  journal = {\araa},
archivePrefix = "arXiv",
   eprint = {1809.07779},
 primaryClass = "astro-ph.SR",
     year = 2019,
    month = aug,
   volume = 57,
    pages = {35-78},
      doi = {10.1146/annurev-astro-091918-104359},
   adsurl = {https://ui.adsabs.harvard.edu/abs/2019ARA%26A..57...35A}
}

@ARTICLE{Rathish2023,
       author = {{Ratnasingam}, R.~P. and {Rogers}, T.~M. and {Chowdhury}, S. and {Handler}, G. and {Vanon}, R. and {Varghese}, A. and {Edelmann}, P.~V.~F.},
        title = "{Internal gravity waves in massive stars. II. Frequency analysis across stellar mass}",
      journal = {\aap},
         year = 2023,
        month = jun,
       volume = {674},
          eid = {A134},
        pages = {A134},
          doi = {10.1051/0004-6361/202245727},
archivePrefix = {arXiv},
       eprint = {2305.06379},
 primaryClass = {astro-ph.SR},
       adsurl = {https://ui.adsabs.harvard.edu/abs/2023A&A...674A.134R}
}

@ARTICLE{Aerts2023,
       author = {{Aerts}, C. and {Molenberghs}, G. and {De Ridder}, J.},
        title = "{Astrophysical properties of 15062 Gaia DR3 gravity-mode pulsators. Pulsation amplitudes, rotation, and spectral line broadening}",
      journal = {\aap},
         year = 2023,
        month = apr,
       volume = {672},
          eid = {A183},
        pages = {A183},
          doi = {10.1051/0004-6361/202245713},
archivePrefix = {arXiv},
       eprint = {2302.07870},
 primaryClass = {astro-ph.SR},
       adsurl = {https://ui.adsabs.harvard.edu/abs/2023A&A...672A.183A}
}

@ARTICLE{Aerts2021-GIW,
       author = {{Aerts}, C. and {Augustson}, K. and {Mathis}, S. and {Pedersen}, M.~G. and {Mombarg}, J.~S.~G. and {Vanlaer}, V. and {Van Beeck}, J. and {Van Reeth}, T.},
        title = "{Rossby numbers and stiffness values inferred from gravity-mode asteroseismology of rotating F- and B-type dwarfs. Consequences for mixing, transport, magnetism, and convective penetration}",
      journal = {\aap},
         year = 2021,
        month = dec,
       volume = {656},
          eid = {A121},
        pages = {A121},
          doi = {10.1051/0004-6361/202142151},
archivePrefix = {arXiv},
       eprint = {2110.06220},
 primaryClass = {astro-ph.SR},
       adsurl = {https://ui.adsabs.harvard.edu/abs/2021A&A...656A.121A}
}

@ARTICLE{GarciaBallot2019,
       author = {{Garc{\'\i}a}, Rafael A. and {Ballot}, J{\'e}r{\^o}me},
        title = "{Asteroseismology of solar-type stars}",
      journal = {Living Reviews in Solar Physics},
         year = "2019",
        month = "Sep",
       volume = {16},
       number = {1},
          eid = {4},
        pages = {4},
          doi = {10.1007/s41116-019-0020-1},
archivePrefix = {arXiv},
       eprint = {1906.12262},
 primaryClass = {astro-ph.SR},
       adsurl = {https://ui.adsabs.harvard.edu/abs/2019LRSP...16....4G}
}

@ARTICLE{patil_2025,
       author = {{Patil}, Aarya A. and {Eadie}, Gwendolyn M. and {Speagle}, Joshua S. and {Thomson}, David J.},
        title = "{Improving Harmonic Analysis Using Multitapering: Precise Frequency Estimation of Stellar Oscillations Using the Harmonic F-test}",
      journal = {\aj},
         year = 2025,
        month = jul,
       volume = {170},
       number = {1},
          eid = {7},
        pages = {7},
          doi = {10.3847/1538-3881/adc9b4},
archivePrefix = {arXiv},
       eprint = {2405.18509},
 primaryClass = {astro-ph.SR},
       adsurl = {https://ui.adsabs.harvard.edu/abs/2025AJ....170....7P}
}

@ARTICLE{thomson_1982,
       author = {{Thomson}, D.~J.},
        title = "{Spectrum Estimation and Harmonic Analysis}",
      journal = {IEEE Proceedings},
         year = 1982,
        month = jan,
       volume = {70},
        pages = {1055-1096},
       adsurl = {https://ui.adsabs.harvard.edu/abs/1982IEEEP..70.1055T}
}

@ARTICLE{springford_2020,
       author = {{Springford}, Aaron and {Eadie}, Gwendolyn M. and {Thomson}, David J.},
        title = "{Improving the Lomb-Scargle Periodogram with the Thomson Multitaper}",
      journal = {\aj},
         year = 2020,
        month = may,
       volume = {159},
       number = {5},
          eid = {205},
        pages = {205},
          doi = {10.3847/1538-3881/ab7fa1},
       adsurl = {https://ui.adsabs.harvard.edu/abs/2020AJ....159..205S}
}

@ARTICLE{slepian_1978,
       author = {{Slepian}, D.},
        title = "{Prolate spheroidal wave functions, Fourier analysis, and uncertainty. V - The discrete case}",
      journal = {AT T Technical Journal},
         year = 1978,
        month = jun,
       volume = {57},
        pages = {1371-1430},
       adsurl = {https://ui.adsabs.harvard.edu/abs/1978ATTTJ..57.1371S}
}

@ARTICLE{Miglio2008,
       author = {{Miglio}, Andrea and {Montalb{\'a}n}, Josefina and {Noels}, Arlette and {Eggenberger}, Patrick},
        title = "{Probing the properties of convective cores through g modes: high-order g modes in SPB and {\ensuremath{\gamma}} Doradus stars}",
      journal = {\mnras},
         year = 2008,
        month = may,
       volume = {386},
       number = {3},
        pages = {1487-1502},
          doi = {10.1111/j.1365-2966.2008.13112.x},
archivePrefix = {arXiv},
       eprint = {0802.2057},
 primaryClass = {astro-ph},
       adsurl = {https://ui.adsabs.harvard.edu/abs/2008MNRAS.386.1487M}
}

@ARTICLE{Bouabid2013,
       author = {{Bouabid}, M. -P. and {Dupret}, M. -A. and {Salmon}, S. and {Montalb{\'a}n}, J. and {Miglio}, A. and {Noels}, A.},
        title = "{Effects of the Coriolis force on high-order g modes in {\ensuremath{\gamma}} Doradus stars}",
      journal = {\mnras},
         year = 2013,
        month = mar,
       volume = {429},
       number = {3},
        pages = {2500-2514},
          doi = {10.1093/mnras/sts517},
       adsurl = {https://ui.adsabs.harvard.edu/abs/2013MNRAS.429.2500B}
}

@ARTICLE{Szewkzuk2022,
       author = {{Szewczuk}, Wojciech and {Walczak}, Przemys{\l}aw and {Daszy{\'n}ska-Daszkiewicz}, Jadwiga and {Mo{\'z}dzierski}, Dawid},
        title = "{Seismic modelling of a very young SPB star - KIC 8264293}",
      journal = {\mnras},
         year = 2022,
        month = mar,
       volume = {511},
       number = {1},
        pages = {1529-1543},
          doi = {10.1093/mnras/stac168},
archivePrefix = {arXiv},
       eprint = {2201.07039},
 primaryClass = {astro-ph.SR},
       adsurl = {https://ui.adsabs.harvard.edu/abs/2022MNRAS.511.1529S}
}

@ARTICLE{Szewczuk2018,
       author = {{Szewczuk}, Wojciech and {Daszy{\'n}ska-Daszkiewicz}, Jadwiga},
        title = "{KIC 3240411 - the hottest known SPB star with the asymptotic g-mode period spacing}",
      journal = {\mnras},
         year = 2018,
        month = aug,
       volume = {478},
       number = {2},
        pages = {2243-2256},
          doi = {10.1093/mnras/sty1126},
archivePrefix = {arXiv},
       eprint = {1805.07100},
 primaryClass = {astro-ph.SR},
       adsurl = {https://ui.adsabs.harvard.edu/abs/2018MNRAS.478.2243S}
}

@PhdThesis{PhD-MayPedersen,
    title = {Interior rotation, mixing and ages of a sample of Slowly Pulsating B stars from gravity-mode asteroseismology},
    author = {May Gade Pedersen},
    year = {2020},
    school = {KU Leuven, Belgium},
    url = {https://fys.kuleuven.be/ster/pub/phd-thesis-may-pedersen/phd-thesis-may-pedersen},
  }

@ARTICLE{Horst2020,
       author = {{Horst}, L. and {Edelmann}, P.~V.~F. and {Andr{\'a}ssy}, R. and {R{\"o}pke}, F.~K. and {Bowman}, D.~M. and {Aerts}, C. and {Ratnasingam}, R.~P.},
        title = "{Fully compressible simulations of waves and core convection in main-sequence stars}",
      journal = {\aap},
         year = 2020,
        month = sep,
       volume = {641},
          eid = {A18},
        pages = {A18},
          doi = {10.1051/0004-6361/202037531},
archivePrefix = {arXiv},
       eprint = {2006.03011},
 primaryClass = {astro-ph.SR},
       adsurl = {https://ui.adsabs.harvard.edu/abs/2020A&A...641A..18H}
}

@ARTICLE{Schultz2022,
       author = {{Schultz}, William C. and {Bildsten}, Lars and {Jiang}, Yan-Fei},
        title = "{Stochastic Low-frequency Variability in Three-dimensional Radiation Hydrodynamical Models of Massive Star Envelopes}",
      journal = {\apjl},
         year = 2022,
        month = jan,
       volume = {924},
       number = {1},
          eid = {L11},
        pages = {L11},
          doi = {10.3847/2041-8213/ac441f},
archivePrefix = {arXiv},
       eprint = {2110.13944},
 primaryClass = {astro-ph.SR},
       adsurl = {https://ui.adsabs.harvard.edu/abs/2022ApJ...924L..11S}
}

@ARTICLE{reeth_2015a,
       author = {{Van Reeth}, T. and {Tkachenko}, A. and {Aerts}, C. and {P{\'a}pics}, P.~I. and {Degroote}, P. and {Debosscher}, J. and {Zwintz}, K. and {Bloemen}, S. and {De Smedt}, K. and {Hrudkova}, M. and {Raskin}, G. and {Van Winckel}, H.},
        title = "{Detecting non-uniform period spacings in the Kepler photometry of {\ensuremath{\gamma}} Doradus stars: methodology and case studies}",
      journal = {\aap},
         year = 2015,
        month = feb,
       volume = {574},
          eid = {A17},
        pages = {A17},
          doi = {10.1051/0004-6361/201424585},
archivePrefix = {arXiv},
       eprint = {1410.8178},
 primaryClass = {astro-ph.SR},
       adsurl = {https://ui.adsabs.harvard.edu/abs/2015A&A...574A..17V}
}

@ARTICLE{breger_1993,
       author = {{Breger}, M. and {Stich}, J. and {Garrido}, R. and {Martin}, B. and {Jiang}, S. -Y. and {Li}, Z. -P. and {Hube}, D.~P. and {Ostermann}, W. and {Paparo}, M. and {Scheck}, M.},
        title = "{Nonradial pulsation of the delta Scuti star BU CANCRI in the Praesepe cluster.}",
      journal = {\aap},
         year = 1993,
        month = apr,
       volume = {271},
        pages = {482-486},
       adsurl = {https://ui.adsabs.harvard.edu/abs/1993A&A...271..482B}
}

@ARTICLE{reeth_2015b,
       author = {{Van Reeth}, T. and {Tkachenko}, A. and {Aerts}, C. and {P{\'a}pics}, P.~I. and {Triana}, S.~A. and {Zwintz}, K. and {Degroote}, P. and {Debosscher}, J. and {Bloemen}, S. and {Schmid}, V.~S. and {De Smedt}, K. and {Fremat}, Y. and {Fuentes}, A.~S. and {Homan}, W. and {Hrudkova}, M. and {Karjalainen}, R. and {Lombaert}, R. and {Nemeth}, P. and {{\O}stensen}, R. and {Van De Steene}, G. and {Vos}, J. and {Raskin}, G. and {Van Winckel}, H.},
        title = "{Gravity-mode Period Spacings as a Seismic Diagnostic for a Sample of {\ensuremath{\gamma}} Doradus Stars from Kepler Space Photometry and High-resolution Ground-based Spectroscopy}",
      journal = {\apjs},
         year = 2015,
        month = jun,
       volume = {218},
       number = {2},
          eid = {27},
        pages = {27},
          doi = {10.1088/0067-0049/218/2/27},
archivePrefix = {arXiv},
       eprint = {1504.02119},
 primaryClass = {astro-ph.SR},
       adsurl = {https://ui.adsabs.harvard.edu/abs/2015ApJS..218...27V}
}

@ARTICLE{gang_2020,
       author = {{Li}, Gang and {Van Reeth}, Timothy and {Bedding}, Timothy R. and {Murphy}, Simon J. and {Antoci}, Victoria and {Ouazzani}, Rhita-Maria and {Barbara}, Nicholas H.},
        title = "{Gravity-mode period spacings and near-core rotation rates of 611 {\ensuremath{\gamma}} Doradus stars with Kepler}",
      journal = {\mnras},
         year = 2020,
        month = jan,
       volume = {491},
       number = {3},
        pages = {3586-3605},
          doi = {10.1093/mnras/stz2906},
archivePrefix = {arXiv},
       eprint = {1910.06634},
 primaryClass = {astro-ph.SR},
       adsurl = {https://ui.adsabs.harvard.edu/abs/2020MNRAS.491.3586L}
}

@book{aerts_2010_book,
  title={Asteroseismology},
  author={Aerts, C. and Christensen-Dalsgaard, J. and Kurtz, D.W.},
  isbn={9781402058035},
  lccn={2009942128},
  series={Astronomy and Astrophysics Library},
  url={https://books.google.ca/books?id=N8pswDrdSyUC},
  year={2010},
  publisher={Springer Netherlands}
}

@ARTICLE{degroote_2010,
       author = {{Degroote}, Pieter and {Aerts}, Conny and {Baglin}, Annie and {Miglio}, Andrea and {Briquet}, Maryline and {Noels}, Arlette and {Niemczura}, Ewa and {Montalban}, Josefina and {Bloemen}, Steven and {Oreiro}, Raquel and {Vu{\v{c}}kovi{\'c}}, Maja and {Smolders}, Kristof and {Auvergne}, Michel and {Baudin}, Frederic and {Catala}, Claude and {Michel}, Eric},
        title = "{Deviations from a uniform period spacing of gravity modes in a massive star}",
      journal = {\nat},
         year = 2010,
        month = mar,
       volume = {464},
       number = {7286},
        pages = {259-261},
          doi = {10.1038/nature08864},
       adsurl = {https://ui.adsabs.harvard.edu/abs/2010Natur.464..259D}
}

@article{rife_1976,
  title={Multiple tone parameter estimation from discrete-time observations},
  author={David C. Rife and Robert R. Boorstyn},
  journal={The Bell System Technical Journal},
  year={1976},
  volume={55},
  pages={1389-1410}
}

@ARTICLE{thomson_2007,
  author={Thomson, David J.},
  journal={IEEE Signal Processing Magazine}, 
  title={Jackknifing Multitaper Spectrum Estimates}, 
  year={2007},
  volume={24},
  number={4},
  pages={20-30},
  doi={10.1109/MSP.2007.4286561}}

@ARTICLE{hekker_2017,
       author = {{Hekker}, S. and {Christensen-Dalsgaard}, J.},
        title = "{Giant star seismology}",
      journal = {\aapr},
         year = 2017,
        month = jun,
       volume = {25},
       number = {1},
          eid = {1},
        pages = {1},
          doi = {10.1007/s00159-017-0101-x},
archivePrefix = {arXiv},
       eprint = {1609.07487},
 primaryClass = {astro-ph.SR},
       adsurl = {https://ui.adsabs.harvard.edu/abs/2017A&ARv..25....1H}
}

@ARTICLE{patil_2024,
       author = {{Patil}, Aarya A. and {Eadie}, Gwendolyn M. and {Speagle}, Joshua S. and {Thomson}, David J.},
        title = "{Improving Power Spectrum Estimation Using Multitapering: Efficient Asteroseismic Analyses for Understanding Stars, the Milky Way, and Beyond}",
      journal = {\aj},
         year = 2024,
        month = nov,
       volume = {168},
       number = {5},
          eid = {193},
        pages = {193},
          doi = {10.3847/1538-3881/ad7029},
archivePrefix = {arXiv},
       eprint = {2209.15027},
 primaryClass = {astro-ph.IM},
       adsurl = {https://ui.adsabs.harvard.edu/abs/2024AJ....168..193P}
}

@ARTICLE{van_beeck_2021,
       author = {{Van Beeck}, J. and {Bowman}, D.~M. and {Pedersen}, M.~G. and {Van Reeth}, T. and {Van Hoolst}, T. and {Aerts}, C.},
        title = "{Detection of non-linear resonances among gravity modes of slowly pulsating B stars: Results from five iterative pre-whitening strategies}",
      journal = {\aap},
         year = 2021,
        month = nov,
       volume = {655},
          eid = {A59},
        pages = {A59},
          doi = {10.1051/0004-6361/202141572},
archivePrefix = {arXiv},
       eprint = {2108.02907},
 primaryClass = {astro-ph.SR},
       adsurl = {https://ui.adsabs.harvard.edu/abs/2021A&A...655A..59V}
}

@ARTICLE{moravveji_2015,
       author = {{Moravveji}, E. and {Aerts}, C. and {P{\'a}pics}, P.~I. and {Triana}, S.~A. and {Vandoren}, B.},
        title = "{Tight asteroseismic constraints on core overshooting and diffusive mixing in the slowly rotating pulsating B8.3V star KIC 10526294}",
      journal = {\aap},
         year = 2015,
        month = aug,
       volume = {580},
          eid = {A27},
        pages = {A27},
          doi = {10.1051/0004-6361/201425290},
archivePrefix = {arXiv},
       eprint = {1505.06902},
 primaryClass = {astro-ph.SR},
       adsurl = {https://ui.adsabs.harvard.edu/abs/2015A&A...580A..27M}
}

@ARTICLE{moravveji_2016,
       author = {{Moravveji}, Ehsan and {Townsend}, Richard H.~D. and {Aerts}, Conny and {Mathis}, St{\'e}phane},
        title = "{Sub-inertial Gravity Modes in the B8V Star KIC 7760680 Reveal Moderate Core Overshooting and Low Vertical Diffusive Mixing}",
      journal = {\apj},
         year = 2016,
        month = jun,
       volume = {823},
       number = {2},
          eid = {130},
        pages = {130},
          doi = {10.3847/0004-637X/823/2/130},
archivePrefix = {arXiv},
       eprint = {1604.02680},
 primaryClass = {astro-ph.SR},
       adsurl = {https://ui.adsabs.harvard.edu/abs/2016ApJ...823..130M}
}

@ARTICLE{papics_2013,
       author = {{P{\'a}pics}, P.~I. and {Tkachenko}, A. and {Aerts}, C. and {Briquet}, M. and {Marcos-Arenal}, P. and {Beck}, P.~G. and {Uytterhoeven}, K. and {Trivi{\~n}o Hage}, A. and {Southworth}, J. and {Clubb}, K.~I. and {Bloemen}, S. and {Degroote}, P. and {Jackiewicz}, J. and {McKeever}, J. and {Van Winckel}, H. and {Niemczura}, E. and {Gameiro}, J.~F. and {Debosscher}, J.},
        title = "{Two new SB2 binaries with main sequence B-type pulsators in the Kepler field}",
      journal = {\aap},
         year = 2013,
        month = may,
       volume = {553},
          eid = {A127},
        pages = {A127},
          doi = {10.1051/0004-6361/201321088},
archivePrefix = {arXiv},
       eprint = {1304.2202},
 primaryClass = {astro-ph.SR},
       adsurl = {https://ui.adsabs.harvard.edu/abs/2013A&A...553A.127P}
}

@ARTICLE{lehmann_2011,
       author = {{Lehmann}, H. and {Tkachenko}, A. and {Semaan}, T. and {Guti{\'e}rrez-Soto}, J. and {Smalley}, B. and {Briquet}, M. and {Shulyak}, D. and {Tsymbal}, V. and {De Cat}, P.},
        title = "{Spectral analysis of Kepler SPB and {\ensuremath{\beta}} Cephei candidate stars}",
      journal = {\aap},
         year = 2011,
        month = feb,
       volume = {526},
          eid = {A124},
        pages = {A124},
          doi = {10.1051/0004-6361/201015769},
archivePrefix = {arXiv},
       eprint = {1009.4316},
 primaryClass = {astro-ph.SR},
       adsurl = {https://ui.adsabs.harvard.edu/abs/2011A&A...526A.124L}
}

@ARTICLE{krticka_2018,
       author = {{Krti{\v{c}}ka}, J. and {Feldmeier}, A.},
        title = "{Light variations due to the line-driven wind instability and wind blanketing in O stars}",
      journal = {\aap},
         year = 2018,
        month = sep,
       volume = {617},
          eid = {A121},
        pages = {A121},
          doi = {10.1051/0004-6361/201731614},
archivePrefix = {arXiv},
       eprint = {1807.09407},
 primaryClass = {astro-ph.SR},
       adsurl = {https://ui.adsabs.harvard.edu/abs/2018A&A...617A.121K}
}

@ARTICLE{krticka_2021,
       author = {{Krti{\v{c}}ka}, J. and {Feldmeier}, A.},
        title = "{Stochastic light variations in hot stars from wind instability: finding photometric signatures and testing against the TESS data}",
      journal = {\aap},
         year = 2021,
        month = apr,
       volume = {648},
          eid = {A79},
        pages = {A79},
          doi = {10.1051/0004-6361/202040148},
archivePrefix = {arXiv},
       eprint = {2103.08755},
 primaryClass = {astro-ph.SR},
       adsurl = {https://ui.adsabs.harvard.edu/abs/2021A&A...648A..79K}
}

@ARTICLE{Blomme_2011,
       author = {{Blomme}, R. and {Mahy}, L. and {Catala}, C. and {Cuypers}, J. and {Gosset}, E. and {Godart}, M. and {Montalban}, J. and {Ventura}, P. and {Rauw}, G. and {Morel}, T. and {Degroote}, P. and {Aerts}, C. and {Noels}, A. and {Michel}, E. and {Baudin}, F. and {Baglin}, A. and {Auvergne}, M. and {Samadi}, R.},
        title = "{Variability in the CoRoT photometry of three hot O-type stars. HD 46223, HD 46150, and HD 46966}",
      journal = {\aap},
         year = 2011,
        month = sep,
       volume = {533},
          eid = {A4},
        pages = {A4},
          doi = {10.1051/0004-6361/201116949},
archivePrefix = {arXiv},
       eprint = {1107.0267},
 primaryClass = {astro-ph.SR},
       adsurl = {https://ui.adsabs.harvard.edu/abs/2011A&A...533A...4B}
}

@ARTICLE{Tkachenko_2014,
       author = {{Tkachenko}, A. and {Degroote}, P. and {Aerts}, C. and {Pavlovski}, K. and {Southworth}, J. and {P{\'a}pics}, P.~I. and {Moravveji}, E. and {Kolbas}, V. and {Tsymbal}, V. and {Debosscher}, J. and {Cl{\'e}mer}, K.},
        title = "{The eccentric massive binary V380 Cyg: revised orbital elements and interpretation of the intrinsic variability of the primary component*}",
      journal = {\mnras},
         year = 2014,
        month = mar,
       volume = {438},
       number = {4},
        pages = {3093-3110},
          doi = {10.1093/mnras/stt2421},
archivePrefix = {arXiv},
       eprint = {1312.3601},
 primaryClass = {astro-ph.SR},
       adsurl = {https://ui.adsabs.harvard.edu/abs/2014MNRAS.438.3093T}
}

@ARTICLE{Ramiaramanantsoa_2018,
       author = {{Ramiaramanantsoa}, Tahina and {Ratnasingam}, Rathish and {Shenar}, Tomer and {Moffat}, Anthony F.~J. and {Rogers}, Tamara M. and {Popowicz}, Adam and {Kuschnig}, Rainer and {Pigulski}, Andrzej and {Handler}, Gerald and {Wade}, Gregg A. and {Zwintz}, Konstanze and {Weiss}, Werner W.},
        title = "{A BRITE view on the massive O-type supergiant V973 Scorpii: hints towards internal gravity waves or sub-surface convection zones}",
      journal = {\mnras},
         year = 2018,
        month = oct,
       volume = {480},
       number = {1},
        pages = {972-986},
          doi = {10.1093/mnras/sty1897},
archivePrefix = {arXiv},
       eprint = {1807.04660},
 primaryClass = {astro-ph.SR},
       adsurl = {https://ui.adsabs.harvard.edu/abs/2018MNRAS.480..972R}
}

@ARTICLE{Aerts_2018,
       author = {{Aerts}, C. and {Bowman}, D.~M. and {S{\'\i}mon-D{\'\i}az}, S. and {Buysschaert}, B. and {Johnston}, C. and {Moravveji}, E. and {Beck}, P.~G. and {De Cat}, P. and {Triana}, S. and {Aigrain}, S. and {Castro}, N. and {Huber}, D. and {White}, T.},
        title = "{K2 photometry and HERMES spectroscopy of the blue supergiant {\ensuremath{\rho}} Leo: rotational wind modulation and low-frequency waves}",
      journal = {\mnras},
         year = 2018,
        month = may,
       volume = {476},
       number = {1},
        pages = {1234-1241},
          doi = {10.1093/mnras/sty308},
archivePrefix = {arXiv},
       eprint = {1802.00621},
 primaryClass = {astro-ph.SR},
       adsurl = {https://ui.adsabs.harvard.edu/abs/2018MNRAS.476.1234A}
}

@ARTICLE{2019ApJ...872L...9P,
       author = {{Pedersen}, May G. and {Chowdhury}, Sowgata and {Johnston}, Cole and {Bowman}, Dominic M. and {Aerts}, Conny and {Handler}, Gerald and {De Cat}, Peter and {Neiner}, Coralie and {David-Uraz}, Alexandre and {Buzasi}, Derek and {Tkachenko}, Andrew and {Sim{\'o}n-D{\'\i}az}, Sergio and {Moravveji}, Ehsan and {Sikora}, James and {Mirouh}, Giovanni M. and {Lovekin}, Catherine C. and {Cantiello}, Matteo and {Daszy{\'n}ska-Daszkiewicz}, Jadwiga and {Pigulski}, Andrzej and {Vanderspek}, Roland K. and {Ricker}, George R.},
        title = "{Diverse Variability of O and B Stars Revealed from 2-minute Cadence Light Curves in Sectors 1 and 2 of the TESS Mission: Selection of an Asteroseismic Sample}",
      journal = {\apjl},
         year = 2019,
        month = feb,
       volume = {872},
       number = {1},
          eid = {L9},
        pages = {L9},
          doi = {10.3847/2041-8213/ab01e1},
archivePrefix = {arXiv},
       eprint = {1901.07576},
 primaryClass = {astro-ph.SR},
       adsurl = {https://ui.adsabs.harvard.edu/abs/2019ApJ...872L...9P}
}

@ARTICLE{2020A&A...639A..81B,
       author = {{Burssens}, S. and {Sim{\'o}n-D{\'\i}az}, S. and {Bowman}, D.~M. and {Holgado}, G. and {Michielsen}, M. and {de Burgos}, A. and {Castro}, N. and {Barb{\'a}}, R.~H. and {Aerts}, C.},
        title = "{Variability of OB stars from TESS southern Sectors 1-13 and high-resolution IACOB and OWN spectroscopy}",
      journal = {\aap},
         year = 2020,
        month = jul,
       volume = {639},
          eid = {A81},
        pages = {A81},
          doi = {10.1051/0004-6361/202037700},
archivePrefix = {arXiv},
       eprint = {2005.09658},
 primaryClass = {astro-ph.SR},
       adsurl = {https://ui.adsabs.harvard.edu/abs/2020A&A...639A..81B}
}

@ARTICLE{2022MNRAS.509.4246E,
       author = {{Elliott}, Ashley and {Richardson}, Noel D. and {Pablo}, Herbert and {Moffat}, Anthony F.~J. and {Bowman}, Dominic M. and {Ibrahim}, Nour and {Handler}, Gerald and {Lovekin}, Catherine and {Popowicz}, Adam and {St-Louis}, Nicole and {Wade}, Gregg A. and {Zwintz}, Konstanze},
        title = "{5 yr of BRITE-Constellation photometry of the luminous blue variable P Cygni: properties of the stochastic low-frequency variability}",
      journal = {\mnras},
         year = 2022,
        month = jan,
       volume = {509},
       number = {3},
        pages = {4246-4255},
          doi = {10.1093/mnras/stab3112},
archivePrefix = {arXiv},
       eprint = {2110.12056},
 primaryClass = {astro-ph.SR},
       adsurl = {https://ui.adsabs.harvard.edu/abs/2022MNRAS.509.4246E}
}

@ARTICLE{2024ApJS..275....2S,
       author = {{Shen}, Dong-Xiang and {Zhu}, Chun-Hua and {L{\"u}}, Guo-Liang and {Lu}, Xi-zhen and {He}, Xiao-long},
        title = "{A Study of Stochastic Low-frequency Variability for Galactic O-type Stars}",
      journal = {\apjs},
         year = 2024,
        month = nov,
       volume = {275},
       number = {1},
          eid = {2},
        pages = {2},
          doi = {10.3847/1538-4365/ad71d3},
archivePrefix = {arXiv},
       eprint = {2408.11082},
 primaryClass = {astro-ph.SR},
       adsurl = {https://ui.adsabs.harvard.edu/abs/2024ApJS..275....2S}
}

@ARTICLE{2024ApJ...969...81Z,
       author = {{Zhang}, Zehao and {Ren}, Yi and {Jiang}, Biwei and {Soszy{\'n}ski}, Igor and {Jayasinghe}, Tharindu},
        title = "{Modeling of Granulation in Red Supergiants in the Magellanic Clouds with the Gaussian Process Regressions}",
      journal = {\apj},
         year = 2024,
        month = jul,
       volume = {969},
       number = {2},
          eid = {81},
        pages = {81},
          doi = {10.3847/1538-4357/ad46fe},
archivePrefix = {arXiv},
       eprint = {2405.01405},
 primaryClass = {astro-ph.SR},
       adsurl = {https://ui.adsabs.harvard.edu/abs/2024ApJ...969...81Z}
}

@ARTICLE{2024A&A...692A..49B,
       author = {{Bowman}, Dominic M. and {Van Daele}, Pieterjan and {Michielsen}, Mathias and {Van Reeth}, Timothy},
        title = "{Photometric detection of internal gravity waves in upper main-sequence stars: IV. Comparable stochastic low-frequency variability in SMC, LMC, and Galactic massive stars}",
      journal = {\aap},
         year = 2024,
        month = dec,
       volume = {692},
          eid = {A49},
        pages = {A49},
          doi = {10.1051/0004-6361/202451419},
archivePrefix = {arXiv},
       eprint = {2410.12726},
 primaryClass = {astro-ph.SR},
       adsurl = {https://ui.adsabs.harvard.edu/abs/2024A&A...692A..49B}
}

@ARTICLE{2025A&A...697A.152K,
       author = {{Kourniotis}, Michalis and {Cidale}, Lydia S. and {Kraus}, Michaela and {Ruiz Diaz}, Matias A. and {Alberici Adam}, Aldana},
        title = "{Variability of Galactic blue supergiants observed with TESS}",
      journal = {\aap},
         year = 2025,
        month = may,
       volume = {697},
          eid = {A152},
        pages = {A152},
          doi = {10.1051/0004-6361/202452360},
archivePrefix = {arXiv},
       eprint = {2503.20860},
 primaryClass = {astro-ph.SR},
       adsurl = {https://ui.adsabs.harvard.edu/abs/2025A&A...697A.152K}
}

@ARTICLE{2010Natur.464..259D,
       author = {{Degroote}, Pieter and {Aerts}, Conny and {Baglin}, Annie and {Miglio}, Andrea and {Briquet}, Maryline and {Noels}, Arlette and {Niemczura}, Ewa and {Montalban}, Josefina and {Bloemen}, Steven and {Oreiro}, Raquel and {Vu{\v{c}}kovi{\'c}}, Maja and {Smolders}, Kristof and {Auvergne}, Michel and {Baudin}, Frederic and {Catala}, Claude and {Michel}, Eric},
        title = "{Deviations from a uniform period spacing of gravity modes in a massive star}",
      journal = {\nat},
         year = 2010,
        month = mar,
       volume = {464},
       number = {7286},
        pages = {259-261},
          doi = {10.1038/nature08864},
       adsurl = {https://ui.adsabs.harvard.edu/abs/2010Natur.464..259D}
}

@ARTICLE{2018ApJ...854..168Z,
       author = {{Zhang}, Chunguang and {Liu}, Chao and {Wu}, Yue and {Luo}, Yangping and {Zhang}, Xiaobin and {Deng}, Licai and {Fu}, Jianning and {Zhang}, Yong and {Hou}, Yonghui and {Wang}, Yuefei},
        title = "{Misclassified B Stars in the Kepler Field}",
      journal = {\apj},
         year = 2018,
        month = feb,
       volume = {854},
       number = {2},
          eid = {168},
        pages = {168},
          doi = {10.3847/1538-4357/aaaae0},
archivePrefix = {arXiv},
       eprint = {1802.08789},
 primaryClass = {astro-ph.SR},
       adsurl = {https://ui.adsabs.harvard.edu/abs/2018ApJ...854..168Z}
}

@ARTICLE{2019A&A...628A..76M,
       author = {{Michielsen}, M. and {Pedersen}, M.~G. and {Augustson}, K.~C. and {Mathis}, S. and {Aerts}, C.},
        title = "{Probing the shape of the mixing profile and of the thermal structure at the convective core boundary through asteroseismology}",
      journal = {\aap},
         year = 2019,
        month = aug,
       volume = {628},
          eid = {A76},
        pages = {A76},
          doi = {10.1051/0004-6361/201935754},
archivePrefix = {arXiv},
       eprint = {1906.05304},
 primaryClass = {astro-ph.SR},
       adsurl = {https://ui.adsabs.harvard.edu/abs/2019A&A...628A..76M}
}

@ARTICLE{2022MNRAS.515..828S,
       author = {{Sharma}, Awshesh N. and {Bedding}, Timothy R. and {Saio}, Hideyuki and {White}, Timothy R.},
        title = "{Pulsating B stars in the Scorpius-Centaurus Association with TESS}",
      journal = {\mnras},
         year = 2022,
        month = sep,
       volume = {515},
       number = {1},
        pages = {828-840},
          doi = {10.1093/mnras/stac1816},
archivePrefix = {arXiv},
       eprint = {2203.02582},
 primaryClass = {astro-ph.SR},
       adsurl = {https://ui.adsabs.harvard.edu/abs/2022MNRAS.515..828S}
}

\begin{appendix}
\onecolumn
\section{Additional plots comparing the F-test detections and period spacing patterns from our work with those in Pedersen et al.\ (2021)}

\begin{figure*}[ht!]
   \centering
   \includegraphics[width=0.85\linewidth]{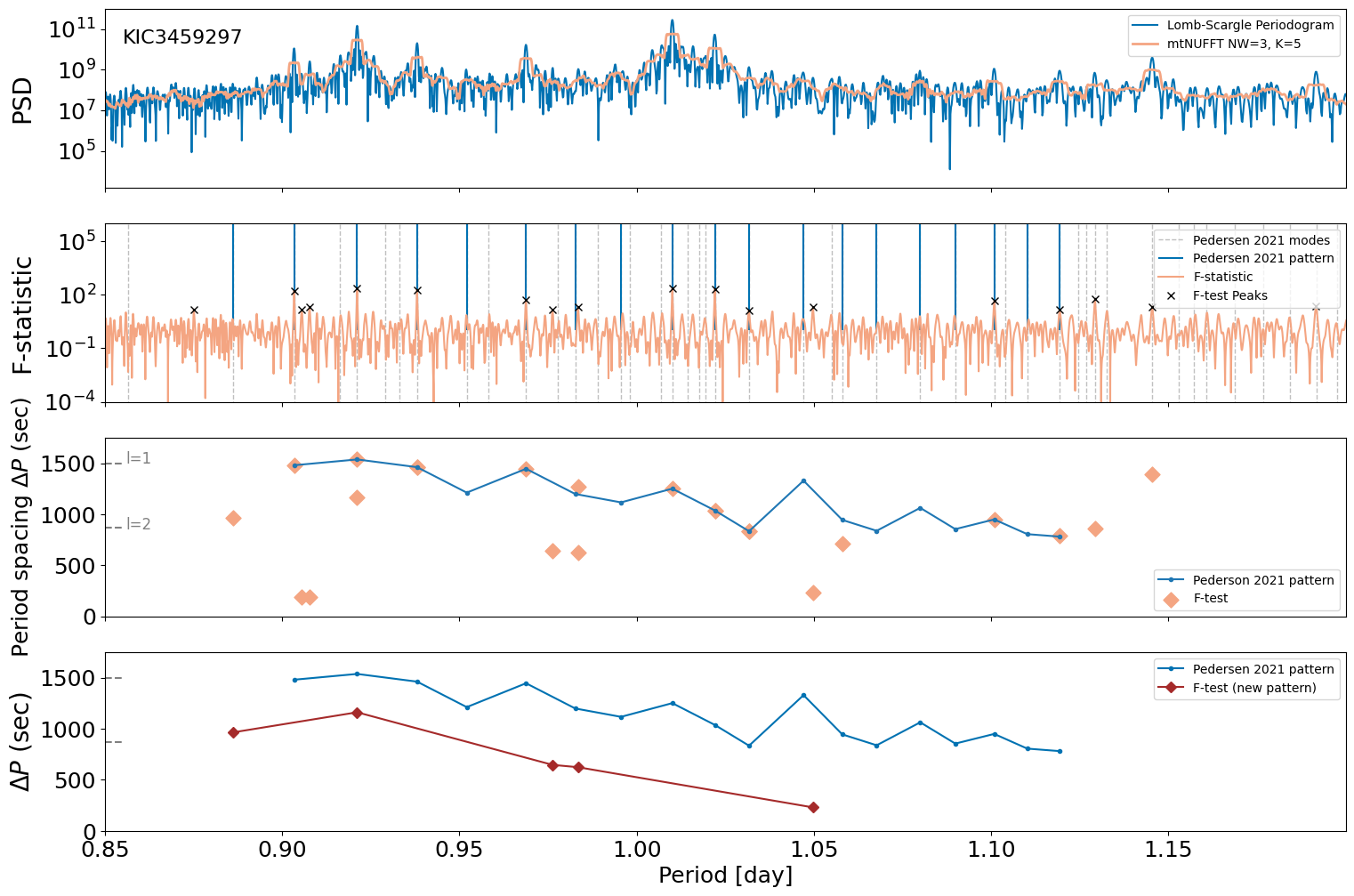}
      \caption{Comparison of our F-test detections with the period spacing patterns in \citetalias{pedersen_2021} for KIC3459297.
              }
         \label{fig:KIC3459297}
\end{figure*}

\begin{figure*}[ht!]
   \centering
   \includegraphics[width=0.85\linewidth]{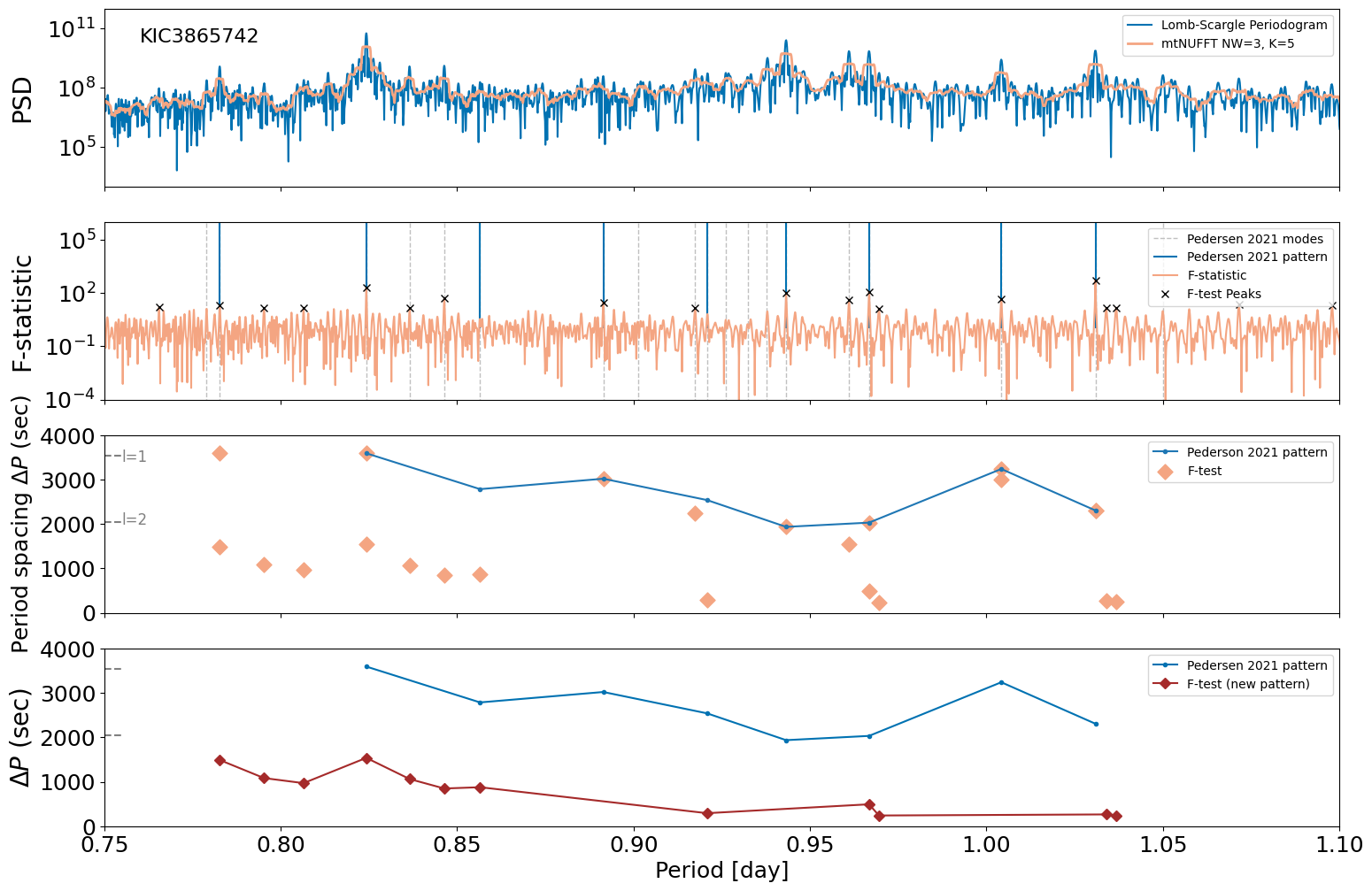}
      \caption{Comparison of our F-test detections with the period spacing patterns in \citetalias{pedersen_2021} for KIC3865742.
              }
         \label{fig:KIC3865742}
\end{figure*}

\begin{figure*}
   \centering
   \includegraphics[width=0.9\linewidth]{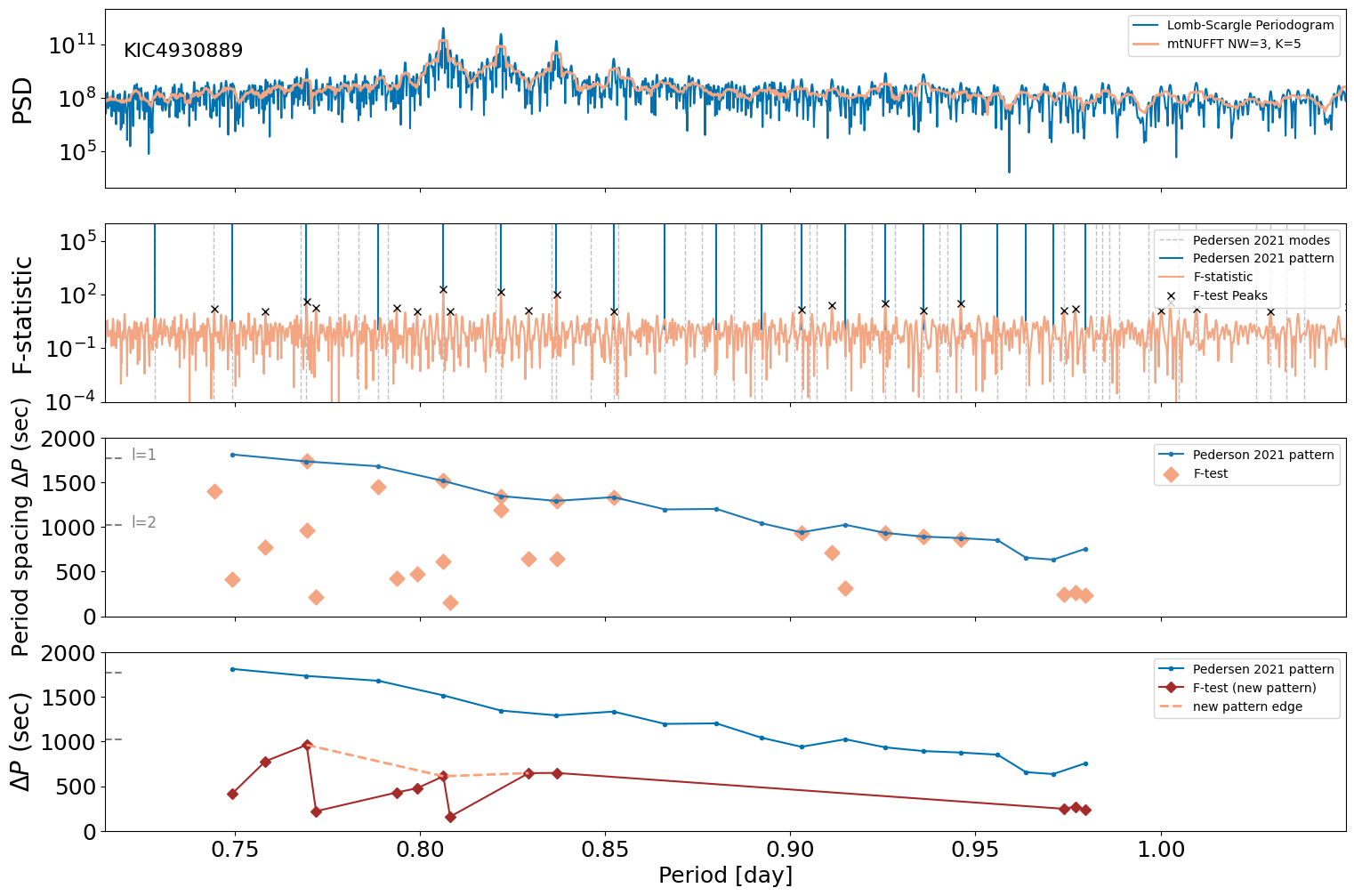}
      \caption{Comparison of our F-test detections with the period spacing patterns in \citetalias{pedersen_2021} for KIC4930889.}
         \label{fig:KIC4930889}
\end{figure*}

\begin{figure*}
   \centering
   \includegraphics[width=0.9\linewidth]{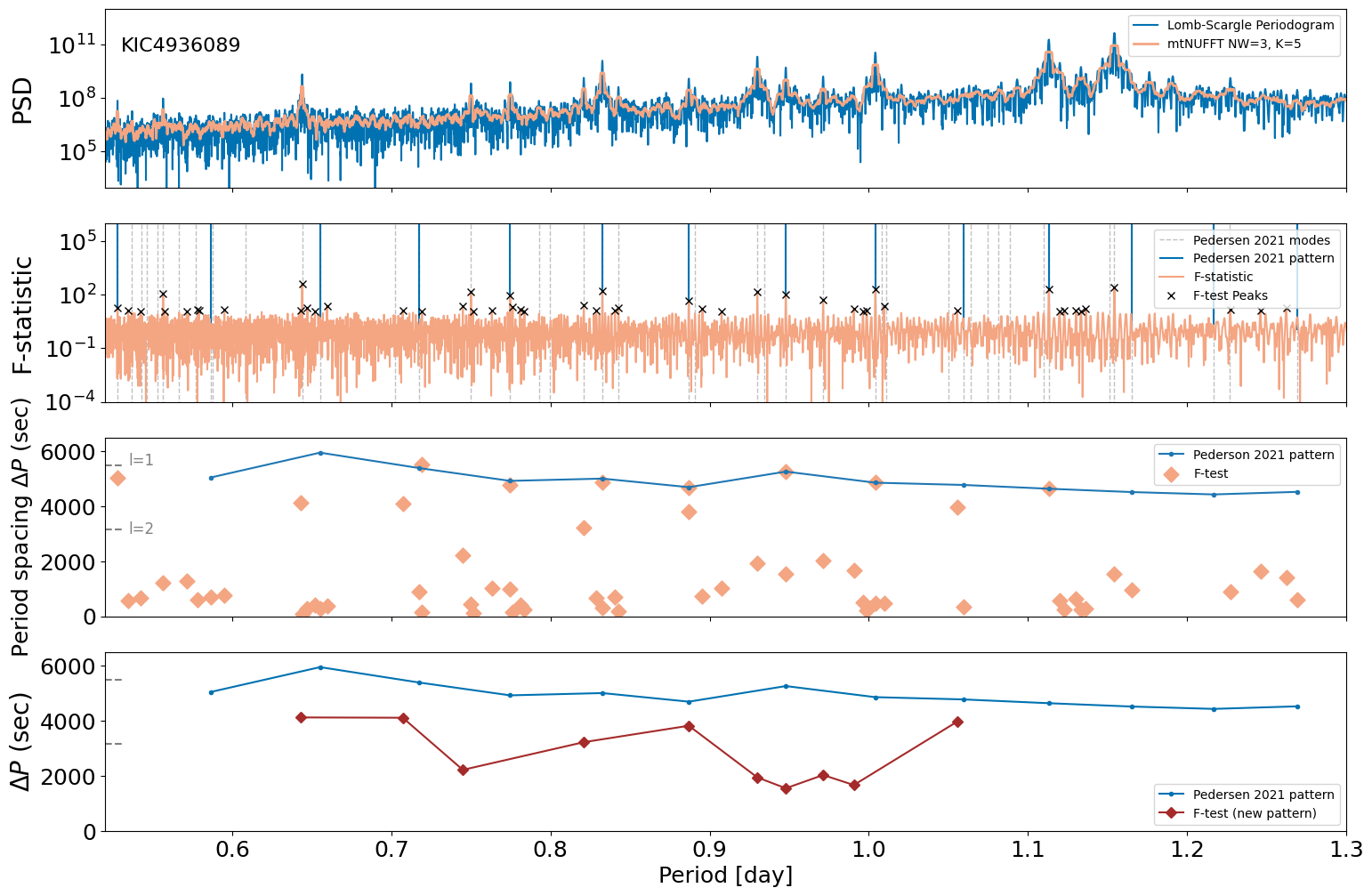}
      \caption{Comparison of our F-test detections with the period spacing patterns in \citetalias{pedersen_2021} for KIC4936089.}
         \label{fig:KIC4936089}
\end{figure*}

\begin{figure*}
   \centering
   \includegraphics[width=0.9\linewidth]{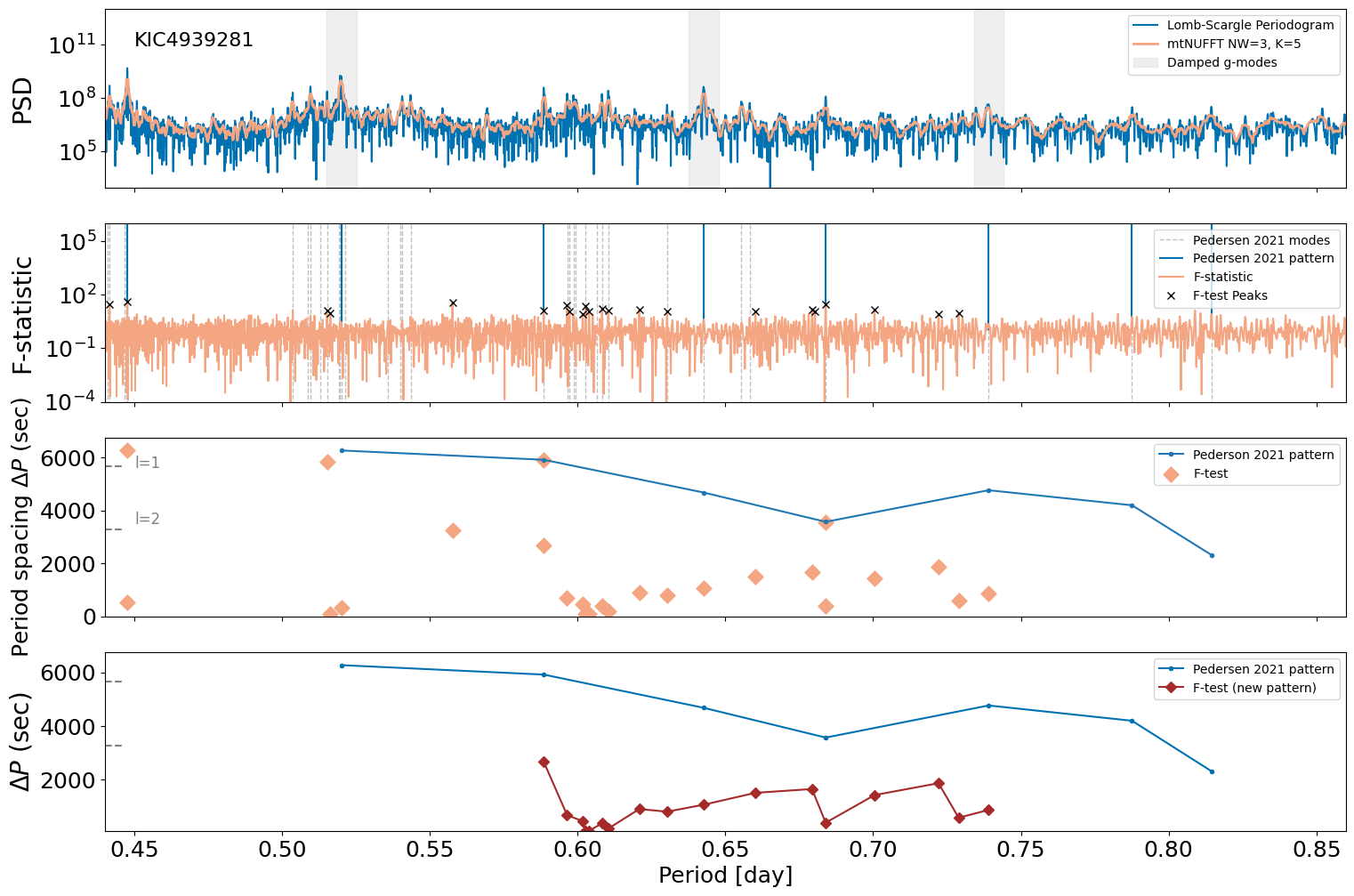}
      \caption{Comparison of our F-test detections with the period spacing patterns in \citetalias{pedersen_2021} for KIC4939281.}
         \label{fig:KIC4939281}
\end{figure*}

\begin{figure*}
   \centering
   \includegraphics[width=0.9\linewidth]{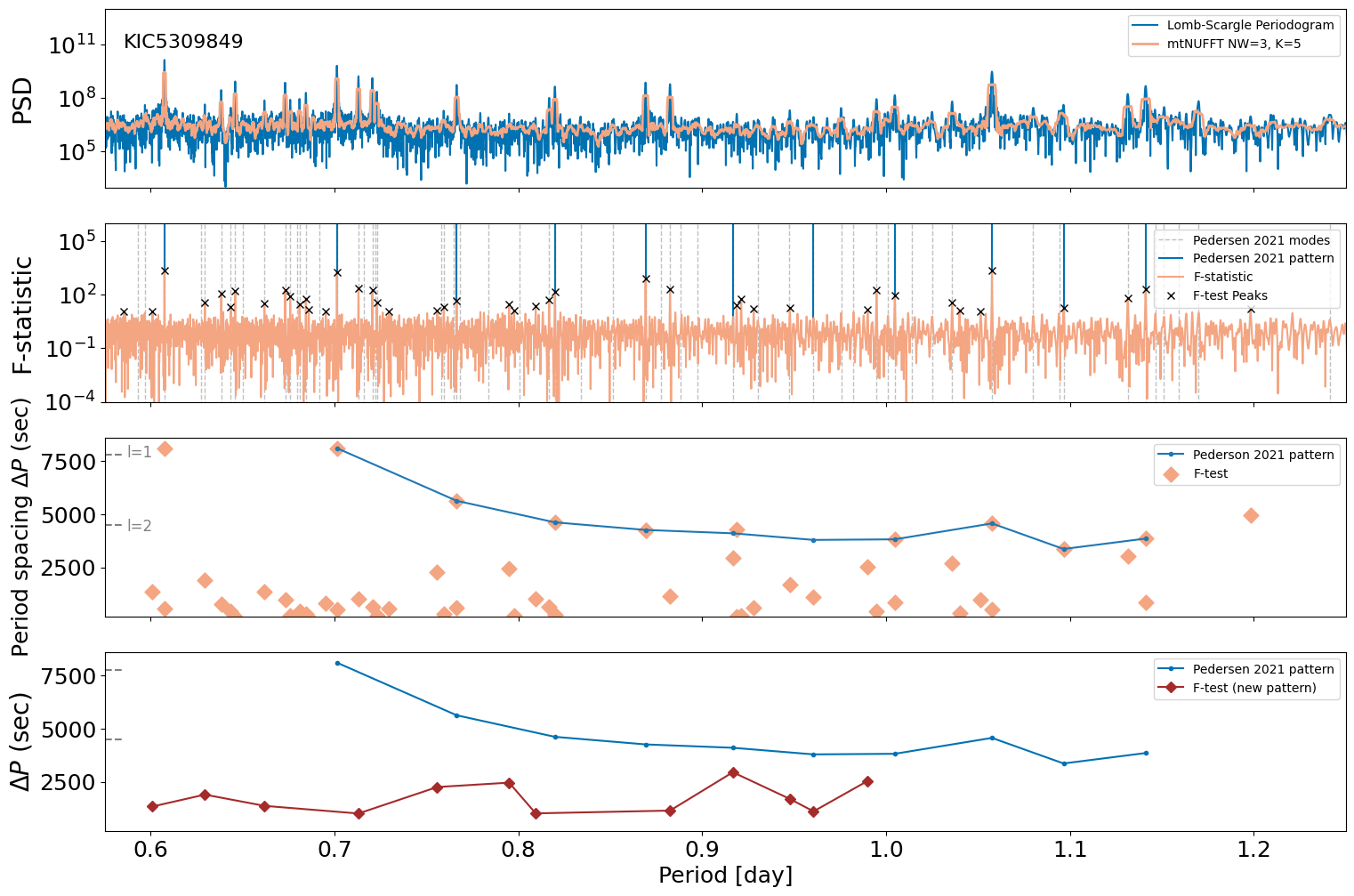}
      \caption{Comparison of our F-test detections with the period spacing patterns in \citetalias{pedersen_2021} for KIC5309849.}
         \label{fig:КIC5309849}
\end{figure*}

\begin{figure*}
   \centering
   \includegraphics[width=0.9\linewidth]{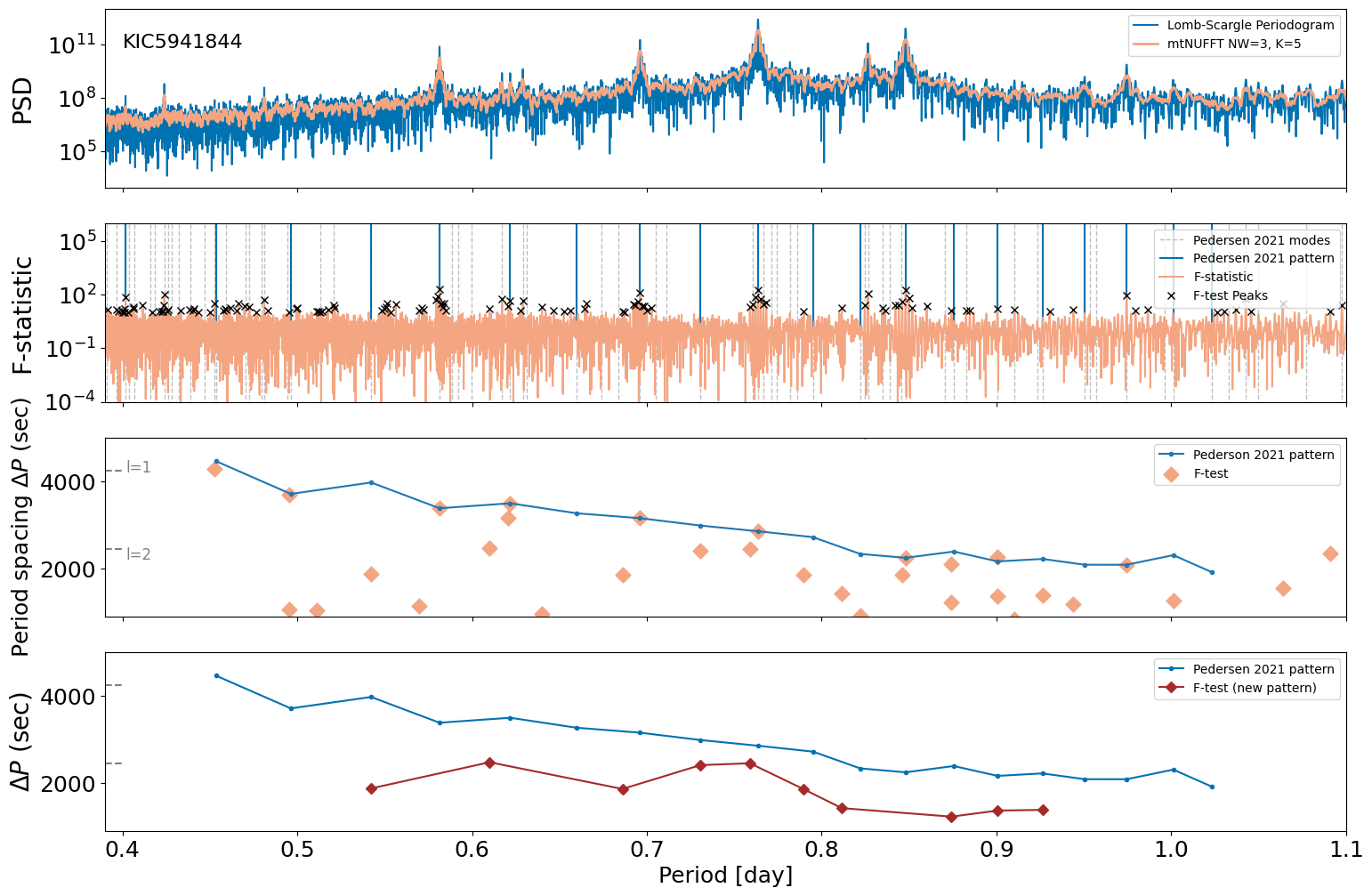}
      \caption{Comparison of our F-test detections with the period spacing patterns in \citetalias{pedersen_2021} for KIC5941844.}
         \label{fig:KIC5941844}
\end{figure*}

\begin{figure*}
   \centering
   \includegraphics[width=0.9\linewidth]{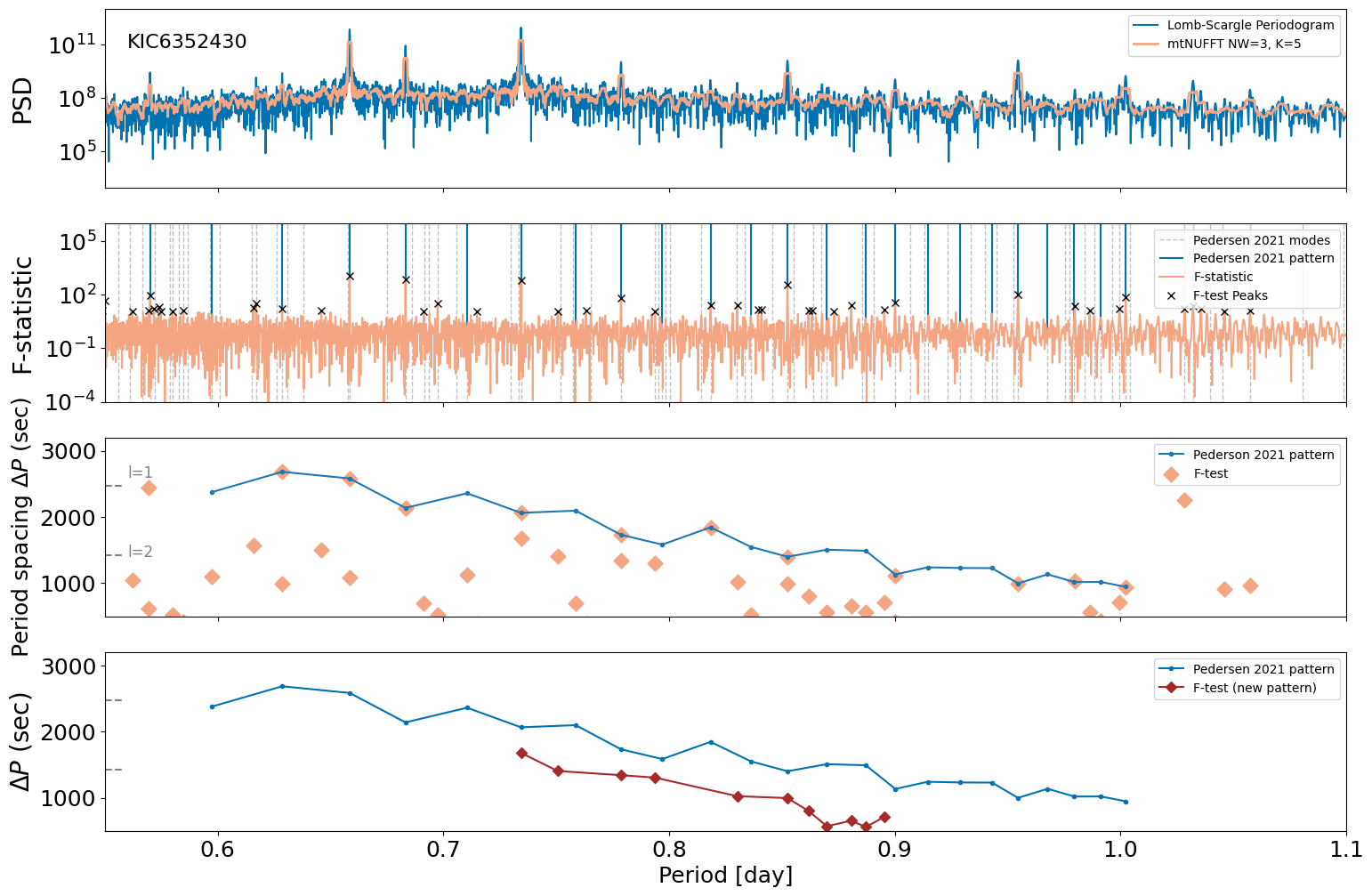}
      \caption{Comparison of our F-test detections with the period spacing patterns in \citetalias{pedersen_2021} for KIC6352430.}
         \label{fig:KIC6352430}
\end{figure*}

\begin{figure*}
   \centering
   \includegraphics[width=0.9\linewidth]{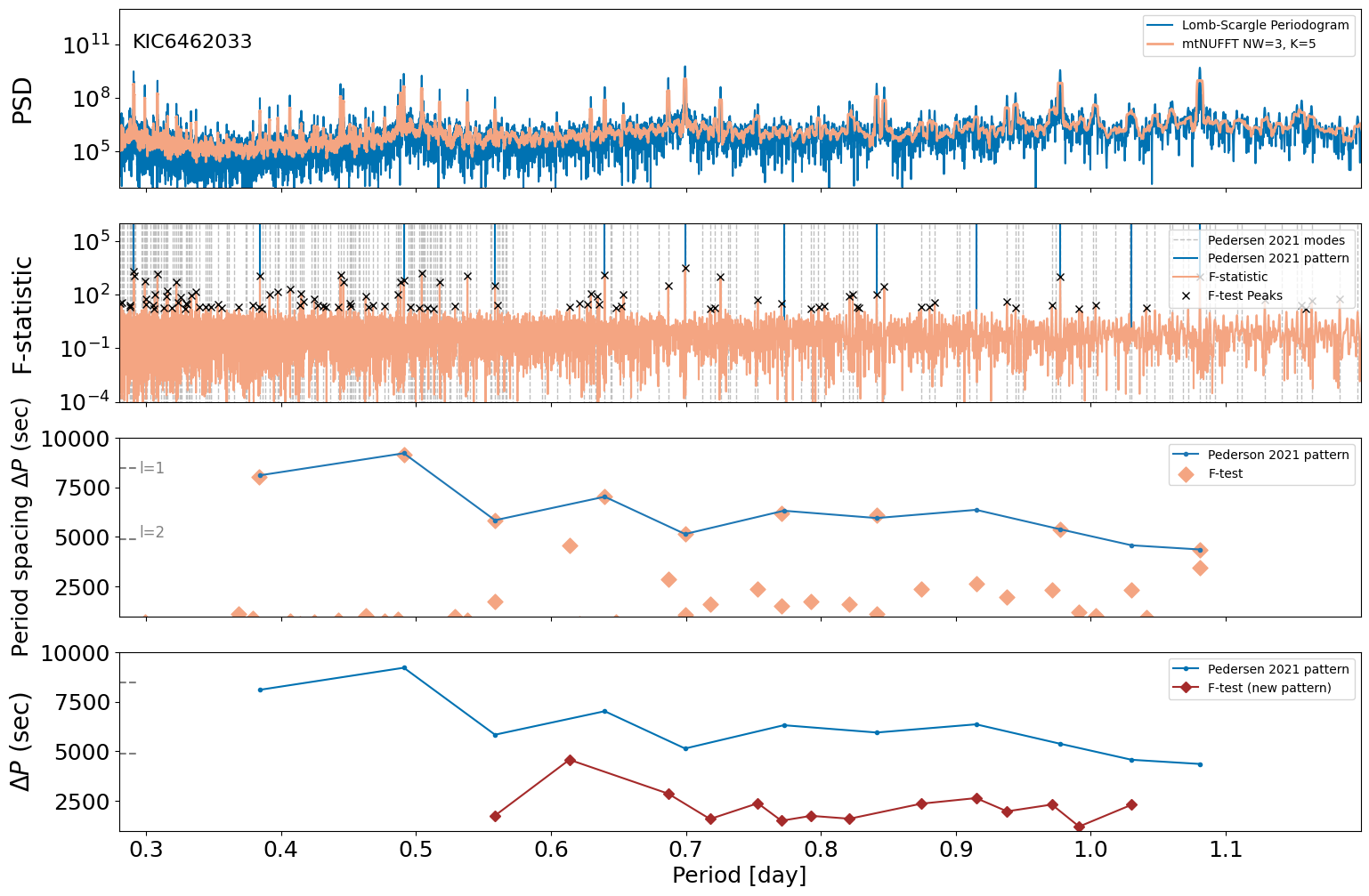}
      \caption{Comparison of our F-test detections with the period spacing patterns in \citetalias{pedersen_2021} for KIC6462033.}
         \label{fig:KIC6462033}
\end{figure*}

\begin{figure*}
   \centering
   \includegraphics[width=0.9\linewidth]{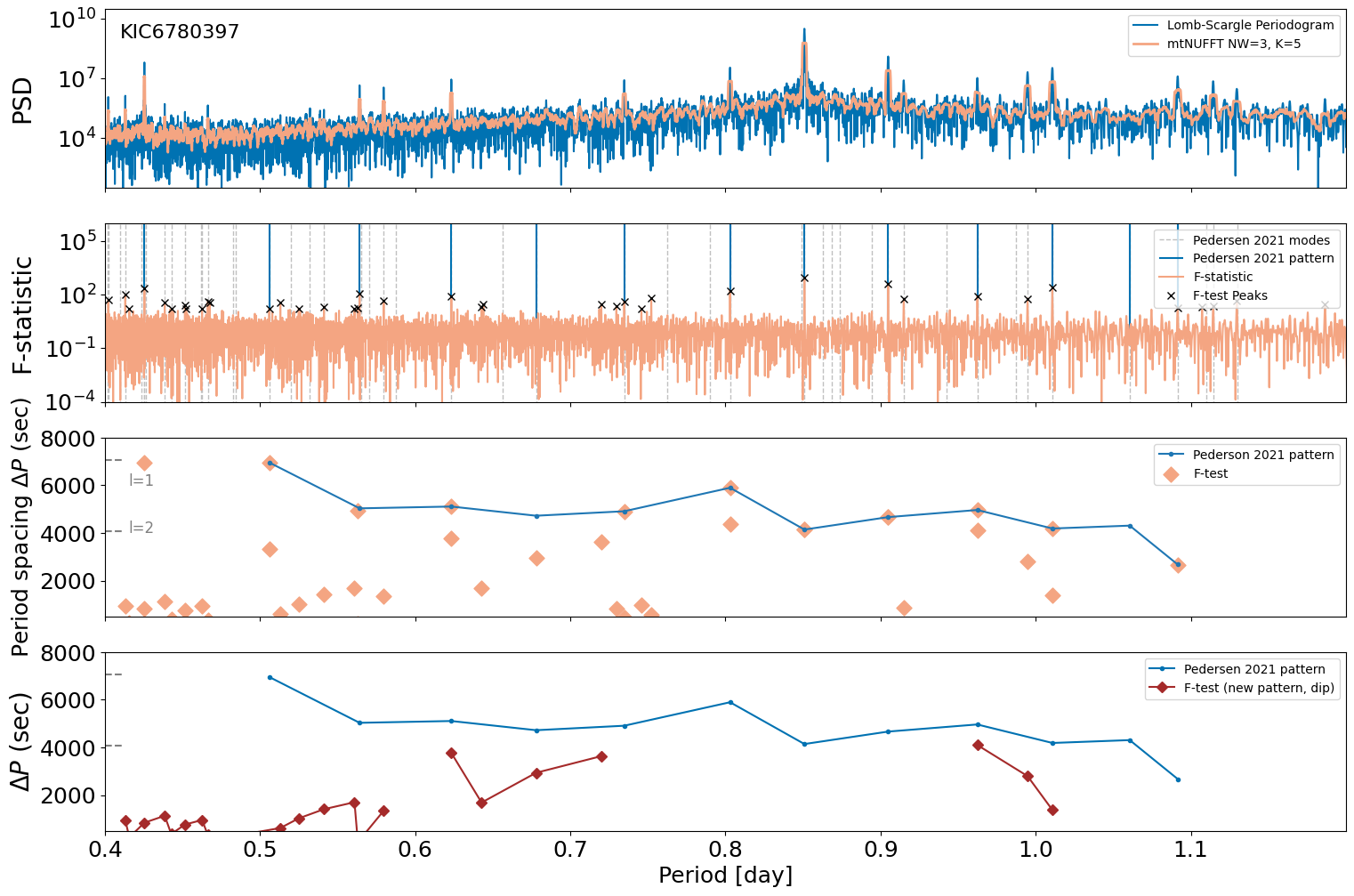}
      \caption{Comparison of our F-test detections with the period spacing patterns in \citetalias{pedersen_2021} for KIC6780397.}
         \label{fig:KIC6780397}
\end{figure*}

\begin{figure*}
   \centering
   \includegraphics[width=0.9\linewidth]{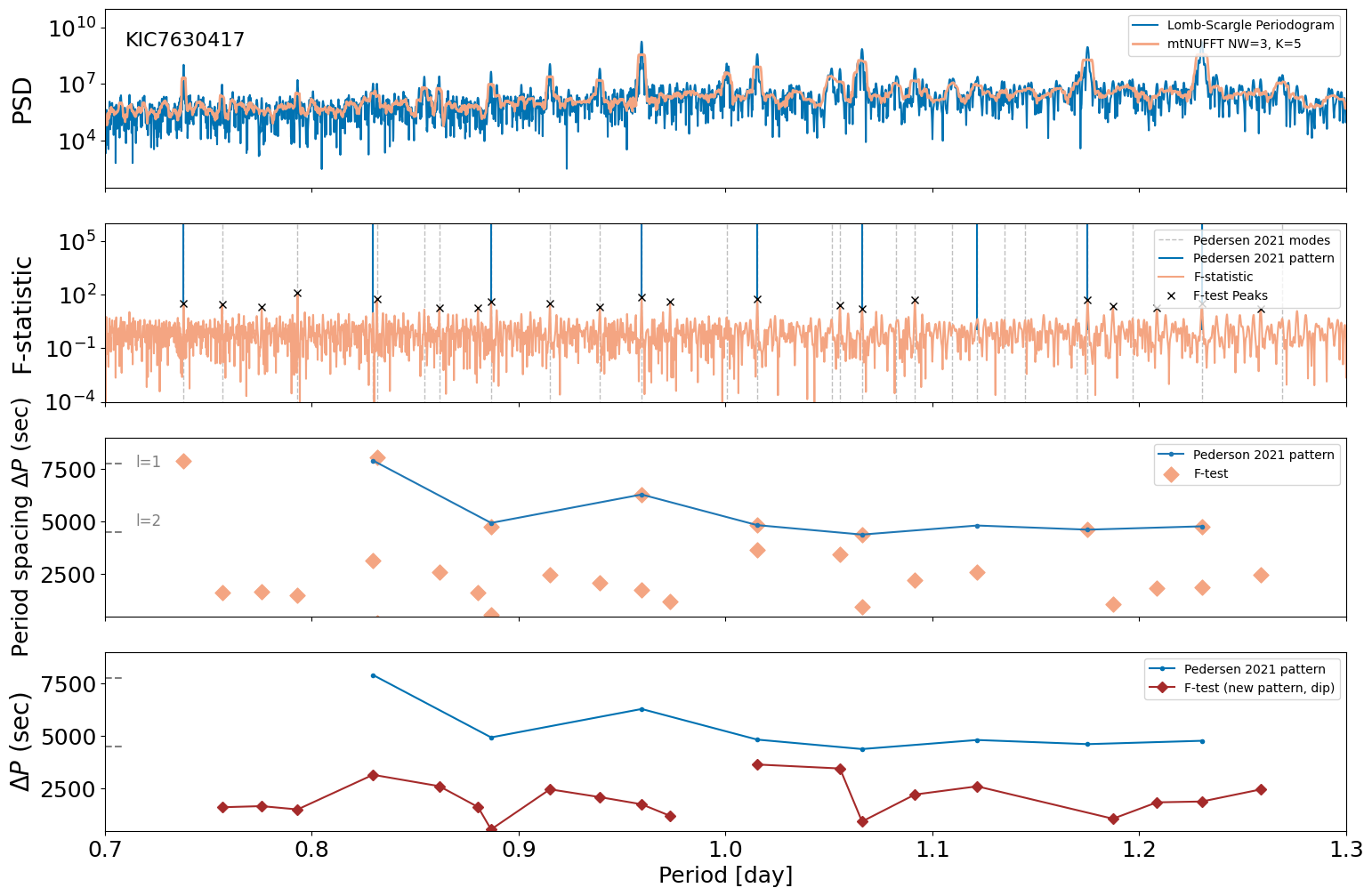}
      \caption{Comparison of our F-test detections with the period spacing patterns in \citetalias{pedersen_2021} for KIC7630417.}
         \label{fig:KIC7630417}
\end{figure*}

\begin{figure*}
   \centering
   \includegraphics[width=0.9\linewidth]{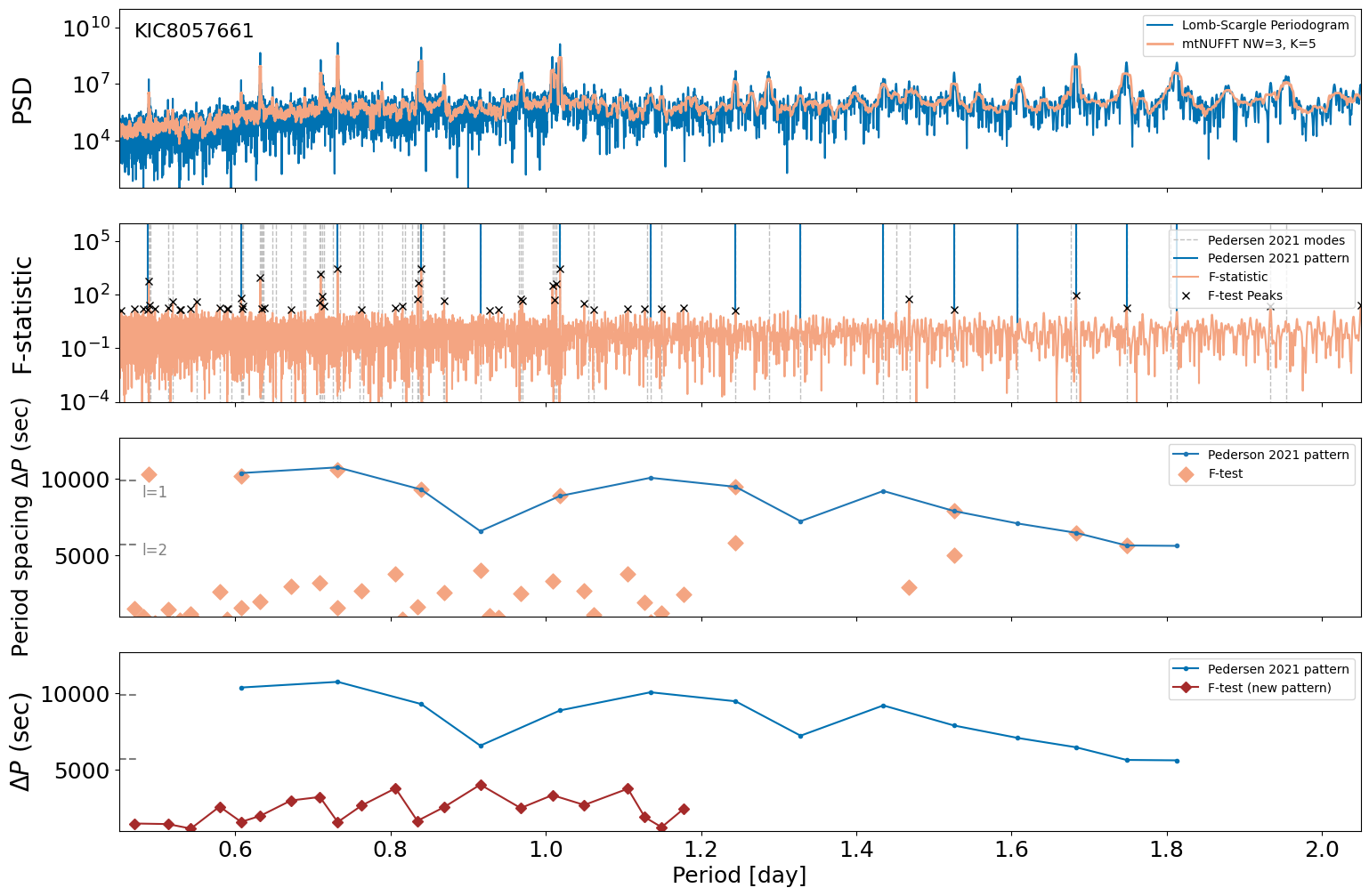}
      \caption{Comparison of our F-test detections with the period spacing patterns in \citetalias{pedersen_2021} for KIC8057661.}
         \label{fig:KIC8057661}
\end{figure*}

\begin{figure*}
   \centering
   \includegraphics[width=0.9\linewidth]{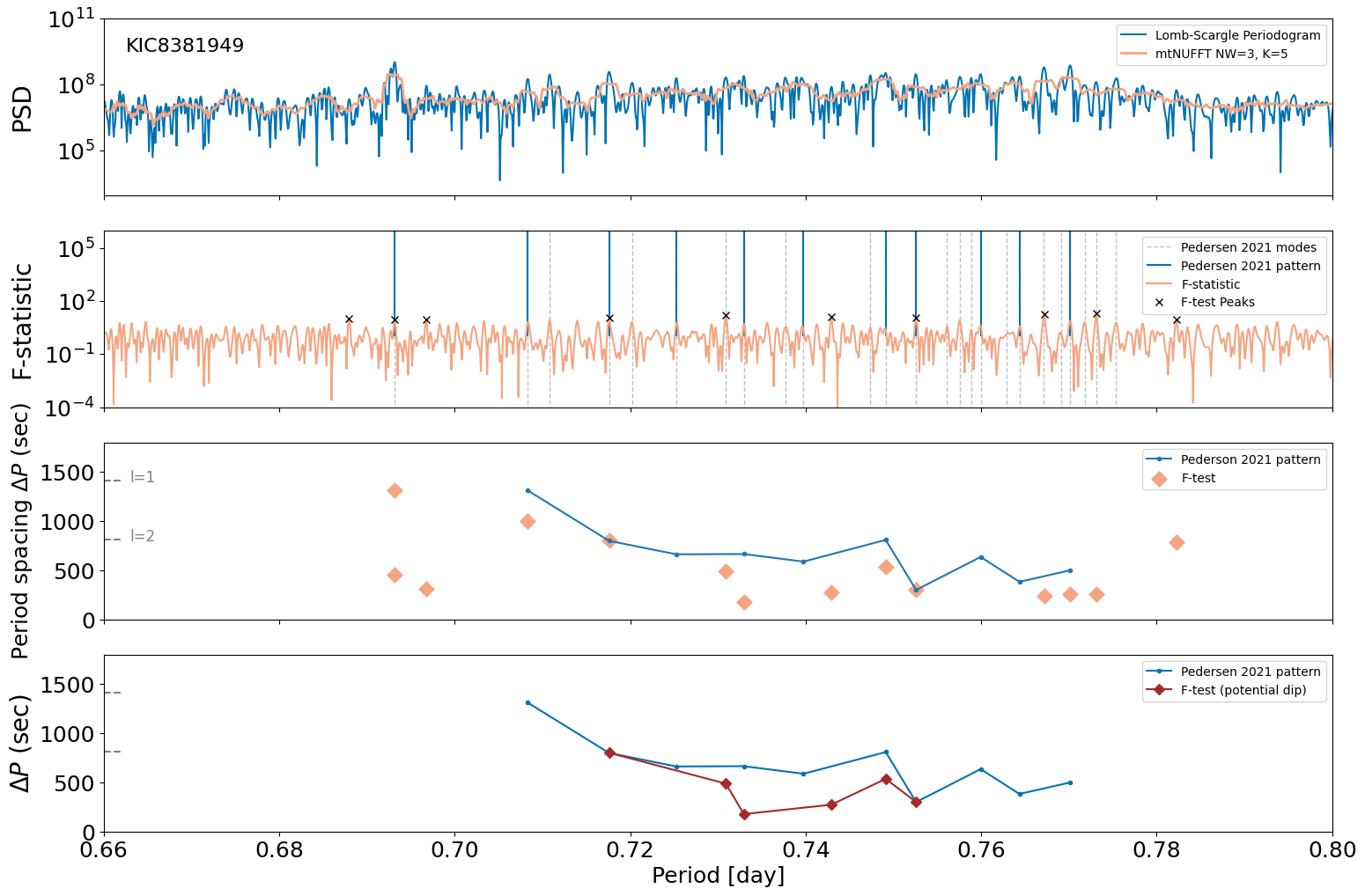}
      \caption{Comparison of our F-test detections with the period spacing patterns in \citetalias{pedersen_2021} for KIC8381949.}
         \label{fig:KIC8381949}
\end{figure*}

\begin{figure*}
   \centering
   \includegraphics[width=0.9\linewidth]{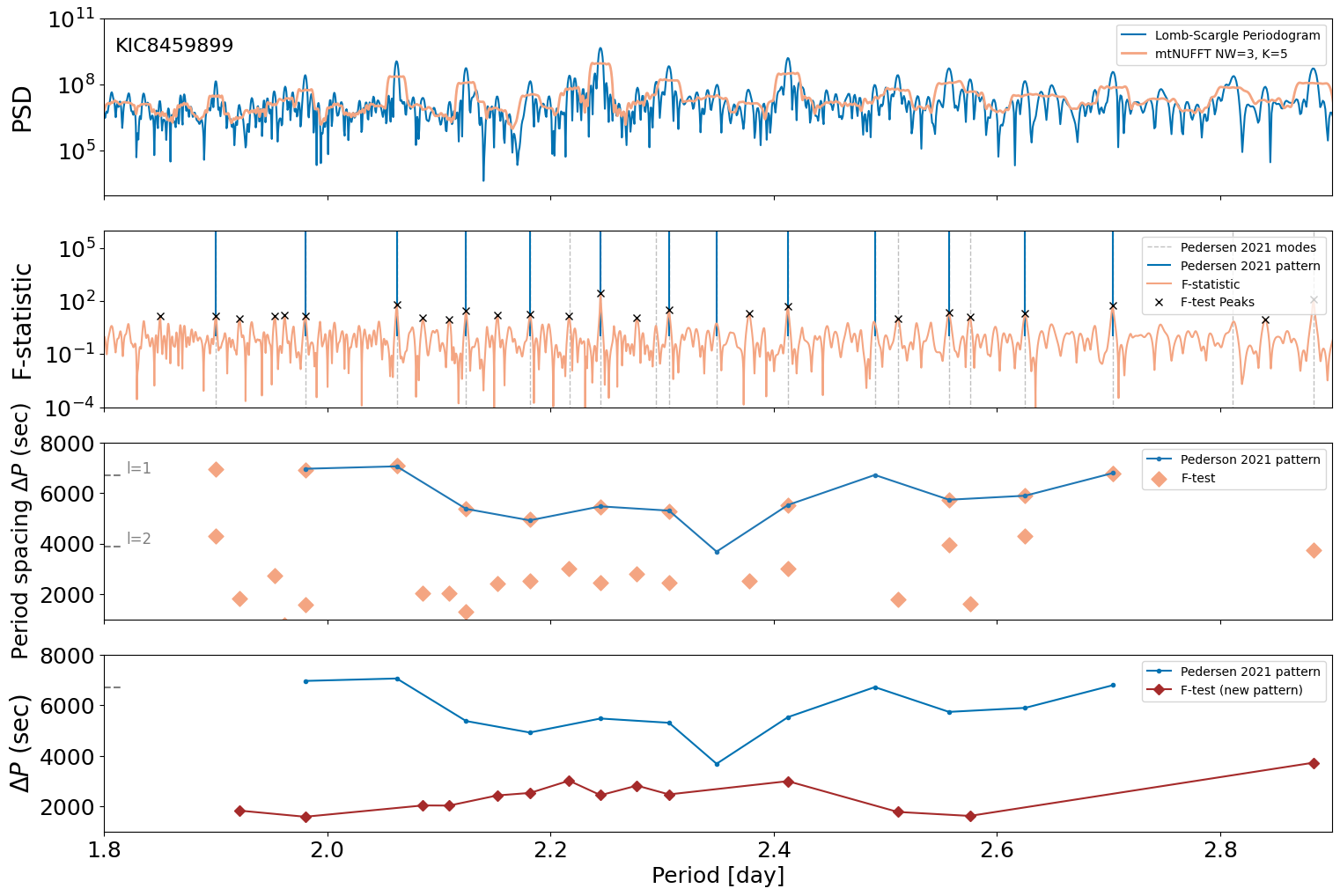}
      \caption{Comparison of our F-test detections with the period spacing patterns in \citetalias{pedersen_2021} for KIC8459899.}
         \label{fig:KIC8459899}
\end{figure*}

\begin{figure*}
   \centering
   \includegraphics[width=0.9\linewidth]{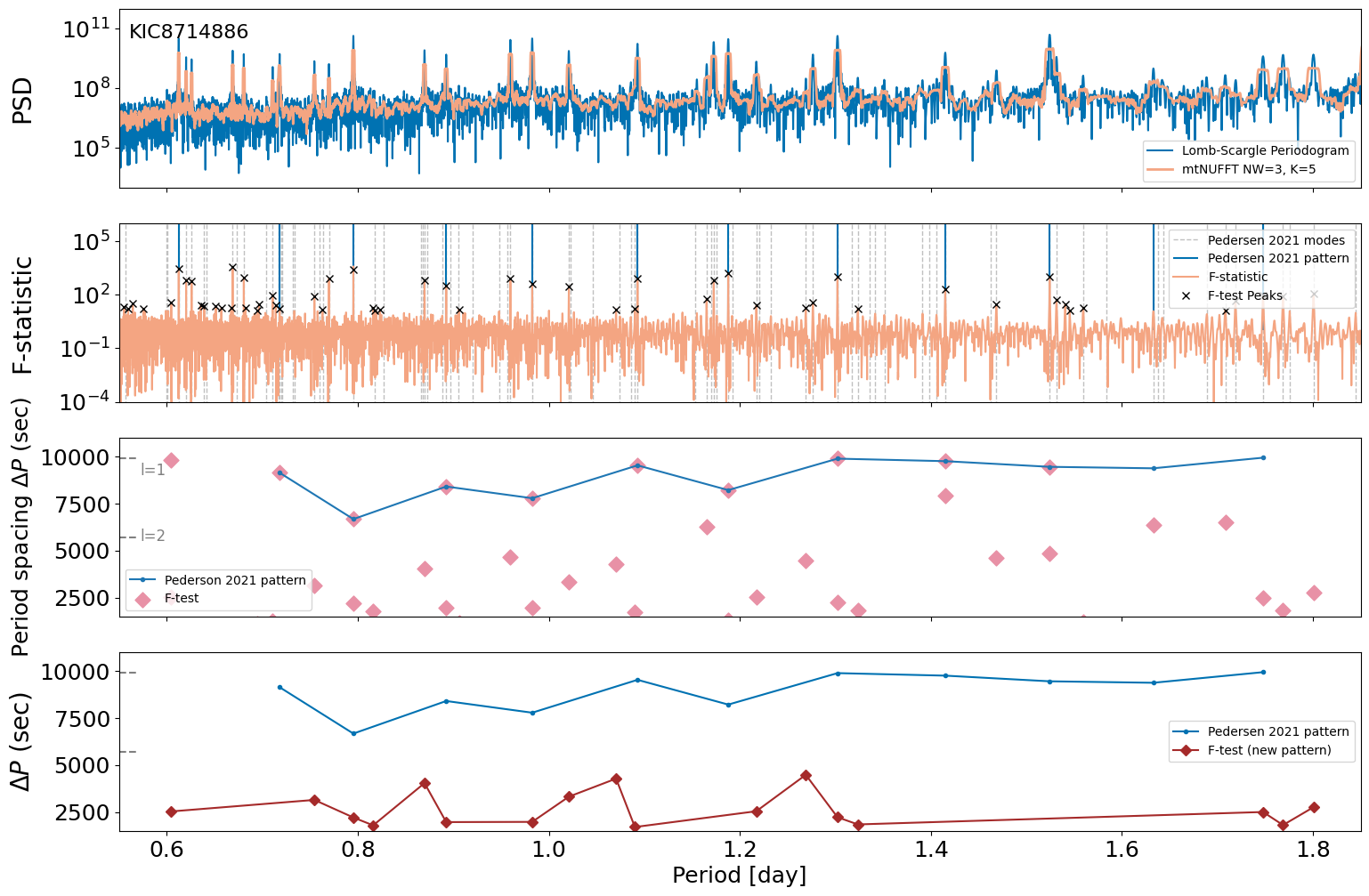}
      \caption{Comparison of our F-test detections with the period spacing patterns in \citetalias{pedersen_2021} for KIC8714886.}
         \label{fig:KIC8714886}
\end{figure*}

\begin{figure*}
   \centering
   \includegraphics[width=0.9\linewidth]{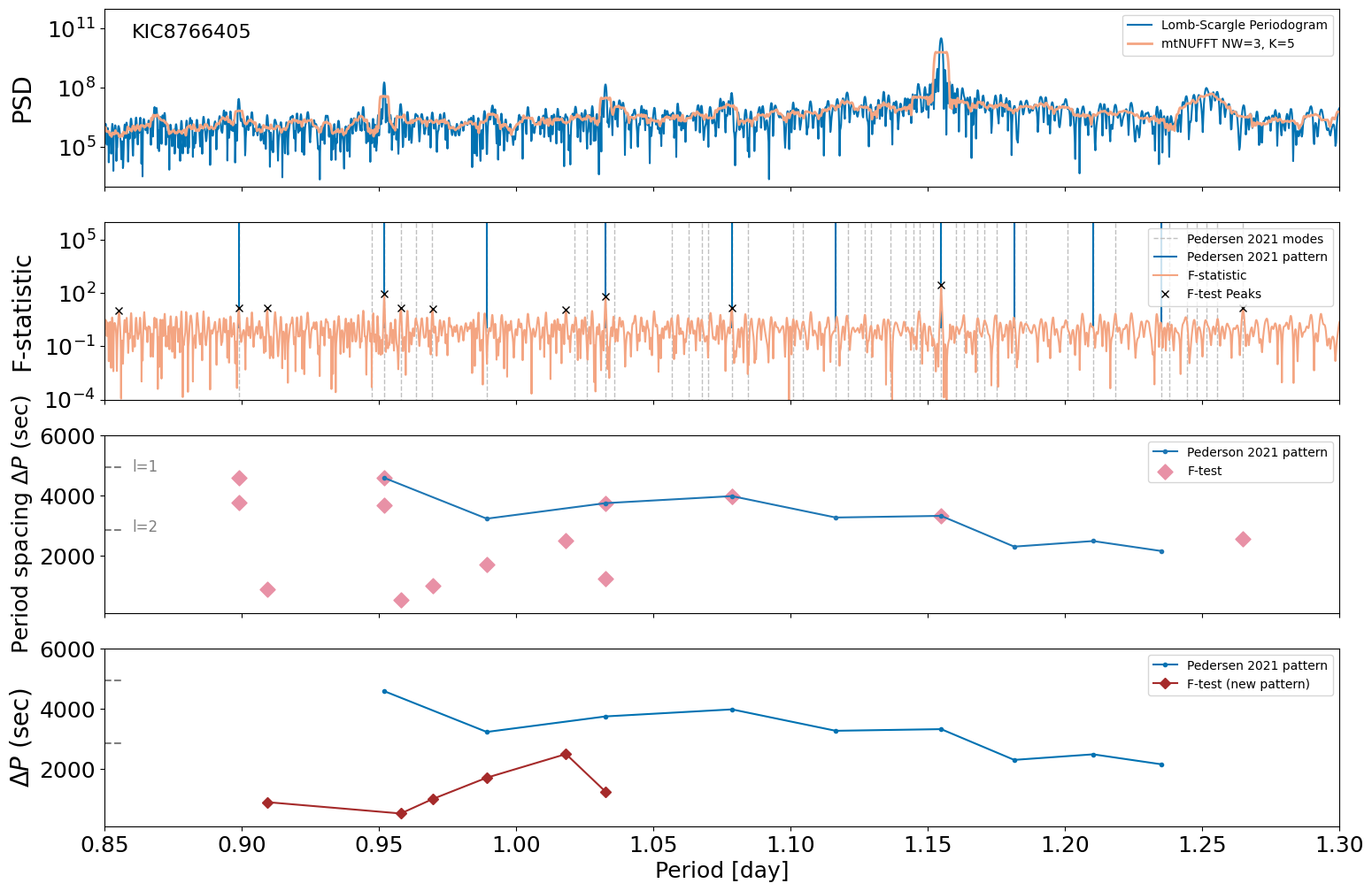}
      \caption{Comparison of our F-test detections with the period spacing patterns in \citetalias{pedersen_2021} for KIC8766405.}
         \label{fig:KIC8766405}
\end{figure*}

\begin{figure*}
   \centering
   \includegraphics[width=0.9\linewidth]{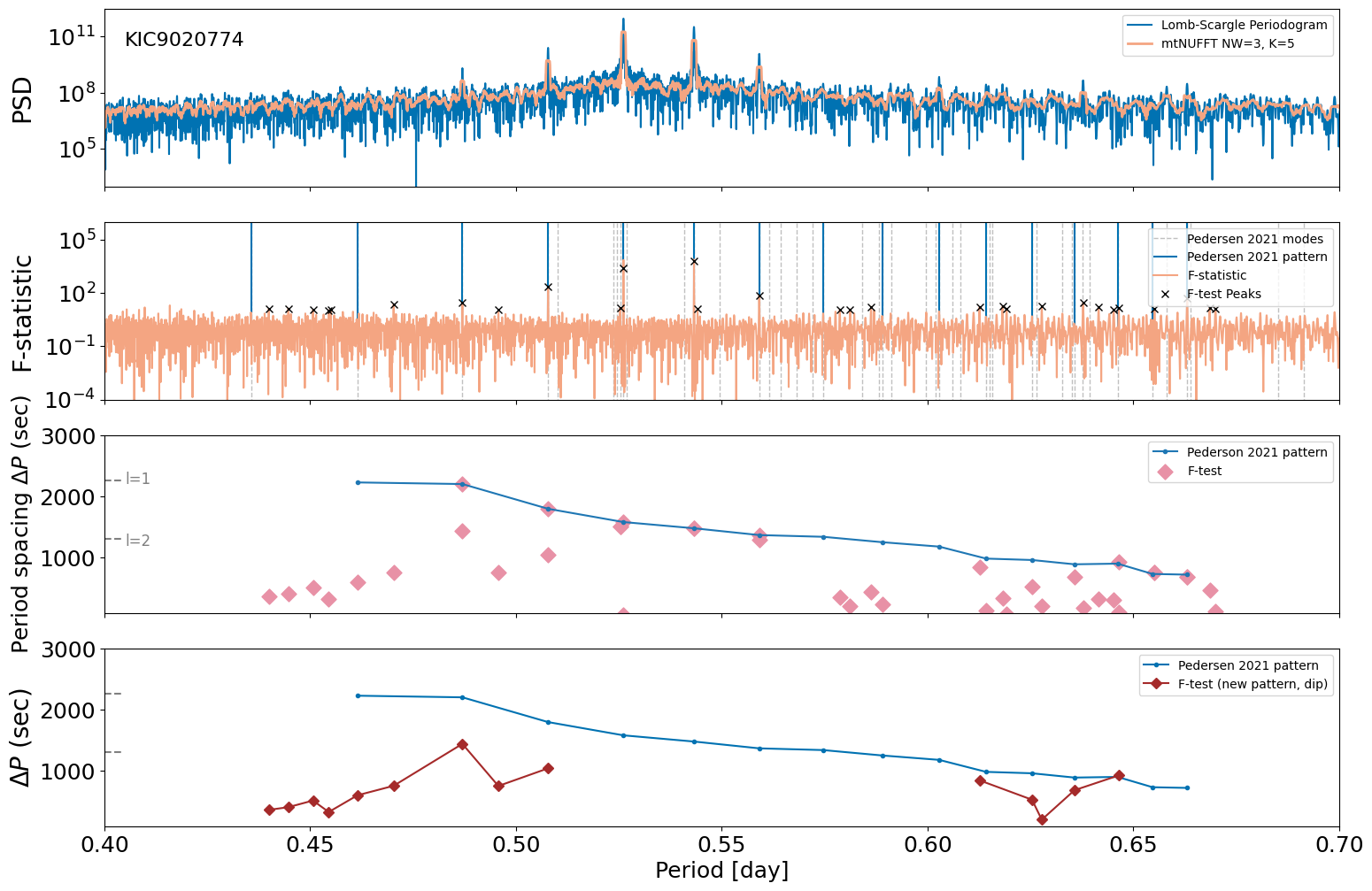}
      \caption{Comparison of our F-test detections with the period spacing patterns in \citetalias{pedersen_2021} for KIC9020774.}
         \label{fig:KIC9020774}
\end{figure*}

\begin{figure*}
   \centering
   \includegraphics[width=0.9\linewidth]{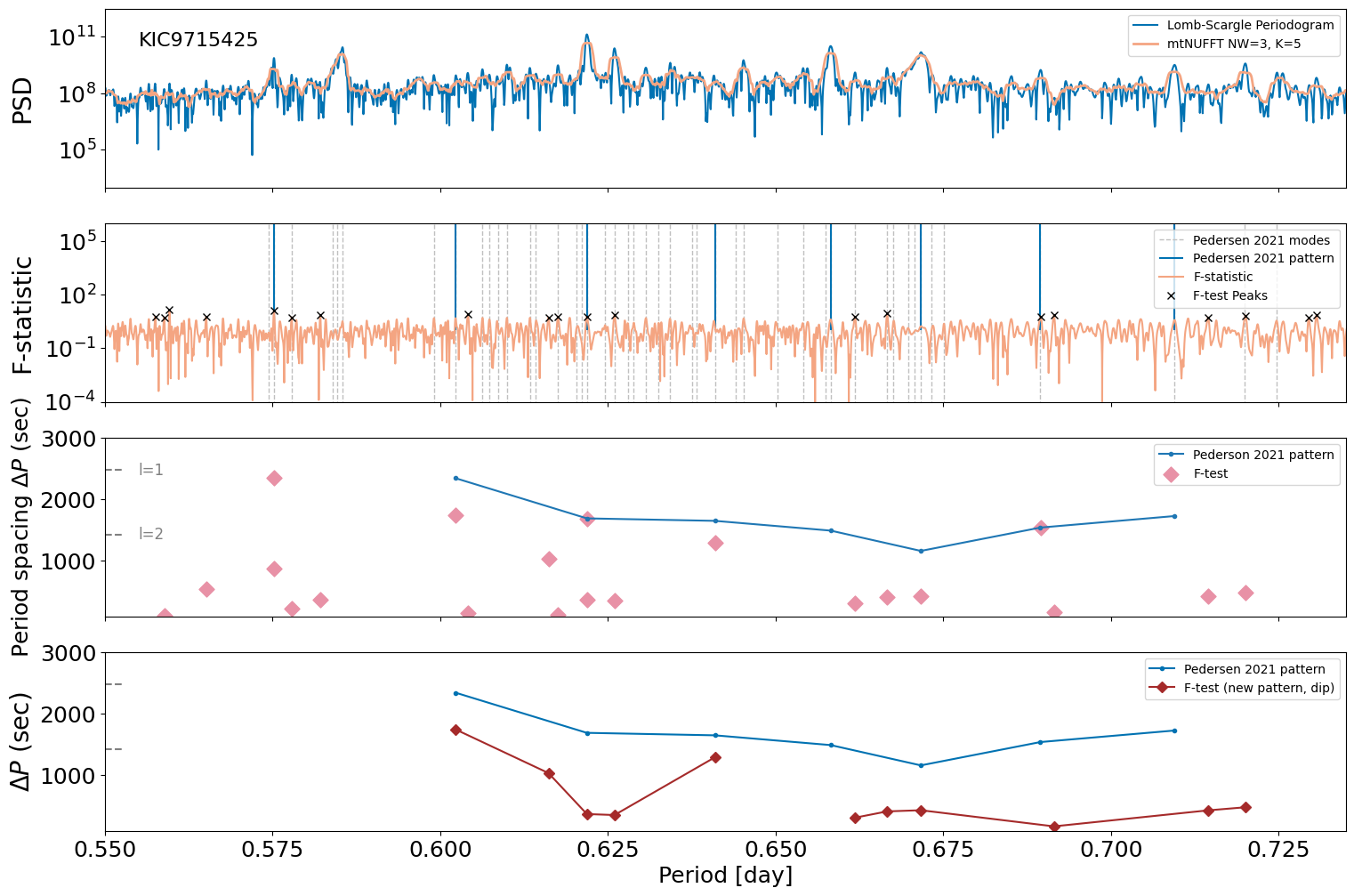}
      \caption{Comparison of our F-test detections with the period spacing patterns in \citetalias{pedersen_2021} for KIC9715425.}
         \label{fig:KIC9715425}
\end{figure*}

\begin{figure*}
   \centering
   \includegraphics[width=0.9\linewidth]{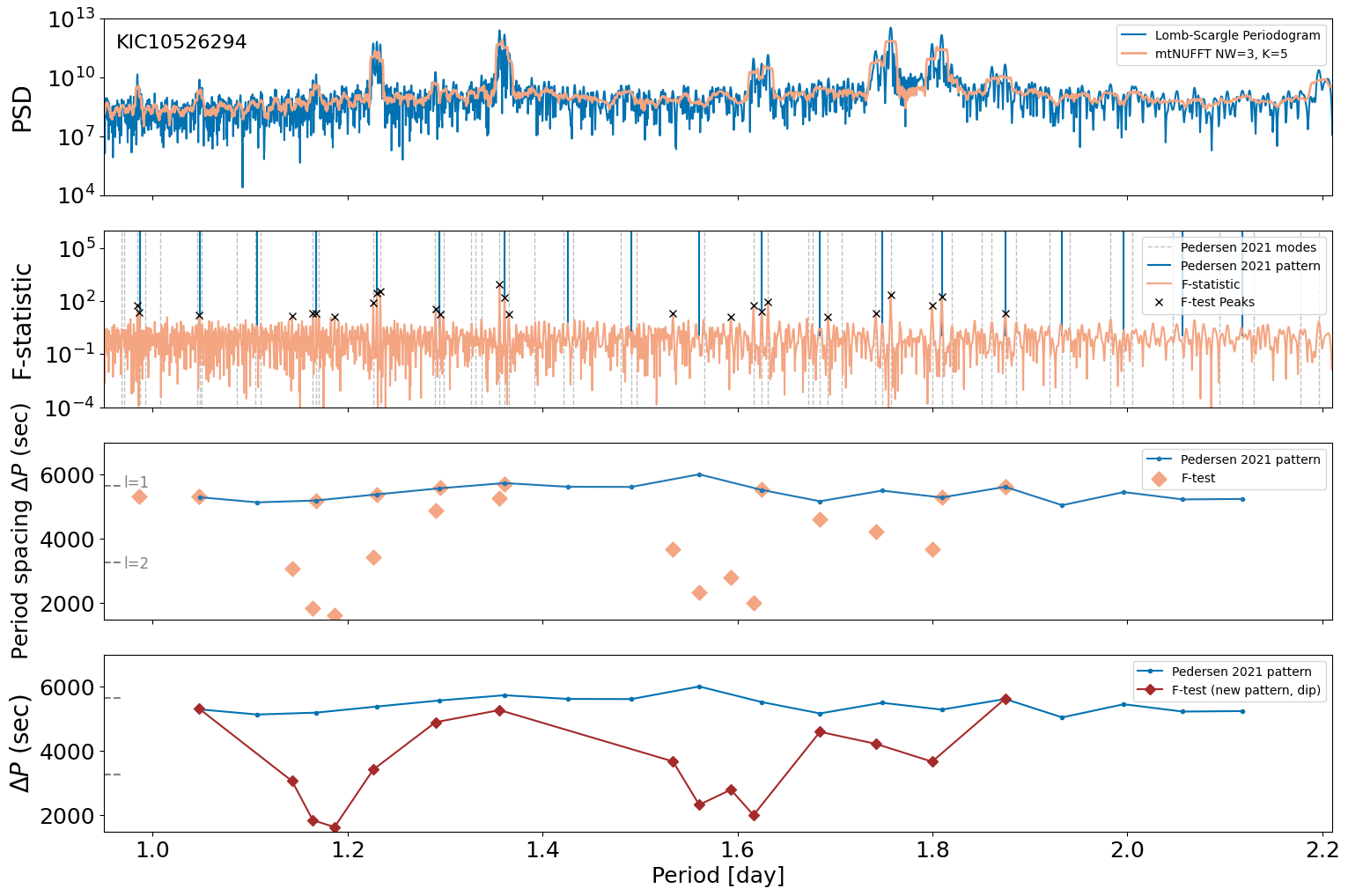}
      \caption{Comparison of our F-test detections with the period spacing patterns in \citetalias{pedersen_2021} for KIC10526294.}
         \label{fig:KIC10526294}
\end{figure*}

\begin{figure*}
   \centering
   \includegraphics[width=0.9\linewidth]{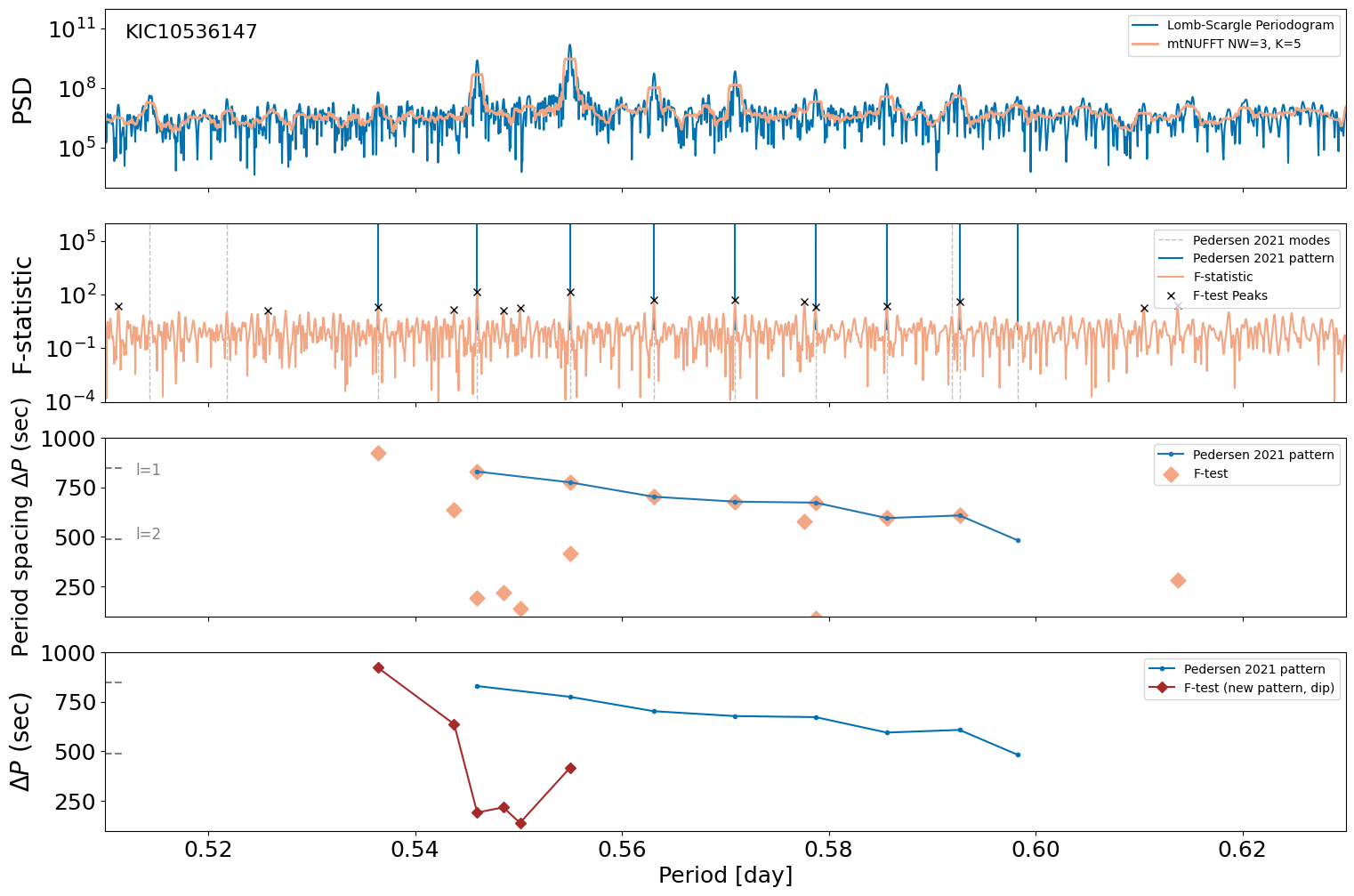}
      \caption{Comparison of our F-test detections with the period spacing patterns in \citetalias{pedersen_2021} for KIC10536147.}
         \label{fig:KIC10536147}
\end{figure*}

\begin{figure*}
   \centering
   \includegraphics[width=0.9\linewidth]{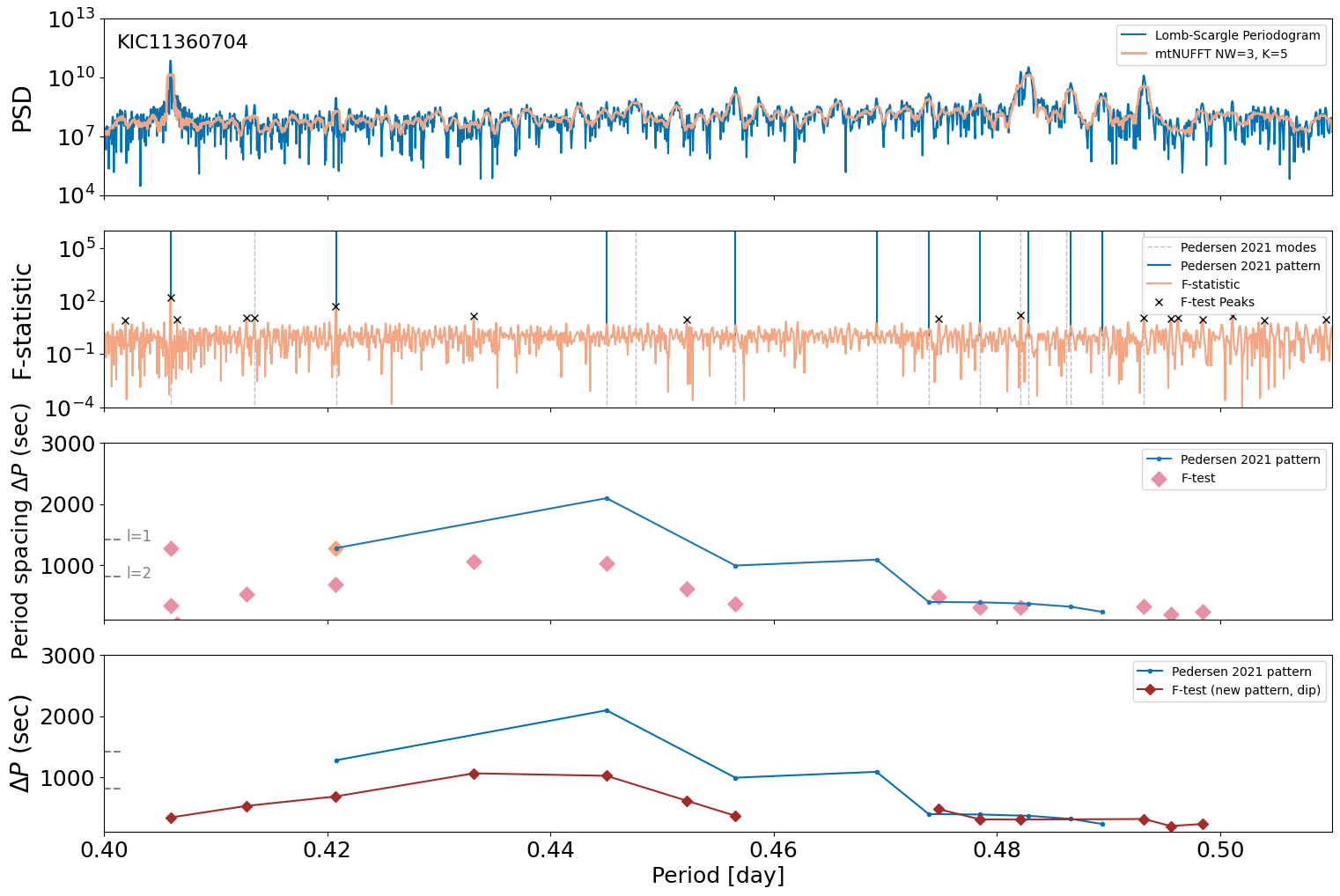}
      \caption{Comparison of our F-test detections with the period spacing patterns in \citetalias{pedersen_2021} for KIC11360704.}
         \label{fig:KIC11360704}
\end{figure*}

\begin{figure*}
   \centering
   \includegraphics[width=0.9\linewidth]{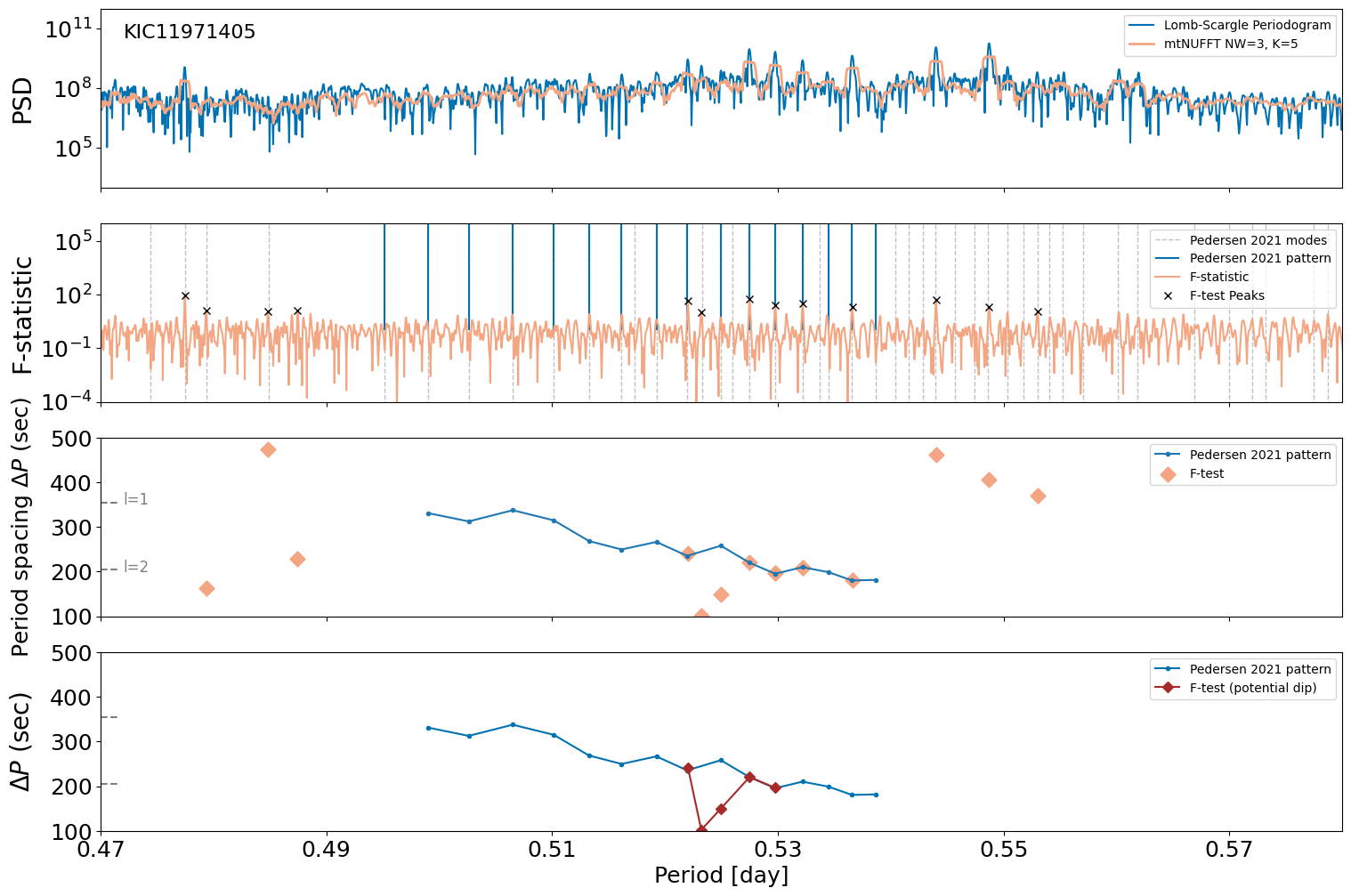}
      \caption{Comparison of our F-test detections with the period spacing patterns in \citetalias{pedersen_2021} for KIC11971405.}
         \label{fig:KIC11971405}
\end{figure*}

\end{appendix}

\end{document}